\documentclass{article}
\usepackage{amssymb}
\usepackage{cite}
\usepackage{geometry}
\usepackage{hyperref}
\usepackage{url,slashed}
\newcommand\fft[2]{\frac{#1}{#2}}
\newcommand\ft[2]{{\textstyle\frac{#1}{#2}}}
\newcommand\Tr{\mathrm{tr\,}}
\newcommand\tr{\mathrm{tr\,}}
\mathchardef\mhyphen="2D

\newcommand{\HH}{\mathcal {H}}
\newcommand{\GG}{\mathcal {G}}
\newcommand{\FF}{\mathcal {F}}
\newcommand{\hGG}{\hat \mathcal {G}}
\newcommand{\hG}{\hat G}
\newcommand{\hnabla}{\hat {\nabla}}

\newcommand{\omp}{\Omega_+}
\newcommand{\omm}{\Omega_-}
\newcommand{\ompm}{\Omega_{\pm}}

\newcommand\beq{\begin{equation}}
\newcommand\eeq{\end{equation}}
\newcommand\bea{\begin{eqnarray}}
\newcommand\eea{\end{eqnarray}}
\newcommand\nn{\nonumber}

\makeatletter
\@addtoreset{equation}{section}
\makeatother

\begin{document}
\setcounter{page}{0}
\begin{titlepage}
\titlepage

\begin{flushright}
IPhT-t13/043 \qquad MCTP-13-09
\end{flushright}

\vspace{30pt}
\begin{center}

{\Large {\bf Higher-derivative couplings in string theory:  \\
\vspace{10pt}  dualities and the $B$-field}}

\vspace{25pt}

James T. Liu$^a$ and Ruben Minasian$^b$

\vspace{20pt}

{${}^a$\it Michigan Center for Theoretical Physics\\
Randall Laboratory of Physics, The University of Michigan\\
Ann Arbor, MI 48109--1040, USA}

\vspace{10pt}

{${}^b$\it Institut de Physique Th\'eorique, CEA/Saclay\\
91191 Gif-sur-Yvette Cedex, France}

\vspace{40pt}

\underline{ABSTRACT}
\end{center}

The first quantum correction to the IIA string effective action arises at the eight-derivative
level and takes the schematic form $(t_8t_8-\fft18\epsilon_{10}\epsilon_{10})R^4+B_2\wedge X_8$.
This correction, however, cannot be complete by itself, as it is neither supersymmetric nor T-duality
covariant.  We reexamine these eight-derivative couplings and conjecture that the simple
replacement $R\to R(\Omega_+)$, where $\Omega_+=\Omega+\fft12\mathcal H$ is the
connection with
torsion, nearly completely captures their dependence on the $B$-field. The exception is in
the odd-odd spin structure sector, where additional terms are needed. We present here a complete result at the level of the five-point function and a partial one for the six-point function.  
Further evidence for this conjecture comes from  considering T-duality as well as heterotic/IIA duality beyond leading order. Finally, we discuss the eleven-dimensional lift of the modified one-loop type IIA couplings.

\vfill
\begin{flushleft}
{April 10, 2013}\\
\vspace{.5cm}
\end{flushleft}
\end{titlepage}

\newpage

\tableofcontents
\pagebreak

\section{Introduction}

Since string theory reduces in its low energy limit to an effective supergravity theory, one
must go beyond this limit in order to capture truly stringy behavior.  Such stringy corrections
may be encoded in both the $\alpha'$ and the genus expansion, with the former generating
higher derivative corrections to the supergravity and the latter corresponding to string quantum
corrections in spacetime.  For the type II string, these corrections start at the eight derivative
level, and were first obtained at the tree level from four-graviton scattering
\cite{Schwarz:1982jn,Gross:1986iv} as well as from the $\sigma$-model beta function approach
\cite{Grisaru:1986px,Grisaru:1986dk,Grisaru:1986kw,Freeman:1986br,Grisaru:1986vi,Freeman:1986zh}.
These tree level corrections take the form
\begin{equation}
e^{-1}\mathcal L\sim e^{-2\phi}(t_8t_8R^4+\ft18\epsilon_{10}\epsilon_{10}R^4).
\end{equation}
The one-loop expression has a similar structure in the CP-even sector \cite{Sakai:1986bi}
\begin{equation}
e^{-1}\mathcal L_{\rm CP\mhyphen even}\sim (t_8t_8R^4\mp\ft18\epsilon_{10}\epsilon_{10}R^4),
\end{equation}
where the top (bottom) sign is for the IIA (IIB) theory.  In addition, the IIA theory has a CP-odd
one-loop term \cite{Vafa:1995fj, DLM}
\begin{equation}
\mathcal L_{\rm CP\mhyphen odd}\sim B_2\wedge[\tr R^4-\ft14(\tr R^2)^2].
\label{eq:Bwed}
\end{equation}

The above expressions, however, cannot be complete, as supersymmetry will necessarily
bring in additional eight-derivative terms built from the other fields in the supergravity
multiplet.  For the type II string, this includes the $B$ field and dilaton in the NSNS sector,
the $p$-form field strengths in the RR sector, and the fermionic superpartners.  Although it
would be desirable to obtain a complete supersymmetric invariant at the eight-derivative level,
the goal of this paper is more modest.  In particular, we reexamine the one-loop $R^4$ terms
in the type II effective theories and attempt to establish the completion of these terms involving
the $B$ field. 

We work in the covariant worldsheet formulation of the type II string, and find that all that is
required in the even-even and the (even-odd $+$ odd-even) spin structure sectors is the
replacement of the Riemann tensor by a four-index tensor computed using a connection
with torsion
\begin{equation}
R_{\mu\nu}{}^{\alpha\beta}\to R_{\mu\nu}{}^{\alpha\beta}(\Omega_+)\equiv
R_{\mu\nu}{}^{\alpha\beta}
+\nabla_{[\mu}H_{\nu]}{}^{\alpha\beta}+\ft12H_{[\mu}{}^{\alpha\gamma}H_{\nu]\gamma}{}^\beta.
\end{equation}
We argue that, in all except the odd-odd spin structure sector, this replacement captures the
full structure of the corrections to all powers in $H=dB$.  A similar replacement is needed in
the odd-odd sector.  However, this is not sufficient, and new kinematic structures beyond the
standard $\epsilon \epsilon R^4$ appear in terms involving higher powers of $H$, such as
$H^2 R^3$, $H^4 R^2$, etc.  We present here a complete result at the level of the five-point
function and a partial one for the six-point function.  Computations involving the one-loop
five-point function can be found in the literature
\cite{Frampton:1986gi,Lam:1986kg,Montag:1992dm,Peeters:2001ub,Richards:2008jg,Richards:2008sa},
but the results pertaining to the six-point function are new.

Stringy higher derivative quantum corrections have been studied for a long time and have
proved to be important in a number of contexts. Yet our knowledge of these terms, even the
simplest one-loop contributions to the effective action, is far from being complete.  It is worth
expanding on some of our motivations:

\vskip 0.3cm
\noindent {\bf Supersymmetry:} While it is believed that different super-invarints made of
curvatures should exist, clearly the $B$ field and dilaton (at least in the Einstein frame)
are needed in order to have minimal supersymmetry in $D=10$. Note that, e.g., the
verification of the fact that $B \wedge X_8(R)$ and $(t_8t_8- \fft18\epsilon_{10} \epsilon_{10} ) R^4$
are related by supersymmetry is done only at the linearized level.  

The $R^4$ couplings in ten dimensions play a crucial role in understanding the quantum
moduli space of $\mathcal{N}=2$ theories, and it can be shown that all $R^2$ terms in
$\mathcal{N}=2$ theories descend from these \cite{AFMN, AMTV}.  The supersymmetry of the former can be
checked explicitly. Indeed, these arise from the expansion of
$F_1 W^2|_{\mathrm{F\mhyphen terms}}$, where $F_1$ is a a function of the chiral vector
superfields, and $W$ is the $\mathcal{N}=2$ chiral Weyl superfield.  Note that in the
four-dimensional  $\mathcal{N}=2$ context, the $B$-field is in the hypermatter and is no
longer in the same multiplet as the graviton, and hence is a priori irrelevant for the
above-mentioned couplings.  However controlling at least the $H^2 R^3$-order terms is
important for better understanding of the hypermultiplet moduli spaces and the
higher-derivative mixing of vector and hypermultiplets.

\vskip 0.3cm
\noindent {\bf Dualities:} The couplings involving only the curvature terms suffer from at
least two types of problems:
\begin{itemize}
\item Clearly expressions made solely out of curvatures cannot have reasonable T-duality
properties; in particular there is no hope of constructing any invariant.
\item Since the duality interchanges Bianchi identities (tree-level) and equations of motion
(one-loop), higher curvature terms are important for $\mathcal{N}=4$ heterotic/Type II duality.
However the heterotic Bianchi identity is highly non-linear, and the curvature terms appearing
there are constructed out of a connection with torsion. So far the duality has been checked at
the linearized level, completely ignoring the dependance of the heterotic Bianchi identity on $H$.
\end{itemize}
These two dualities will provide important tests for our results. 

\vskip 0.3cm
\noindent {\bf Flux backgrounds:} Even if higher curvature terms are often our best hope for
avoiding the no-go theorems for flux compactifications, they have remained largely unexplored.
Understanding the $H$-flux in one-loop terms has already helped with some AdS results, such
as the stringy calculation of $c-a$, but clearly more remains to be done. This will eventually be
important for both Minkowski and AdS compactifications.

\vskip 0.3cm
\noindent {\bf Testing generalised geometry:} Generalised geometry determined by the
$B$-field has proven to be extremely useful in understanding the (on and off shell) classical
superymmetry. In particular, the generalised Levi-Civita connection $\nabla^{B}$ is used
to form Dirac operators. On the mathematical side, the appropriate local index theorem with
torsionful connections have been known for some time \cite{BisTor}. It is a natural question to ask how much
generalised geometry captures the systematics of the string perturbation theory (and maybe
the so called exceptional generalised geometry of the full string theory). Understanding $R^4$
corrections would be the first step in this direction.

\vskip 0.3cm
The $B_2 \wedge X_8$ term, (\ref{eq:Bwed}), seems to be a good starting point for studying
the effects of the $B$-field, since it is related to anomalies and is robust. It is clear right away
that the only possible modification of $X_8$ is an addition of a closed term
$X_8 \, \mapsto X_8 + Y$ with $d Y =0$. Indeed, in the absence of NS5 branes, the coupling
should be invariant under the $B$ field gauge transformation $B \, \mapsto \, B + d \Lambda$. Moreover, the shift $Y$ must be exact. This is required by the invariance of the coupling under diffeomorphisms (in the absence of NS5). Moreover, a non-exact $Y$ would modify the anomaly cancellation mechanism since $X_8$ and the fivebrane worldvolume anomaly are related via the inflow mechanism. The spinorial cohomology arguments might provide another argument for $Y$ being exact.

We shall argue that not only is $Y$ exact, but that this exact shift is induced by the shift in the
connection, and the complete $X_8$ should be computed using a connection with torsion.
Replacing Riemann curvatures by torsionful ones is known not to work for the full set of
quantum corrections (or even at the leading two-derivative order) beyond the linearised level.
Once more, the special nature of $B_2 \wedge X_8$ plays an important role here, and as we
shall see, replacing the Levi-Civita connection by a torsional one
$\Omega \longrightarrow \Omega_{\pm }=\Omega \pm \ft12\mathcal H$ where
$\mathcal H^{\alpha\beta}=H_\mu{}^{\alpha\beta}dx^\mu$ works. 
Note that $X_8$ computed from the torsional connection will have all powers of torsion which
would clash with the CP-invariance of the effective action.  Fortunately, the GSO projection for
the IIA string simply restores the symmetry of the theory under the reversal of $B$, and in fact 
\beq
X_8(\Omega^{LC}) \,\, \longrightarrow \,\, \fft12 \left[X_8(\omp) + X_8(\omm) \right].
\eeq
One expects a relative minus sign for the IIB string, but as we shall argue, there the entire
coupling must vanish on-shell. This claim is tested by T-duality and heterotic/Type II duality
(with 16 supercharges), and backed by string calculations. This is the story of the contribution
from the even-odd $\pm$ odd-even worldsheet spin structures. 
 
How about the even-even and odd-odd spin structures (which we sometimes call ``CP-even"
contributions)? Note that the structure of these is such that they are automatically even in
powers of $H$ under the replacement $\Omega \longrightarrow \Omega_{\pm }$. Moreover,
as we shall show in the even-even sector, this replacement is sufficient and captures the
full string theoretic result. The odd-odd structure is more subtle, and a combination of
worldsheet calculation with different duality arguments is needed in order to (at least
partially) fix it.
  
Putting our findings together, one may conjecture how the two $\mathcal{N}=1$ superinvariants
will be completed with the $B$-field as well:
\bea
\label{superinv}
J_0(\Omega) = \left(t_8 t_8 + \ft18 \epsilon_{10} \epsilon_{10}\right)R^4 &\longrightarrow&
\left(t_8 t_8 + \ft18 \epsilon_{10} \epsilon_{10}\right)R^4(\Omega_+)
+ \ft 13 \epsilon_{10} \epsilon_{10} H^2 R^3 (\Omega_+) + ...  \nn \\
&& = J_0 (\Omega_+) + \Delta J_0(\Omega_+, H) \nn \\
J_1(\Omega) = t_8 t_8 R^4  -  \ft14 \epsilon_{10} t_8 B R^4 &\longrightarrow&
t_8 t_8 R^4 (\Omega_+)  -  \ft18 \epsilon_{10} t_8 B \left(R^4 (\Omega_+)
+ R(\omm) \right)= J_1(\Omega_+).\qquad
\eea
Note that $J_0 (\Omega_+) + \Delta J_0(\Omega_+, H)$ appears at tree level both in IIA and IIB
and at one loop in IIB, while $J_0 (\Omega_+) - 2 J_1(\Omega_+) + \Delta J_0(\Omega_+, H)$
appears at one loop in IIA. The detailed structure of $\Delta J_0(\Omega_+, H)$ is explained
in the body of the paper (notably in Section~\ref{sec:corrrr} and Section~\ref{sec:morecor}) and is only
partially fixed by our computations.  A summary of our results in ten dimensions is given in
Table~\ref{tbl:res}.

\begin{table}[t]
\centering
{\renewcommand{\arraystretch}{1.5}
\begin{tabular}{|c|l|l|}
\hline
&\hss No $B$  &\hss With $B$ \\ \hline
{\bf e-o} &  $\fft18 (t_8 \epsilon_{10} + \epsilon_{10} t_8) B R^4 $& $\fft18(t_8 \epsilon_{10}
+ \epsilon_{10} t_8) B R^4 (\Omega_+) $\\
{\bf+}&$=B\wedge X_8 (\Omega^{LC}) $&$ =  \fft18t_8 \epsilon_{10} B ( R^4 (\Omega_+)
+ R^4 (\Omega_-))$\\
{\bf o-e}& $=\fft1{ 192 (2\pi)^4} B\wedge \left(\tr R^4-\fft14(\tr R^2)^2\right)$  &
$= \fft12 B \wedge \left[X_8(\omp) + X_8(\omm) \right]$   \\ 
&&$=\fft1{ 192 \cdot (2\pi)^4} B\wedge \left(\tr R^4-\fft14(\tr R^2)^2 + \mbox{exact terms} \right)$
\\[4pt] \hline
{\bf e-e} &$t_8 t_8 R^4 $ & $t_8 t_8 R^4 (\Omega_+) = t_8 t_8 R^4 (\Omega_-) $ \\ \hline
{\bf o-o} &$\fft18\epsilon_{10} \epsilon_{10} R^4$ &
$\ft18\epsilon_{10}\epsilon_{10}\left(R(\Omega_+)^4+\ft83H^2 R(\Omega_+)^3 + \cdots \right)$ \\
& &  $= \ft18\epsilon_{10}\epsilon_{10}\left(R(\Omega_-)^4
+\ft83H^2 R(\Omega_-)^3 + \cdots \right)$  \\ \hline
\end{tabular}
}
\caption{The modification of the $R^4$ terms upon inclusion of the $B$-field.  We have
defined $\Omega_{\pm}=\Omega \pm\ft12\HH$. The precise form of the terms appearing
in the odd-odd spin structure is explained in Section~\ref{sec:corrrr} and Section~\ref{sec:morecor}.}
\label{tbl:res}
\end{table}

T-duality provides one of the motivations for studying the inclusion of the $B$-field in
$R^4$ couplings, and at the same time provides a very stringent test of the proposed
modifications.  In Section~\ref{sec:tdual} we concentrate on showing that the more
topological part of the $\alpha'^3$ couplings, namely  $B_2 \wedge [X_8(R^+) + X_8(R^-)]$,
transforms correctly under T-duality. A brief discussion about T-duality in the CP-even sector
is discussed in Appendix~\ref{app:T-CP-even}.

As already mentioned above, lower dimensional $R^2$ terms descend from the
ten-dimensional $R^4$ couplings. In particular, the $K3$ reduction of IIA theory yields a
six-dimensional theory with 16 supercharges in six dimensions dual to heterotic strings
compactified on $\mathbb{T}^4$. Here a combination of string four-point function computation
(which is a moral equivalent of a six-point computation in ten dimensions) together with the
duality considerations allow us to fix the six-dimensional one-loop four-derivative correction
completely. The result is:
\begin{eqnarray}
\label{eq:666ddd}
e^{-1}\mathcal L_{\rm IIA}^{d=6}&=&e^{-2\varphi}[R+4\partial\varphi^2-\ft1{12}H_{\mu\nu\rho}^2]
\nonumber\\
&&+\ft1{16}\alpha'[t_4t_4R(\Omega_+)^2
-\ft18\epsilon_6\epsilon_6R(\Omega_+)^2-\ft16\epsilon_6\epsilon_6H^2R(\Omega_+)
-\ft1{72}\epsilon_6\epsilon_6H^4]\nonumber\\
&&+\ft1{64}\alpha'\epsilon^{\alpha\beta\mu\nu\rho\sigma}B_{\alpha\beta}
[R_{\mu\nu}{}^{ab}(\Omega_+)R_{\rho\sigma}{}^{ab}(\Omega_+)+
R_{\mu\nu}{}^{ab}(\Omega_-)R_{\rho\sigma}{}^{ab}(\Omega_-)],
\end{eqnarray}
where $\varphi$ is the IIA dilaton, and where the CP-even expressions are defined below in
(\ref{eq:t4t4Rplus2}), (\ref{eq:E4def}), (\ref{eq:e4e4Rplus2}), (\ref{eq:e5e5H2Rplus})
and (\ref{eq:e4e4H4}).

As one would expect from heterotic/type II duality, two different formulations of $(1,1)$ theory
exist in six dimensions. One, the ``heterotic" $(1,1)$, has no Green-Schwarz type couplings,
but has a non-trivial Bianchi identity for the NSNS 3-form $H$. The ``type II" $(1,1)$ theory, on
the contrary, has a GS-type coupling and a closed 3-form $H$. This has been worked out (without
the additional higher derivative gravitational couplings) by Romans in \cite{Romans:1985tw}.
The two theories are related via strong-weak coupling duality, and the equations of motion of
one map to the Bianchi identities of the other. What turns out to be important in our context is
that the structure of the fermonic operators is very different in the two versions of the $(1,1)$
theory. The ``heterotic" $(1,1)$ has a {\sl single} Dirac operator, where the torsional connection
$\omm$ appears. The Bianchi identity for the $H$ field and the $\alpha'R^2$ expressions are
written in terms of a torsional connection with opposite sign on the torsion $\omp$. The
``type II" $(1,1)$ has two different Dirac operators, and both $\omp$ and $\omm$ appear there.
The averaging of $H$ in the twisted $B \wedge \tr R^2$ coupling can be seen as the
consequence of this.

Thinking of (\ref{eq:666ddd}) as the reduction of ten-dimensional type IIA action on $K3$ is
instructive and  provides additional information about the five and six-point function results
in ten dimensions.  The lift of (\ref{eq:666ddd}) to ten dimensions confirms our claim that for
all terms of order $\alpha'^3$, except the odd-odd spin structure sector, the replacement 
$R\to R(\Omega_+)$ is sufficient. It also fully agrees with the $H^2 R^3$ and
$(\nabla H)^2 R^2$ terms. Finally, the lift allows us to identify six candidates for the
ten-dimensional $H^4 R^2$ terms. This ambiguity can be resolved by explicit six-point function
computation. Note that $H^6 R$ and $H^8$ terms do not affect lower-dimensional theories
obtained via compactifications (without flux) on Ricci-flat manifolds.  Finally, one may
contemplate the lift of $\mathcal O(\alpha'^3)$ couplings to M theory. Note that the tree-level
coupling gets suppressed in the eleven-dimensional limit, while the lift of the one-loop couplings
involving the curvatures and $H$ should produce a {\sl complete} set of couplings at eight
derivatives involving the curvature the eleven-dimensional four-form $\hG$. Unfortunately, once more the lift involves
some ambiguities. Note that the reduction on a nontrivial graviphoton
background with a generic $\hG$, while reproducing the NSNS sector, yields a IIA one-loop
effective action with up to eight derivatives and with all the RR fields.

The structure of the paper is as follows. In Section~\ref{sec:stam} we looks at the five-point string
amplitude calculation and see the emergence of the structure outlined in Table~\ref{tbl:res}.
Section~\ref{sec:tdual} (as well as Appendix~\ref{app:T-CP-even}) is devoted to T-duality tests
of the proposed modifications of the $R^4$ couplings. In Section~\ref{sec:sixd} we use
heterotic/type II duality to fix completely the structure of $\alpha'$ terms at the one-loop
four-derivative level. In particular, we observe that the heterotic/type II duality map does not receive
corrections at $\mathcal O(\alpha')$.  In Section~\ref{sec:morecor}, we use this result to determine,
up to slight ambiguity, the structure of $\alpha'^3 H^4 R^2$ terms in ten dimensions, that would
otherwise require a six-point amplitude computation. Finally, in Section~\ref{sec:morecor11}, we
discuss the lifting of the modified $R^4$ couplings to eleven dimensions which leads to
incorporation of the four-form $\hG$.   We provide a summary of our key notation in
Appendix~\ref{app:conv}.

\section{String amplitude calculations}
\label{sec:stam}

We begin with a review of the Type II string effective action up to the $R^4$ terms.
We work in the covariant world-sheet formalism, and restrict our attention to external NSNS
states, namely the metric, antisymmetric tensor and dilaton.  As demonstrated by KLT
\cite{Kawai:1985xq}, closed string amplitudes may be rewritten in terms of
products of left- and right-moving open string amplitudes.  Thus the basic components of
the string computations are the chiral vertex operators
\begin{eqnarray}
V^{(-1)}(k,\zeta)&=&\zeta_\mu\delta(\gamma)\psi^\mu e^{ik\cdot X},\nonumber\\
V^{(0)}(k,\zeta)&=&\zeta_\mu(i\partial X^\mu+\ft12\alpha'k\cdot\psi\psi^\mu)e^{ik\cdot X},
\end{eqnarray}
given in the $-1$ and $0$ pictures, respectively.  Closed string vertex operators may be
built as a combination of left- and right-moving components, and in particular the closed
string polarization may be taken as a tensor product
\begin{equation}
\theta_{\mu\nu}=\zeta_\mu\otimes\bar\zeta_\nu.
\end{equation}
Furthermore, the various NSNS fields are be encoded as
\begin{equation}
\theta^{\mu\nu}=h^{\mu\nu}+b^{\mu\nu}+\ft12(\eta^{\mu\nu}-\bar k^\mu k^\nu-k^\mu\bar k^\nu)\phi.
\label{eq:thetapol}
\end{equation}
Here $\tilde g_{\mu\nu}=\eta_{\mu\nu}+h_{\mu\nu}$, where $h_{\mu\nu}$ is taken to be transverse
and traceless ($k^\mu h_{\mu\nu}=0$, and $h^\mu_\mu=0$).  The antisymmetric tensor is
given by $b_{\mu\nu}$ with $H=db$ and $k^\mu b_{\mu\nu}=0$.  Finally, $\phi$ is the
conventionally normalized dilaton.  Note that $\bar k$ is defined according to $\bar k^2=0$
and $k\cdot\bar k=1$, so that the transversality condition $k_\mu\theta^{\mu\nu}=0$ is
enforced for the dilaton.

Before proceeding, it is worth noting that when discussing effective string actions, the issue
of string frame versus Einstein frame invariably arises.  The string frame is perhaps the most
natural, since in this frame the dilaton explicitly counts string loops.  However, the choice
of polarizations encoded in (\ref{eq:thetapol}) actually leads to Einstein-frame scattering
amplitudes.  The reason for this is that the trace-free metric fluctuation $h_{\mu\nu}$ is
strictly orthogonal to the dilaton fluctuation.  This ensures that there is no $h-\phi$
propagator mixing, as would happen in the string frame.  As a matter of notation,
$\tilde g_{\mu\nu}$ will denote the Einstein-frame metric, while $g_{\mu\nu}$ will denote the
string-frame metric.  These are simply related by the standard Weyl scaling
\begin{equation}
\tilde g_{\mu\nu}=e^{-\phi/2}g_{\mu\nu}.
\label{eq:Weyl}
\end{equation}
Our notation is thus to use tildes in the Einstein frame and none in the string frame.  We
find this more convenient, as we will mainly emphasize string-frame expressions.

\subsection{The three-point function}

We first review the familiar result for the tree-level three point function.  For each massless
open string vector, we associate its polarization index with $\mu_i$ and
its momentum with $k_i$.  The open superstring three-point function then reproduces the
trilinear gluon coupling
\begin{equation}
\mathcal A(1,2,3)=\eta^{\mu_1\mu_2}k_{12}^{\mu_3}+\eta^{\mu_2\mu_3}k_{23}^{\mu_1}
+\eta^{\mu_3\mu_1}k_{31}^{\mu_2},
\end{equation}
where we have ignored overall coupling and normalization factors.  The KLT relation for
the three-point function is trivial, and the closed superstring amplitude is
\begin{equation}
\mathcal M(1,2,3)=\bar\mathcal A(1,2,3)\mathcal A(1,2,3).
\end{equation}
This can be explicitly written as
\begin{equation}
\mathcal M(1,2,3)=4\theta_1^{\mu_1\nu_1}k_1^{\mu_3}k_1^{\nu_3}\theta_2^{\mu_1\nu_1}
\theta_3^{\mu_3\nu_3}-4\theta_1^{\mu_1\nu_1}\theta_2^{\mu_1\nu_2}k_2^{\mu_3}k_2^{\nu_1}
\theta_3^{\mu_3\nu_2}-4\theta_1^{\mu_1\nu_1}k_1^{\mu_3}k_1^{\nu_2}\theta_2^{\mu_1\nu_2}
\theta_3^{\mu_3\nu_1}+\mbox{cyclic},
\label{eq:c3pt}
\end{equation}
and is identical for both the IIA and IIB string.

Expanding the amplitude, we find non-vanishing contributions for
\begin{equation}
h\mathrm-h\mathrm-h,\quad b\mathrm-b\mathrm-h,\quad b\mathrm-b\mathrm-\phi,\quad
\phi\mathrm-\phi\mathrm-h.
\end{equation}
These three-point functions may be obtained from the Einstein-frame action
\begin{equation}
\mathcal L=\sqrt{-\tilde g}[\tilde R-\ft12\partial\phi^2-\ft1{12}e^{-\phi}H^2],
\label{eq:EinsAct}
\end{equation}
which is a long-familiar result.  Using (\ref{eq:Weyl}) to transform to the string frame,
we find that the curvature scalar becomes
\begin{equation}
\tilde R=e^{\phi/2}[R+\ft12(d-1)\Box \phi-\ft1{16}(d-1)(d-2)\partial\phi^2].
\end{equation}
Inserting this into (\ref{eq:EinsAct}), setting $d=10$ and integrating $\Box\phi$ by parts then
gives the standard string frame NSNS action
\begin{equation}
\mathcal L=e^{-2\phi}\sqrt{-g}[R+4\partial\phi^2-\ft1{12}H^2].
\label{eq:Ltree0}
\end{equation}

\subsection{The tree-level four-point function}

The tree-level four-point function was investigated in 
\cite{Gross:1986iv,Gross:1986mw},
and at the eight-derivative level gives rise to the well-known $\alpha'^3R^4$ correction in
the effective action.  At the linearized level, it is easy to see that string amplitude is kinematically
built out of the gauge invariant combination
\begin{equation}
\bar{\tilde R}^{\mu_1\mu_2}{}_{\nu_1\nu_2}=4\theta^{[\mu_1}{}_{[\nu_1}k^{\mu_2]}k_{\nu_2]},
\label{eq:Ramp}
\end{equation}
where $\mu_1\mu_2$ are associated with the left-movers, while $\nu_1\nu_2$ are associated
with the right-movers.  Since the polarization $\theta_{\mu\nu}$ encodes all NSNS fields,
$\bar{\tilde R}^{\mu_1\mu_2}{}_{\nu_1\nu_2}$ generalizes the ordinary Riemann tensor.  Making
use of (\ref{eq:thetapol}), we find the linearized expression
\begin{equation}
\bar{\tilde R}_{\mu_1\mu_2}{}^{\nu_1\nu_2}=R_{\mu_1\mu_2}{}^{\nu_1\nu_2}
+e^{-\phi/2}\nabla_{[\mu_1}H_{\mu_2]}{}^{\nu_1\nu_2}
-\delta_{[\mu_1}{}^{[\nu_1}\nabla_{\mu_2]}\nabla^{\nu_2]}\phi,
\label{eq:Rbar}
\end{equation}
which is suggestive of a connection with torsion.  Note that, strictly speaking, the $e^{-\phi/2}$
factor in front of $\nabla H$ as absent at the linearized level.  However, it must be present at
the nonlinear order to ensure that $H$ has the proper weight.  Furthermore, while the $\nabla H$
term appears to differentiate between left- and right-movers, this distinction is artificial, as closure
of $H$ ensures that
\begin{equation}
\bar{\tilde R}_{\mu_1\mu_2\nu_1\nu_2}(H)=\bar{\tilde R}_{\nu_1\nu_2\mu_1\mu_2}(-H),
\label{eq:Rflip}
\end{equation}
which is a manifestation of worldsheet parity.

The linearized curvature tensor with torsion, (\ref{eq:Rbar}), takes a particularly simple form
when Weyl scaled to the string frame
\begin{equation}
\bar R_{\mu_1\mu_2}{}^{\nu_1\nu_2}=R_{\mu_1\mu_2}{}^{\nu_1\nu_2}
+\nabla_{[\mu_1}H_{\mu_2]}{}^{\nu_1\nu_2}.
\label{eq:hatRlin}
\end{equation}
In particular, the term linear in the dilaton drops out.  In the string frame, the resulting tree-level
$R^4$ addition to the effective action takes the form
\begin{equation}
\mathcal L_{\rm tree}^{\alpha'^3}=e^{-2\phi}\sqrt{-\tilde g}\left[\fft{\zeta(3)}{3\cdot2^{11}}\alpha'^3
(t_8t_8\bar R^4+\ft18\epsilon_{10}\epsilon_{10}\bar R^4)\right].
\end{equation}
We have included the $\epsilon_{10}\epsilon_{10}R^4$ term, even though strictly speaking
it does not arise at the level of the four-point function.  It is, however, implied in the
Green-Schwarz formalism, and moreover appears in the $\sigma$-model calculations
\cite{Grisaru:1986vi,Freeman:1986zh}.
At tree level, this correction is common to both the IIA and IIB superstring.

\subsection{The one-loop $\alpha'^3R^4$ corrections}
\label{sec:corrrr}

In the RNS formalism for the Type II string, the one-loop string amplitude involves a sum
over spin structures for both left- and right-moving sectors.  The even-even and odd-odd
spin structures contribute to the CP-even sector, while the even-odd and odd-even spin
structures contribute to the CP-odd sector.  Because of maximal supersymmetry, the loop
correction starts at the level of the four-point function.  However, the odd spin structure
will not contribute until the five-point function, because of the necessity of soaking up
ten fermion zero modes.

Before discussing the closed string corrections, we make a few observations for the open string.
For a one-loop open string amplitude in the even spin structure sector, all vertex operators
are taken in the 0 picture.  In particular, the four-point function involves the contraction
\begin{equation}
\left\langle\sum_{i=1}^4 V^{(0)}(k^i,\zeta^{(i)})\right\rangle=
\zeta^{(1)}_{\mu_1}\cdots\zeta^{(4)}_{\mu_4}
\sum_a\left\langle\prod_{i=1}^4
(i\partial X^{\mu_i}+\ft12\alpha'k_i\cdot\psi\psi^{\mu_i})
e^{ik_i\cdot X}\right\rangle_{\!\!a},
\end{equation}
where the sum is over the three even spin structures.  Breaking up the vertex operators into
individual worldsheet bosons and fermions, we see that the amplitude is a sum of terms involving
contractions between an even number of worldsheet fermions.  The terms with fewer than
four worldsheet fermions vanish when summed over spin structures.  (This is closely related
to spacetime supersymmetry.)  Hence the kinematic structure of this amplitude starts with a
minimum of two momentum factors.  The actual details are somewhat obscure in
the covariant approach, but can be more easily seen in the Green-Schwarz formulation.  The
resulting kinematical factor is simply
\begin{equation}
\mathcal A_{\rm even}\sim t_8^{\mu_1\cdots\mu_8}\,(k^1_{\mu_1}\zeta^{(1)}_{\mu_2})\cdots
(k^4_{\mu_7}\zeta^{(4)}_{\mu_8}),
\label{eq:Aeven}
\end{equation}
and is identical to the tree-level case.

Turning to the odd spin structure sector, for the one-loop amplitude, we include one picture
changing operator $\delta(\beta)\psi\cdot\partial X$, take one vertex operator in the $-1$
picture, and put the remaining vertex operators in the $0$ picture.  The five-point function
then involves the contraction
\begin{eqnarray}
\left\langle V^{(-1)}(k^1,\zeta^{(1)})\sum_{i=2}^5V^{(0)}(k^i,\zeta^{(i)})\right\rangle
&=&\zeta^{(1)}_{\mu_1}\cdots\zeta^{(5)}_{\mu_5}\left\langle\psi\cdot\partial X(0)\psi^{\mu_1}
\prod_{i=2}^5(i\partial X^{\mu_i}+\ft12\alpha'k_i\cdot\psi\psi^{\mu_i})
\prod_{i=1}^5e^{ik_i\cdot X}\right\rangle_{\!\!\rm odd}.\nn\\
\label{eq:5pointodd}
\end{eqnarray}
However, since we need to soak up ten fermion zero modes in order to get a non-vanishing
amplitude, we see that each vertex operator in the 0 picture must contribute two worldsheet
fermions, so that
\begin{equation}
\mathcal A_{\rm odd}\sim
\epsilon^{\alpha\mu_0\mu_1\cdots\mu_8}\,\zeta^{(1)}_{\mu_0}
(k^2_{\mu_1}\zeta^{(2)}_{\mu_2})\cdots(k^5_{\mu_7}\zeta^{(5)}_{\mu_8})
\left\langle\partial X_\alpha(0)\sum_{i=1}^5e^{ik_i\cdot X}\right\rangle.
\label{eq:Aodd}
\end{equation}
(Note that the bosonic contractions vanish in the zero momentum limit.)  Since we have
placed the first vertex operator in the $-1$ picture, this amplitude seems to single out
the first particle for special treatment.  However, by using momentum conservation, it is
easy to see that this amplitude is fully bose symmetric under interchange of any of the
five particles.

\subsubsection{The even-even amplitude}

We now return to the closed string, where the amplitudes are built out of a combination of
left and right movers.  In the CP-even sector, the even-even four-point function is a combination
of (\ref{eq:Aeven}) on the left and on the right.  Taking (\ref{eq:Ramp}) into account
directly gives $t_8t_8\bar R^4$ with a one-loop coefficient arising from performing the
modular integrals \cite{Sakai:1986bi}.%
\footnote{The linearized connection with torsion $\bar R$ also appears in the heterotic
one-loop computation \cite{Ellis:1987dc,Abe:1987ud,Abe:1988cq}.}
Transforming to the string frame, and writing (\ref{eq:hatRlin})
schematically as $\bar R=R+\nabla H$, we see that this term has an expansion of the form
\begin{equation}
t_8t_8\bar R^4=t_8t_8[R^4+6(\nabla H)^2R^2+(\nabla H)^4].
\end{equation}
In particular, terms odd in $H$ vanish by worldsheet parity.

\subsubsection{The odd-odd amplitude}

The odd-odd contribution is somewhat more unusual, and first arises at the level of the five-point
function.  Although the amplitude is insensitive to this choice, we single out the first vertex operator
to be in the $(-1,-1)$ picture.  In this case, we find
\begin{equation}
\mathcal A\sim
\epsilon^{\alpha\mu_0\mu_1\cdots\mu_8}\epsilon^{\beta\nu_0\nu_1\cdots\nu_8}
\theta^{(1)}_{\mu_0\nu_0}(k^2_{\mu_1}k^2_{\nu_1}\theta^{(2)}_{\mu_2\nu_2})
\cdots(k^5_{\mu_7}k^5_{\nu_7}\theta^{(5)}_{\mu_8\nu_8})
\left\langle\partial X_\alpha(0)\bar\partial X_\beta(0)\sum_{i=1}^5e^{ik_i\cdot X}\right\rangle.
\end{equation}
At the eight-derivative level, we only take the boson zero mode contraction
\begin{equation}
\langle\overline\partial X^\alpha\partial X^\beta\rangle=
-\fft{\alpha'}{8\pi\tau_2}\eta^{\alpha\beta},
\label{eq:nczero}
\end{equation}
so that
\begin{equation}
\mathcal A\sim
\epsilon^{\alpha\mu_0\mu_1\cdots\mu_8}\epsilon_\alpha{}^{\nu_0\nu_1\cdots\nu_8}
\theta^{(1)}_{\mu_0\nu_0}(k^2_{\mu_1}k^2_{\nu_1}\theta^{(2)}_{\mu_2\nu_2})
\cdots(k^5_{\mu_7}k^5_{\nu_7}\theta^{(5)}_{\mu_8\nu_8}).
\label{eq:Aoo}
\end{equation}
Focusing on gravitons and antisymmetric tensors only (since the dilaton is unimportant in
the string frame), we see that the amplitude contributes to five graviton scattering, two $B$
and three graviton scattering, and four $B$ and one graviton scattering.  We now look at
these three possibilities in turn.

For five gravitons, although the amplitude is bose symmetric, it is convenient to think a particular
permutation where the first graviton is singled out.  Using the identity
\begin{equation}
\epsilon^{\alpha\mu_0\mu_1\cdots\mu_8}\epsilon_\alpha{}^{\nu_0\nu_1\cdots\nu_8}
=\ft12\eta^{\mu_0\nu_0}\epsilon^{\alpha\beta\mu_1\cdots\mu_8}
\epsilon_{\alpha\beta}{}^{\nu_1\cdots\nu_8}
-4\epsilon^{\alpha\beta\mu_1\cdots\mu_8}
\eta^{\mu_0[\nu_1}\epsilon_{\alpha\beta}{}^{|\nu_0|\nu_2\cdots\nu_8]},
\label{eq:e9e8ident}
\end{equation}
along with tracelessness of the on-shell graviton, $\theta^{(1)}_{\mu_0\nu_0}
=h^{(1)}_{\mu_0\nu_0}$, we see that the odd-odd amplitude has the form
\begin{equation}
\mathcal A_{h^5}\sim\epsilon^{\alpha\beta\mu_1\cdots\mu_8}
\epsilon_{\alpha\beta}{}^{\nu_1\cdots\nu_8}
[(h^{(1)\,\lambda}_{\nu_1}
R^{(2)}_{\mu_1\mu_2\lambda\nu_2})
R^{(3)}_{\mu_3\mu_4\nu_3\nu_4}
R^{(4)}_{\mu_5\mu_6\nu_5\nu_6}
R^{(5)}_{\mu_7\mu_8\nu_7\nu_8}+\mbox{3 more}].
\label{eq:oohR4}
\end{equation}
Here $R^{(i)}_{\mu_1\mu_2\nu_1\nu_2}$ represents the linearized Riemann tensor
given by substituting $\theta^{(i)}_{\mu\nu}=h^{(i)}_{\mu\nu}$ into (\ref{eq:Ramp}).
It is now clear that the odd-odd five-graviton amplitude is an expansion of
$\epsilon_8\epsilon_8R^4$, which may be written covariantly as
$-(1/2)\epsilon_{10}\epsilon_{10}R^4$.

The amplitude for two antisymmetric tensor and three graviton scattering was computed
in \cite{Peeters:2001ub} (and in the Green-Schwarz formulation in \cite{Richards:2008sa})
by assigning $\theta^{(1)}_{\mu_0\nu_0}=b^{(1)}_{\mu_0\nu_0}$
and $\theta^{(2)}_{\mu_2\nu_2}=b^{(2)}_{\mu_2\nu_2}$.  In this case, we may rewrite
(\ref{eq:Aoo}) using momentum conservation as
\begin{equation}
\mathcal A_{b^2h^3}\sim
-\epsilon^{\alpha\mu_0\mu_1\cdots\mu_8}\epsilon_\alpha{}^{\nu_0\nu_1\cdots\nu_8}
(k^1_{\mu_1}b^{(1)}_{\mu_2\nu_0})(k^2_{\nu_1}b^{(2)}_{\nu_2\mu_0})
(k^3_{\mu_3}k^3_{\nu_3}\theta^{(3)}_{\mu_4\nu_4})
\cdots(k^5_{\mu_7}k^5_{\nu_7}\theta^{(5)}_{\mu_8\nu_8}).
\label{eq:Aoore}
\end{equation}
The graviton expressions recreate linearized Riemann tensors, while 
$(k^1_{\mu_1}b^{(1)}_{\mu_2\nu_0})$ and $(k^2_{\nu_1}b^{(2)}_{\nu_2\mu_0})$ recreate
the antisymmetric tensor field strength $H=dB$.  The resulting amplitude is of the form
\cite{Peeters:2001ub}
\begin{equation}
\mathcal A_{b^2h^3}\sim
-\epsilon_{\alpha\mu_0\mu_1\cdots\mu_8}\epsilon^{\alpha\nu_0\nu_1\cdots\nu_8}
(H^{\mu_1\mu_2}{}_{\nu_0}H_{\nu_1\nu_2}{}^{\mu_0}-\ft19
H^{\mu_1\mu_2\mu_0}H_{\nu_1\nu_2\nu_0})R^{\mu_3\mu_4}{}_{\nu_3\nu_4}
R^{\mu_5\mu_6}{}_{\nu_5\nu_6}R^{\mu_7\mu_8}{}_{\nu_7\nu_8}.
\label{eq:oob2r3}
\end{equation}
The two $H^2$ combinations, along with the relative factor of $-1/9$ arise in relating the
full antisymmetrized $H_{\mu_1\mu_2\nu_0}\to k_{[\mu_1}b_{\mu_2\nu_0]}$ to the kinematics
of the string amplitude (\ref{eq:Aoore}), where only $\mu_1\mu_2$ are antisymmetrized.
However, these two kinematical combinations are in fact related by first rewriting
$\epsilon\epsilon$ in terms of a fully antisymmetrized $\delta$-function, and then making use
of the symmetry properties of the Riemann tensor.  With the assistance of the computer
algebra system Cadabra \cite{Peeters:2006kp,Peeters:2007wn}, we find
\begin{eqnarray}
&&\epsilon_{\alpha\mu_0\mu_1\cdots\mu_8}\epsilon^{\alpha\nu_0\nu_1\cdots\nu_8}
H^{\mu_1\mu_2\mu_0}H_{\nu_1\nu_2\nu_0}R^{\mu_3\mu_4}{}_{\nu_3\nu_4}
R^{\mu_5\mu_6}{}_{\nu_5\nu_6}R^{\mu_7\mu_8}{}_{\nu_7\nu_8}\nonumber\\
&&\kern4em=
3\epsilon_{\alpha\mu_0\mu_1\cdots\mu_8}\epsilon^{\alpha\nu_0\nu_1\cdots\nu_8}
H^{\mu_1\mu_2}{}_{\nu_0}H_{\nu_1\nu_2}{}^{\mu_0}
R^{\mu_3\mu_4}{}_{\nu_3\nu_4}
R^{\mu_5\mu_6}{}_{\nu_5\nu_6}R^{\mu_7\mu_8}{}_{\nu_7\nu_8},
\label{eq:cadaid}
\end{eqnarray}
in which case the $b^2h^3$ amplitude may be written as
\begin{equation}
\mathcal A_{b^2h^3}\sim
-\ft23\epsilon_{\alpha\mu_0\mu_1\cdots\mu_8}\epsilon^{\alpha\nu_0\nu_1\cdots\nu_8}
H^{\mu_1\mu_2}{}_{\nu_0}H_{\nu_1\nu_2}{}^{\mu_0}
R^{\mu_3\mu_4}{}_{\nu_3\nu_4}
R^{\mu_5\mu_6}{}_{\nu_5\nu_6}R^{\mu_7\mu_8}{}_{\nu_7\nu_8}.
\label{eq:oob2r3n}
\end{equation}

The amplitude with four antisymmetric tensors and one graviton is similar, except two of the
three Riemann tensors will be replaced by $\nabla H$
\begin{eqnarray}
\mathcal A_{b^4h}&\sim&
-\epsilon_{\alpha\mu_0\mu_1\cdots\mu_8}\epsilon^{\alpha\nu_0\nu_1\cdots\nu_8}
(H^{\mu_1\mu_2}{}_{\nu_0}H_{\nu_1\nu_2}{}^{\mu_0}-\ft19
H^{\mu_1\mu_2\mu_0}H_{\nu_1\nu_2\nu_0})
\nabla^{\mu_3}H^{\mu_4}{}_{\nu_3\nu_4}
\nabla^{\mu_5}H^{\mu_6}{}_{\nu_5\nu_6}R^{\mu_7\mu_8}{}_{\nu_7\nu_8}.\nn\\
\label{eq:oob4r}
\end{eqnarray}
Note that this substitution of $R\to\nabla H$ essentially incorporates the linearized
connection with torsion $\bar R=R+\nabla H$, although the combinatorial factors do
not correspond to this direct substitution.  While it may be possible to simplify this
expression, as we did for the $b^2h^3$ amplitude, we have unfortunately been unable
to generalize the identity (\ref{eq:cadaid}) where two Riemann tensors have been replaced
by $\nabla H$.

Combining (\ref{eq:oohR4}), (\ref{eq:oob2r3n}) and (\ref{eq:oob4r}) gives us an
effective Lagrangian for the odd-odd sector
\begin{eqnarray}
e^{-1}\mathcal L_{\rm odd\mhyphen odd}&\sim&
-\ft18\epsilon_{\alpha\beta\mu_1\cdots\mu_8}\epsilon^{\alpha\beta\nu_1\cdots\nu_8}
R^{\mu_1\mu_2}{}_{\nu_1\nu_2}R^{\mu_3\mu_4}{}_{\nu_3\nu_4}
R^{\mu_5\mu_6}{}_{\nu_5\nu_6}R^{\mu_7\mu_8}{}_{\nu_7\nu_8}\nonumber\\
&&-\ft1{12}\epsilon_{\alpha\mu_0\mu_1\cdots\mu_8}\epsilon^{\alpha\nu_0\nu_1\cdots\nu_8}
H^{\mu_1\mu_2}{}_{\nu_0}H_{\nu_1\nu_2}{}^{\mu_0}
R^{\mu_3\mu_4}{}_{\nu_3\nu_4}
R^{\mu_5\mu_6}{}_{\nu_5\nu_6}R^{\mu_7\mu_8}{}_{\nu_7\nu_8},\nonumber\\
&&-\ft1{16}\epsilon_{\alpha\mu_0\mu_1\cdots\mu_8}\epsilon^{\alpha\nu_0\nu_1\cdots\nu_8}
(H^{\mu_1\mu_2}{}_{\nu_0}H_{\nu_1\nu_2}{}^{\mu_0}-\ft19
H^{\mu_1\mu_2\mu_0}H_{\nu_1\nu_2\nu_0})\nonumber\\
&&\kern12em\times\nabla^{\mu_3}H^{\mu_4}{}_{\nu_3\nu_4}
\nabla^{\mu_5}H^{\mu_6}{}_{\nu_5\nu_6}R^{\mu_7\mu_8}{}_{\nu_7\nu_8},
\label{eq:ooLag0}
\end{eqnarray}
which reproduces all odd-odd five-point functions in the NSNS sector.  Note that the
sign of this contribution is opposite for the IIA and IIB string because of the opposite
GSO projections.  Following \cite{Gross:1986mw}, we may expect this combination
to simplify when written in terms of $\bar R$.  However, it is not obvious that the linearized
curvature (\ref{eq:hatRlin}) is the best object to use, since the first line of (\ref{eq:ooLag0})
is sensitive to the nonlinear part of the Riemann tensor when computing the five-point function.

Of course, the natural non-linear completion of $\bar R$ is to introduce the full connection with
torsion $\Omega_+=\Omega+\ft12\mathcal H$, which reads in components
\begin{equation}
\Omega_{+\,\mu}{}^{\alpha\beta}=\Omega_\mu{}^{\alpha\beta}+\ft12H_\mu{}^{\alpha\beta}.
\label{eq:Omegatorsion}
\end{equation}
The curvature computed out of $\Omega_+$ is then
\begin{equation}
R(\Omega_+)=R+\ft12d\mathcal H+\ft14\mathcal H\wedge\mathcal H,\qquad
\mathcal H^{\alpha\beta}=H_\mu{}^{\alpha\beta}dx^\mu,
\label{eq:Romegaplus}
\end{equation}
or in components
\begin{equation}
R(\Omega_+)_{\mu\nu}{}^{\alpha\beta}=R_{\mu\nu}{}^{\alpha\beta}
+\nabla_{[\mu}H_{\nu]}{}^{\alpha\beta}+\ft12H_{[\mu}{}^{\alpha\gamma}H_{\nu]\gamma}{}^\beta.
\end{equation}
Replacing $R$ by $R(\Omega_+)$ in (\ref{eq:ooLag0}), and working only to the level of the
five-point function, we find
\begin{eqnarray}
e^{-1}\mathcal L_{\rm odd\mhyphen odd}&\!\!\!\sim\!\!\!&
-\ft18\epsilon_{\alpha\beta\mu_1\cdots\mu_8}\epsilon^{\alpha\beta\nu_1\cdots\nu_8}
R^{\mu_1\mu_2}{}_{\nu_1\nu_2}(\Omega_+)R^{\mu_3\mu_4}{}_{\nu_3\nu_4}(\Omega_+)
R^{\mu_5\mu_6}{}_{\nu_5\nu_6}(\Omega_+)R^{\mu_7\mu_8}{}_{\nu_7\nu_8}(\Omega_+)
\nonumber\\
&&-\ft13\epsilon_{\alpha\mu_0\mu_1\cdots\mu_8}\epsilon^{\alpha\nu_0\nu_1\cdots\nu_8}
H^{\mu_1\mu_2}{}_{\nu_0}H_{\nu_1\nu_2}{}^{\mu_0}
R^{\mu_3\mu_4}{}_{\nu_3\nu_4}(\Omega_+)
R^{\mu_5\mu_6}{}_{\nu_5\nu_6}(\Omega_+)R^{\mu_7\mu_8}{}_{\nu_7\nu_8}(\Omega_+)
\nonumber\\
&&
+\ft1{16}\epsilon_{\alpha\mu_0\mu_1\cdots\mu_8}\epsilon^{\alpha\nu_0\nu_1\cdots\nu_8}
(9H^{\mu_1\mu_2}{}_{\nu_0}H_{\nu_1\nu_2}{}^{\mu_0}+\ft19
H^{\mu_1\mu_2\mu_0}H_{\nu_1\nu_2\nu_0})\nonumber\\
&&\kern9em
\times\nabla^{\mu_3}H^{\mu_4}{}_{\nu_3\nu_4}
\nabla^{\mu_5}H^{\mu_6}{}_{\nu_5\nu_6}R^{\mu_7\mu_8}{}_{\nu_7\nu_8}(\Omega_+).
\label{eq:ooLaghat}
\end{eqnarray}
We have had to use (\ref{eq:e9e8ident}) as well as on-shell integration by parts in order
to obtain this expression.  Note that the five-point function does not distinguish between
the use of $\bar R$ and $R(\Omega_+)$ in the mixed terms with $H$.  However, we expect
that $R(\Omega_+)$ is in fact the correct quantity for extending beyond the five-point function.

\subsubsection{The odd-even and even-odd amplitudes}

We now turn to the CP-odd sector which is responsible for the $B\wedge X_8$ term.  In the
RNS formalism, this term arises from a combination of odd-even and even-odd spin structure
contributions.  Because of the presence of the odd spin structure, this amplitude starts at the
five-point level, and will involve a single epsilon tensor.  Since worldsheet parity relates
the odd-even and even-odd contributions, we only need to consider, say, the odd-even
sector explicitly.  In this case, the one-loop amplitude involves one vertex operator in the $(-1,0)$
picture and the remaining four in the $(0,0)$ picture.  In anticipation of the $B\wedge X_8$
structure, we take the $(-1,0)$ picture vertex to encode the antisymmetric tensor $b_{\mu\nu}$
and give the remaining four vertex operators general polarizations $\theta^{(i)}_{\mu\nu}$.

After soaking up ten fermion zero modes in the odd spin structure sector and using a boson
zero mode contraction to remove $\partial X_\alpha(0)$ in (\ref{eq:Aodd}), we find that the
five-point amplitude reduces to the chiral contraction
\begin{equation}
\mathcal A_{\rm odd\mhyphen even}\sim b^{(1)}_{\alpha\beta}
\epsilon^{\alpha\beta\mu_1\cdots\mu_8}
(k^1_{\mu_1}\theta^{(1)}_{\mu_2\lambda_1})\cdots(k^4_{\mu_7}\theta^{(4)}_{\mu_8\lambda_4})
\sum_a\left\langle\prod_{i=1}^4
(i\overline\partial X^{\lambda_i}+\ft12\alpha'k_i\cdot\bar\psi\bar\psi^{\lambda_i})
e^{ik_i\cdot X}\right\rangle_{\!\!a}.
\label{eq:cpoddamp}
\end{equation}
Of course, the worldsheet integrals must still be performed.  In fact, this amplitude essentially
computes the elliptic genus \cite{Lerche:1987sg,Lerche:1987qk,Witten:1986bf,Vafa:1995fj},
and hence we end up with the odd-even contribution
\begin{equation}
\mathcal A_{\rm odd\mhyphen even}\sim t^8_{\mu_1\cdots\mu_8}B_2\wedge\bar R_2^{\mu_1\mu_2}
\wedge\cdots\wedge\bar R_2^{\mu_7\mu_8}=24B_2\wedge(\Tr\bar R^4-\ft14(\Tr\bar R^2)^2).
\end{equation}

The full CP-odd amplitude is a sum of odd-even and even-odd spin structures.  These contributions
are identical except that worldsheet parity flips the sign of $B_{\mu\nu}$.  As a result, combining
the spin structures picks out terms odd in $B_{\mu\nu}$ for the IIA string and even in $B_{\mu\nu}$
for the IIB string.  This can be seen explicitly by writing the CP-odd amplitude schematically as
\begin{equation}
\mathcal A_{\rm CP\mhyphen odd}\sim(t_8\epsilon_{10}\pm\epsilon_{10}t_8)B\bar R(H)^4.
\end{equation}
The sign choice is a consequence of the GSO projection, and the positive sign is taken for
the IIA string while the negative sign is taken for the IIB string.  We now flip the order of
$\epsilon_{10}t_8$ in the second term so that it matches the first.  Making use of (\ref{eq:Rflip})
then gives
\begin{equation}
\mathcal A_{\rm CP\mhyphen odd}\sim t_8B\wedge(\bar R(H)^4\pm\bar R(-H)^4),
\end{equation}
where $\epsilon_{10}$ has been made implicit in the wedge product.

\subsubsection{The one-loop effective Lagrangian}

We may combine the contributions from the various spin structure sectors to arrive at the
one-loop effective Lagrangian (in the string frame)
\begin{eqnarray}
\mathcal L_{\rm loop}^{\alpha'^3}&=&\sqrt{-g}\left[\fft{(2\pi)^2}{3^2\cdot2^{13}}\alpha'^3
\left(t_8t_8\bar R^4\mp\ft18\epsilon_{10}\epsilon_{10}\left(R(\Omega_+)^4
+\ft83H^2 R(\Omega_+)^3-\ft12H^2(\nabla H)^2 R(\Omega_+)\right)\right)\right]\nonumber\\
&&-\fft{(2\pi)^2}{3\cdot2^6}\alpha'^3B_2\wedge
\left[\Tr \bar R^4-\ft14(\Tr\bar R^2)^2\right]_{\mathrm{even(odd)\>in}\>B_2\>
\mathrm{for\>IIA(IIB)}}.
\label{eq:olLaglin}
\end{eqnarray}
The odd-odd contribution is written schematically here; the explicit form is given in
(\ref{eq:ooLaghat}).  Note that the CP-odd contribution is often given in terms of the
eight-form
\begin{equation}
X_8(R)=\fft1{(2\pi)^43\cdot2^6}\left(\Tr R^4-\fft14(\Tr R^2)^2\right).
\label{eq:X8def}
\end{equation}
Using $X_8$, the one-loop CP-odd contribution may equivalently be written as
\begin{equation}
\mathcal L_{\rm CP\mhyphen odd}=-(2\pi)^6\alpha'^3B_2\wedge\ft12[X_8(\bar R(H))
\pm X_8(\bar R(-H))],
\end{equation}
where the top sign is for the IIA string.  In particular, this `averages' or `anti-averages' in $B_2$
for the IIA and IIB string, respectively.

It is important to understand that the effective Lagrangian (\ref{eq:olLaglin}) is only designed to
reproduce the eight-derivative one-loop four-point function in the even-even sector and the
five-point function in the remaining sectors.  To the lowest order in $B$ (none in the CP-even
sector and linear in the CP-odd sector), this reduces to the familiar
\begin{equation}
\left.\mathcal L_{\rm loop}^{\alpha'^3}\right|_{{\rm lowest\>in}\>B}
=\sqrt{-g}\left[\fft{(2\pi)^2}{3^2\cdot2^{13}}\alpha'^3
(t_8t_8R^4\mp\ft18\epsilon_{10}\epsilon_{10}R^4)\right]
-(2\pi)^6\alpha'^3B_2\wedge X_8(R)\Bigr|_{\rm IIA\>only}.
\end{equation}
Because of general covariance, this eight-derivative purely gravitational contribution is
essentially complete.  (Although terms proportional to the Ricci tensor are undetermined
at this order, they are unimportant as they may be absorbed by field redefinition.)  However,
if we are interested in the full eight-derivative action including the $B$-field, then terms up to
$\mathcal O(H^8)$ may appear, and these will require the calculation of an eight-point function
to pin down.

Although in principle there is no obstruction to the computation of the one-loop eight-point
function, in practice it is beyond our technical means.  However, we conjecture that the full
eight-derivative action at the non-linear level involves the replacement of the linearized
curvature $\bar R$ by the complete curvature with torsion $R(\Omega_+)$ given in
(\ref{eq:Romegaplus}).  We thus postulate that (\ref{eq:olLaglin}) should be extended to
\begin{eqnarray}
\mathcal L_{\rm loop}^{\alpha'^3}&=&\sqrt{-g}\biggl[\fft{(2\pi)^2}{3^2\cdot2^{13}}\alpha'^3
\Bigl(t_8t_8R(\Omega_+)^4
\mp\ft18\epsilon_{10}\epsilon_{10}\left(R(\Omega_+)^4
+\ft83H^2 R(\Omega_+)^3-\ft12H^2(\nabla H)^2 R(\Omega_+)+\cdots\right)\Bigr)\biggr]\nn\\
&&-(2\pi)^6\alpha'^3B_2\wedge\ft12[X_8(R(\Omega_+))\pm X_8(R(\Omega_-))].
\label{eq:olLaghat}
\end{eqnarray}
The even-even spin structure term $t_8t_8R(\Omega_+)^4$ agrees with five-point
function computation of \cite{Richards:2008jg,Richards:2008sa}.%
\footnote{Although the loop computation of \cite{Richards:2008jg,Richards:2008sa} was
performed in the Green-Schwarz formalism, there is nevertheless an obvious
split between the $t_8t_8$ and $\epsilon_8\epsilon_8$ contributions.}
However, there appears to be a disagreement in the odd-odd spin structure sector, as
\cite{Richards:2008sa} only found the $\epsilon_{10}\epsilon_{10}R(\Omega_+)^4$
combination, and not the remaining odd-odd terms.  We will present evidence below
that these additional terms are in fact present in the odd-odd sector.  In particular, the
six-dimensional completion of $\epsilon_6\epsilon_6R^2$ will include terms of the form
$\epsilon_6\epsilon_6H^2R$ and $\epsilon_6\epsilon_6H^4$.  The former lifts to
$\epsilon_{10}\epsilon_{10}H^2R^3$, while the latter lifts to
$\epsilon_{10}\epsilon_{10}H^4R^2$, which
cannot be seen below the level of the six-point function.
The lift of the six-dimensional Lagrangian also strongly suggests that the remaining
sectors, namely the even-even part of the CP-even sector and the entire CP-odd sector,
are completed by the simple replacement $R\to R(\Omega_+)$, as indicated in (\ref{eq:olLaghat}).

\subsection{The vanishing of CP-odd couplings in IIB}
\label{sec:cp-oddB}

Before turning to the T-duality and six-dimensional heterotic/IIA tests of (\ref{eq:olLaghat}),
we first comment on the CP-odd contribution in the IIB case.  In particular, while the
$B\wedge X_8$ term plays an important role in the IIA string and its lift to eleven-dimensions,
it is known to be absent in the IIB string.  Of course, the sign choice in the CP-odd term in
(\ref{eq:olLaghat}) is designed to project $X_8$ onto terms even in $B$ for the IIA string, and
odd in $B$ for the IIB string.  As a result, this is consistent with the absence of
$B_2\wedge X_8(R)$ in the IIB case.  However, terms of the form $B_2\wedge R^3\wedge\nabla H$
and $B_2\wedge R\wedge(\nabla H)^3$ appear to survive.  It would be puzzling to expect
that the IIB string has such unusual CP-odd interactions.  Thus we anticipate that these
unwanted terms in fact vanish on-shell, and hence do not contribute to the IIB effective action.

Although we have been unable to prove the vanishing of the full ten-dimensional CP-odd coupling
in IIB, as a first step we may show that the corresponding term in six dimensions vanishes at the
linearized level.  The six-dimensional CP-odd term that we are interested in takes the form
\begin{equation}
\mathcal I=B\wedge\Tr\left[R(\Omega_+)\wedge R(\Omega_+)\right]_{\mathrm{odd\>in}\>B_2}
=B\wedge\Tr\left[d\mathcal H\wedge(R+\ft14\mathcal H^2)\right].
\label{eq:IIBCPodd6}
\end{equation}
At the linearized level, we take
\begin{equation}
\mathcal I=B\wedge\Tr(d\mathcal H\wedge R)=-H\wedge\Tr(\mathcal H\wedge R),
\end{equation}
where we have integrated by parts.  In components, this takes the form
\begin{equation}
\mathcal I=\epsilon^{\mu_1\cdots\mu_6}H_{\mu_1\mu_2\mu_3}H_{\mu_4\nu_1\nu_2}
R_{\mu_5\mu_6}{}^{\nu_1\nu_2},
\label{eq:Icomp}
\end{equation}
which matches the expression obtained in \cite{Gregori:1997hi} from the scattering of two
antisymmetric tensors and one graviton.

We now demonstrate that this vanishes at the level of the three point amplitude.
To see this, we rewrite $\mathcal I$ as
\begin{equation}
\mathcal I=\epsilon^{\mu_1\cdots\mu_6}\delta^\alpha_\beta
H_{\mu_1\mu_2\mu_3}H_{\mu_4\nu_1\alpha} R_{\mu_5\mu_6}{}^{\nu_1\beta},
\end{equation}
and make the substitution
\begin{equation}
\delta_{\beta}^{\alpha}=-\fft1{5!}\epsilon_{\beta\sigma_1\cdots\sigma_5}
\epsilon^{\alpha\sigma_1\cdots\sigma_5}.
\end{equation}
We may rearrange the $\epsilon$-tensors to obtain
\begin{eqnarray}
\mathcal I&=&-\fft1{5!}\epsilon^{\mu_1\cdots\mu_6}
\epsilon_{\beta\sigma_1\cdots\sigma_5}
\epsilon^{\alpha\sigma_1\cdots\sigma_5}
H_{\mu_1\mu_2\mu_3}H_{\mu_4\nu_1\alpha}R_{\mu_5\mu_6}{}^{\nu_1\beta}\nonumber\\
&=&\fft1{5!}\delta_{\beta\sigma_1\cdots\sigma_5}^{\mu_1\cdots\mu_6}
\epsilon^{\alpha\sigma_1\cdots\sigma_5}
H_{\mu_1\mu_2\mu_3}H_{\mu_4\nu_1\alpha}R_{\mu_5\mu_6}{}^{\nu_1\beta}\nonumber\\
&=&\epsilon^{\alpha\sigma_1\cdots\sigma_5}(
3H_{\beta\sigma_1\sigma_2}H_{\sigma_3\nu_1\alpha}R_{\sigma_4\sigma_5}{}^{\nu_1\beta}
-H_{\sigma_1\sigma_2\sigma_3}H_{\beta\nu_1\alpha}R_{\sigma_4\sigma_5}{}^{\nu_1\beta}
+2H_{\sigma_1\sigma_2\sigma_3}H_{\sigma_4\nu_1\alpha}R_{\beta\sigma_5}{}^{\nu_1\beta}).
\nonumber\\
\end{eqnarray}
The first term vanishes because the $H^2$ expression is symmetric under
$\nu_1\leftrightarrow\beta$, while the Riemann tensor is antisymmetric under this
exchange.  A simple relabeling of indices now gives
\begin{equation}
\mathcal I=\epsilon^{\mu_1\cdots\mu_6}
(-H_{\mu_1\mu_2\mu_3}H_{\mu_4\nu_1\nu_2}R_{\mu_5\mu_6}{}^{\nu_1\nu_2}
+2H_{\mu_1\mu_2\mu_3}H_{\mu_4\mu_5\nu_1}R_{\mu_6}^{\nu_1}).
\end{equation}
Comparing with (\ref{eq:Icomp}), we see that
\begin{equation}
\mathcal I=-\mathcal I+2\epsilon^{\mu_1\cdots\mu_6}
H_{\mu_1\mu_2\mu_3}H_{\mu_4\mu_5\nu_1}R_{\mu_6}^{\nu_1},
\end{equation}
or
\begin{equation}
\mathcal I=\epsilon^{\mu_1\cdots\mu_6}
H_{\mu_1\mu_2\mu_3}H_{\mu_4\mu_5\nu_1}R_{\mu_6}^{\nu_1}.
\end{equation}
This vanishes identically at the level of the $b^2h$ three-point function, since the Ricci tensor
vanishes on-shell for a single external graviton.

Ideally, we would like to demonstrate that the full six-dimensional CP-odd term (\ref{eq:IIBCPodd6})
vanishes on-shell at the non-linear level, and that the same holds for the ten-dimensional
IIB term
\begin{equation}
B_2\wedge\ft12[X_8(R(\Omega_+))-X_8(R(\Omega_-))].
\end{equation}
Unfortunately, the ten-dimensional expressions are quite cumbersome to manipulate, and
we have been unable to prove that this contribution vanishes on-shell.  Nevertheless, we have
seen that it indeed vanishes at the level of the $b^2h^3$ five-point function for a few sample
cases we have looked at.  This vanishing depends crucially on the relative factor of $-1/4$ in
the expression (\ref{eq:X8def}) for $X_8$, which suggests that any proof will have to emphasize
the structure of the $t_8$ tensor.

\section{Testing the couplings: T-duality}
\label{sec:tdual}

T-duality provides a very stringent test of the proposed couplings. Indeed it is quite unlikely
that higher derivative terms made solely out of curvatures will have correct T-duality properties;
fixing this problem is one of the consequences of inclusion of the $B$-field. In this section we
shall concentrate on showing that  the more topological part of the $\alpha'^3$ couplings,
namely  $B_2 \wedge [X_8(R^+) + X_8(R^-)]$, transforms correctly under T-duality. The
rest will be discussed in Appendix \ref{app:T-CP-even}.

For our purposes it should be sufficient to consider U(1) fibered backgrounds:
$U(1) \hookrightarrow M \stackrel{\pi}{\longrightarrow}M_B$. There exists on $M$ a globally
defined smooth one-form $e=e^9$ such that its curvature is a horizontal two-form
$d e = \pi^* T$ and $T \in H^2(M_B, \mathbb{Z})$.
For deriving  local formulae when considering a circle reduction, we shall simply take a
ten-dimensional metric of the form
\begin{equation}
ds^2=\eta_{\alpha\beta}e^\alpha e^\beta+(d\psi+\mathcal A)^2.
\end{equation}
Let
\begin{equation}
e^9=d\psi+\mathcal A,\qquad de^9=d\mathcal A=T,
\end{equation}
where $T$ denotes the field strength of the circle U(1). For simplicity we take the U(1) radius
to be constant and hence do not consider the scalar mode in the reduction (although
we shall discuss it later on).

Let us denote the isometry generator $k$ and assume that the three-form $H$ respects the
isometry, i.e.\ $\mathcal{L}_k H =0$. Note that, due to closure of $H$, we automatically have a
closed two-form (with integral periods), $\tilde T = \imath_k H$. Just like the curvature of
the circle bundle $T$, this form is in the second cohomology of the base manifold
$M_B$, namely $\tilde T \in H^2(M_B, \mathbb{Z})$.
T-duality  exchanges these two forms $T$ and $\tilde T$ (and the corresponding topological
numbers, $c_1$ and $\int \tilde T$ respectively).

It will be expedient to write the local expressions for  the antisymmetric tensor as well (we are tacitly assuming $\mathcal{L}_k B =0$):
\begin{equation}
\label{B-redu}
B_2=b_2+b_1\wedge e^9,\qquad H_3=dB_2=db_2-b_1\wedge T+\tilde T\wedge e^9,
\end{equation}
where $\tilde T=db_1$ is the winding U(1).  
(With some assumption on the $B$-field, the generalisation to an arbitrary $\mathbb{T}^n$
bundle, i.e.\ $n$ isometry vectors, is straightforward.).
Now write
\begin{equation}
H_3=h_3+\tilde T\wedge e^9,
\label{eq:H3red}
\end{equation}
where $h_3=db_2-b_1\wedge T$, so that $dh_3=-\tilde T\wedge T$.
Before going on, let us note that the two-form $b_2$ in the expansion (\ref{B-redu}) is not
invariant under T-duality. Instead, $b_2 \longrightarrow b_2 + b_1 \wedge \mathcal A$.
One can check that the horizontal three-form $h_3$ in (\ref{eq:H3red}) is T-duality invariant.
This will prove to be of crucial importance.

With these preliminaries out of the way, recall that what we wish to show is that the
CP-odd term in (\ref{eq:olLaghat}) transforms correctly under T-duality. But first we need
to work out some useful expressions for the curvatures on U(1) fibered manifolds. In
Sections~\ref{sec:x4red} and \ref{sec:x8red} we shall study the four-form and eight-form
polynomials in curvature that appear in six- and ten-dimensional effective actions respectively.
We shall then be ready to address the T-duality of the even-odd and odd-even IIA one-loop
amplitude contributions in Section~\ref{sec:doublet}. The CP-even parts of the one-loop
effective action (or to be more precise the equivalent terms in the simpler six-dimensional
case) are discussed Appendix \ref{app:T-CP-even}.

\subsection{Curvatures: The circle reduction}
\label{sec:circle}

We start by reducing the curvature $R(\Omega_+)$ where the connection with torsion is
$\Omega_+=\Omega+\fft12\mathcal H$ with $\mathcal H^{ab}=H_c{}^{ab}e^c$.  (Note that
we can obtain $\Omega_-$ by taking $H\to-H$.)  The reduction of $\Omega_+$ is
straightforward
\begin{eqnarray}
\Omega_+^{\alpha\beta}&=&\omega_+^{\alpha\beta}-\ft12T_-^{\alpha\beta}e^9,\nonumber\\
\Omega_+^{\alpha9}&=&-\ft12T_+^\alpha{}_\gamma e^\gamma,
\end{eqnarray}
where we have defined
\begin{equation}
T_\pm=T\pm\tilde T,
\end{equation}
and where $\omega_+=\omega+\fft12h$.

We now compute the curvature according to $R(\Omega_+)=d\Omega_++\Omega_+\wedge
\Omega_+$.  The result is
\begin{eqnarray}
R^{\alpha\beta}(\Omega_+)&=&R^{\alpha\beta}(\omega_+)
-\ft14T_+^\alpha{}_\gamma T_+^\beta{}_\delta e^\gamma e^\delta-\ft12T_-^{\alpha\beta}T
\nonumber\\
&&-\ft12(\nabla_\gamma T_-^{\alpha\beta}+\ft12h_\gamma{}^{\alpha\delta}T_{-\delta}{}^\beta
-\ft12T_-^{\alpha\delta}h_{\gamma\delta}{}^\beta)e^\gamma\wedge e^9,\nonumber\\
R^{\alpha9}(\Omega_+)&=&-\ft12(\nabla_\gamma T_+^\alpha{}_\delta
+\ft12h_\gamma{}^{\alpha\beta}T_{+\beta\delta})e^\gamma e^\delta
-\ft14T_-^{\alpha\delta}T_{+\delta\gamma}e^\gamma\wedge e^9,
\label{eq:Rfred}
\end{eqnarray}
where
\begin{equation}
R^{\alpha\beta}(\omega_+)=R^{\alpha\beta}+\ft12\nabla_\gamma h_\delta{}^{\alpha\beta}
e^\gamma\wedge e^\delta+\ft14h_\gamma{}^{\alpha\epsilon}h_{\delta\epsilon}{}^\beta
e^\gamma\wedge e^\delta.
\end{equation}
For convenience in notation, we introduce the one-form $T^\alpha$
\begin{equation}
T^\alpha\equiv T_\gamma{}^\alpha e^\gamma.
\end{equation}
Furthermore, the covariant derivative with torsion acts as
\begin{eqnarray}
D_+T^{\alpha\beta}&=&[\nabla_\gamma T^{\alpha\beta}+\ft12h_\gamma{}^\alpha{}_\delta
T^{\delta\beta}+\ft12h_\gamma{}^\beta{}_\delta T^{\alpha\delta}]e^\gamma,\nonumber\\
D_+T^\alpha&=&[\nabla_\gamma T_\delta{}^\alpha+\ft12h_\gamma{}^\alpha{}_\beta
T_\delta{}^\beta]e^\gamma\wedge e^\delta.
\end{eqnarray}
In this case, we have
\begin{eqnarray}
R^{\alpha\beta}(\Omega_+)&=&[R^{\alpha\beta}(\omega_+)-\ft14T_+^\alpha\wedge T_+^\beta
-\ft12T_-^{\alpha\beta}T]-\ft12D_+T_-^{\alpha\beta}e^9,\nonumber\\
R^{\alpha9}(\Omega_+)&=&\ft12D_+T_+^\alpha+\ft14T_-^\alpha{}_\beta T_+^\beta\wedge e^9.
\end{eqnarray}
Note that we have used the Bianchi identity on $T_+$ in order to rewrite $R^{\alpha9}(\Omega_+)$.

In components, (\ref{eq:Rfred}) becomes
\begin{eqnarray}
R_{\alpha\beta}{}^{\gamma\delta}(\Omega_+)&=&R_{\alpha\beta}{}^{\gamma\delta}(\omega_+)
-\ft14(T_{+\alpha}{}^\gamma T_{+\beta}{}^\delta-T_{+\alpha}{}^\delta T_{+\beta}{}^{\gamma})
-\ft12T_{\alpha\beta}T_-^{\gamma\delta},\nonumber\\
R_{\alpha9}{}^{\gamma\delta}(\Omega_+)&=&-\ft12(\nabla_\alpha T_-^{\gamma\delta}
+\ft12h_\alpha{}^{\gamma\beta}T_{-\beta}{}^\delta-\ft12T_-^{\gamma\beta}h_{\alpha\beta}{}^\delta),
\nonumber\\
R_{\alpha\beta}{}^{\gamma9}(\Omega_+)&=&-(\nabla_{[\alpha}T_+^\gamma{}_{\beta]}
+\ft12h_{[\alpha}{}^{\gamma\delta}T_{+|\delta|\beta]}),\nonumber\\
R_{\alpha9}{}^{\gamma9}(\Omega_+)&=&\ft14T_{+\alpha\delta}T_-^{\gamma\delta}.
\label{eq:riemanns1}
\end{eqnarray}
Note that the symmetry $R_{ABCD}(\Omega_+)=R_{CDAB}(\Omega_-)$ is hidden in these
expressions.  In particular, the relation on $R_{\alpha\beta\gamma9}$ requires use of the Bianchi identity $dT_\pm=0$, while the relation on $R_{\alpha\beta\gamma\delta}$ requires use of the
Bianchi identity $dh_3=-\tilde T\wedge T$.  This symmetry can be made more explicit by
writing
\begin{eqnarray}
R_{\alpha\beta}{}^{\gamma\delta}(\Omega_+)&=&\ft12[R_{\alpha\beta}{}^{\gamma\delta}
(\omega_+)+R^{\gamma\delta}{}_{\alpha\beta}(\omega_-)]\nonumber\\
&&-\ft18(T_{+\alpha}{}^\gamma T_{+\beta}{}^\delta-T_{+\alpha}{}^\delta T_{+\beta}{}^{\gamma})
-\ft18(T_{-\alpha}{}^\gamma T_{-\beta}{}^\delta-T_{-\alpha}{}^\delta T_{-\beta}{}^{\gamma})
\nonumber\\
&&-\ft18(T_{+\alpha\beta}T_+^{\gamma\delta}+T_{-\alpha\beta}T_-^{\gamma\delta}
+2T_{+\alpha\beta}T_-^{\gamma\delta}).
\end{eqnarray}
Curiously, note that all except for the last ($T_{+\alpha\beta}T_-^{\gamma\delta}$) term is
symmetrical under the interchange $T\leftrightarrow \tilde T$.

One contraction yields the Ricci tensor
\begin{eqnarray}
R_{\alpha\beta}(\Omega_+)&=&R_{\alpha\beta}(\omega_+)-\ft14T_{+\alpha\gamma}
T_{+\beta}{}^\gamma-\ft14T_{-\alpha\gamma}T_{-\beta}{}^\gamma,\nonumber\\
R_{\alpha9}(\Omega_+)&=&-\ft12(\nabla^\gamma T_{+\gamma\alpha}
+\ft12h_{\alpha\gamma\delta}T_+^{\gamma\delta}),\nonumber\\
R_{9\alpha}(\Omega_+)&=&-\ft12(\nabla^\gamma T_{-\gamma\alpha}
-\ft12h_{\alpha\gamma\delta}T_-^{\gamma\delta}),\nonumber\\
R_{99}(\Omega_+)&=&\ft14T_{+\alpha\beta}T_-^{\alpha\beta}.
\label{eq:riccis1}
\end{eqnarray}
The symmetry between $T$ and $\tilde T$ is broken in the Ricci component $R_{99}(\Omega_+)$.
An additional contraction yields the Ricci scalar
\begin{equation}
R(\Omega_+)=R(\omega_+)-\ft14T_+^2-\ft14T_-^2+\ft14T_+T_-
=R(\omega_+)-\ft14T^2-\ft34\tilde T^2.
\end{equation}
Note that
\begin{eqnarray}
R_{\alpha\beta}(\omega_+)&=&R_{\alpha\beta}-\ft12\nabla^\gamma h_{\alpha\beta\gamma}
-\ft14h_{\alpha\gamma\delta}h_\beta{}^{\gamma\delta},\nonumber\\
R(\omega_+)&=&R-\ft14h^2,
\end{eqnarray}
so in particular
\begin{equation}
R(\Omega_+)=R-\ft14h^2-\ft14T^2-\ft34\tilde T^2.
\label{eq:Ricciscalar+}
\end{equation}
This expression is not symmetrical between $T$ and $\tilde T$.

Reduction of the two-derivative action is somewhat interesting.  Ignoring the dilaton, we start
with the simple action
\begin{equation}
e^{-1}\mathcal L=R(\Omega_+)+\ft16H^2=R-\ft1{12}H^2.
\end{equation}
Note that this is not just $R(\Omega+)$.  Reduction of $R(\Omega_+)$ is given in
(\ref{eq:Ricciscalar+}), and reduction of $H^2$ follows from (\ref{eq:H3red})
\begin{equation}
H^2=h^2+3\tilde T^2.
\end{equation}
Putting this together gives
\begin{equation}
e^{-1}\mathcal L=R(\Omega_+)+\ft16H^2
=[R-\ft14h^2-\ft14T^2-\ft34\tilde T^2]+\ft16[h^2+3\tilde T^2]
=R-\ft1{12}h^2-\ft14T^2-\ft14\tilde T^2,
\end{equation}
which now is symmetrical between $T$ and $\tilde T$.

\subsection{The reduction of $X_4$}
\label{sec:x4red}

As we had already seen in Section \ref{sec:cp-oddB}, our couplings  have six-dimensional
counterparts with very closely related properties that, however, are lighter and easier to
deal with. Hence as a warm-up we shall start from $X_4$, a four-form quadratic in curvature,
and then turn to $X_8$.

We now take
\begin{equation}
X_4(\Omega_+)\equiv- \fft{1}{8 \pi^2} R^{ab}(\Omega_+)\wedge R^{ab}(\Omega_+),
\label{eq:X4Omega+}
\end{equation}
and reduce along $e^9$.  The result is
\begin{equation}
X_4(\Omega_+)=\tilde X_4+\tilde X_3\wedge e^9,
\end{equation}
where $\tilde X_3$ is a closed form ($d \tilde X_3= 0$) but $\tilde X_4$ is not
($d \tilde X_4 = \tilde X_3 \wedge T$).
We can now construct
\begin{eqnarray}
 {- 8 \pi^2} \left( \tilde X_4-\tilde X_2\wedge T \right) &=&R^{\alpha\beta}(\omega_+)\wedge R^{\alpha\beta}(\omega_+)
-\ft12R^{\alpha\beta}(\omega_+)\wedge T_+^\alpha{}_\gamma T_+^\beta{}_{\delta}
e^\gamma\wedge e^\delta  \nonumber\\
&&+\ft12(\nabla_\gamma T_+^\alpha{}_\delta+\ft12h_\gamma{}^{\alpha\rho}T_+^\rho{}_\delta)
(\nabla_\epsilon T_+^\alpha{}_{\iota}+\ft12h_\epsilon{}^{\alpha\sigma}T_+^\sigma{}_\iota)
e^\gamma\wedge e^\delta\wedge e^\epsilon\wedge e^\iota  ,\nonumber\\
\tilde X_3&=&d\tilde X_2,\nonumber\\
 {- 8 \pi^2} \tilde X_2&=&[-R^{\alpha\beta}(\omega_+)+\ft14T_+^\alpha{}_\gamma T_+^\beta{}_\delta
e^\gamma\wedge e^\delta+\ft14T_-^{\alpha\beta}T]T_-^{\alpha\beta}.
\label{eq:X4X2}
\end{eqnarray}

In particular, note that the combination $\tilde X_4-\tilde X_2\wedge T$  in (\ref{eq:X4X2}) is
not only closed but is naturally built out of the combination $T_+$.  As a result, it is invariant
under $T\leftrightarrow\tilde T$.%
\footnote{Since $\tilde X_2$ is defined as an inverse derivative on the closed form
$\tilde X_3$ (formally $\tilde X_2 = d^{-1} \imath_k X_4 $, where $\imath_k$ denotes the
contraction with the isometry generator $k$), it is ambiguous up to closed forms. As a
consequence, there is an ambiguity in the definition of the closed four-form
$ \tilde X_4-\tilde X_2\wedge T$. Demanding that the latter is globally-defined and
T-duality invariant fixes the ambiguity.  In this sense, (\ref{eq:X4X2}) gives the unique
expressions for $\tilde X_2$ and  $\tilde X_4-\tilde X_2\wedge T$ that appear in the
effective action.}
Since this term arises from the even-odd plus the
odd-even spin structure sector of the IIA loop amplitude, the actual contribution we
are interested in is $\fft12[X_4(\Omega_+)+X_4(\Omega_-)]$, which picks out terms
even in $H$.  Nevertheless, $\tilde X_4-\tilde X_2\wedge T\big|_{\mathrm{even\,in}\,H}$
remains invariant under the interchange $T\leftrightarrow\tilde T$.%
\footnote{We shall return to this point in Section~\ref{sec:hetBI}.}

We can explicitly ``average" (\ref{eq:X4X2}) over the even-odd and odd-even spin structure
sectors.  For $\tilde X_4-\tilde X_2\wedge T$, we first note that
\begin{equation}
\label{eq:X_4-nonave}
 {- 8 \pi^2} \left(  \tilde X_4-\tilde X_2\wedge T \right) =R^{\alpha\beta}(\omega_+)
 \wedge R^{\alpha\beta}(\omega_+)
-\ft12R^{\alpha\beta}(\omega_+)\wedge T_+^\alpha{}_\gamma T_+^\beta{}_{\delta}
e^\gamma\wedge e^\delta
+\ft18D_\alpha(\omega_-)T_+\wedge D^\alpha(\omega_-)T_+,
\end{equation}
where
\begin{equation}
D_\mu(\Omega_\pm)V^\alpha{}_\beta
=\nabla_\mu V^\alpha{}_\beta\pm\ft12H_\mu{}^\alpha{}_\lambda V^\lambda{}_\beta
\pm\ft12H_{\mu\beta}{}^\lambda V^\alpha{}_\lambda,
\end{equation}
and similarly for $\omega_\pm$.  As a result, we have
\begin{eqnarray}
\label{eq:ave1}
 {- 8 \pi^2} \left(  \tilde X_4-\tilde X_2\wedge T \right) \Big|_{\mathrm{averaged}}&=&
\ft12[R^{\alpha\beta}(\omega_+)\wedge R^{\alpha\beta}(\omega_+)
+R^{\alpha\beta}(\omega_-)\wedge R^{\alpha\beta}(\omega_-)]\nonumber\\
&&-\ft14[R^{\alpha\beta}(\omega_+)T_+^\alpha{}_\gamma T_+^\beta{}_{\delta}
+R^{\alpha\beta}(\omega_-)T_-^\alpha{}_\gamma T_-^\beta{}_{\delta}]
\wedge e^\gamma\wedge e^\delta\nonumber\\
&&+\ft1{16}[D_\alpha(\omega_-)T_+\wedge D^\alpha(\omega_-)T_+
+D_\alpha(\omega_+)T_-\wedge D^\alpha(\omega_+)T_-],\nn\\
\end{eqnarray}
which is invariant under T-duality. From other side
\begin{eqnarray}
\label{eq:ave2}
 {- 8 \pi^2}  \tilde X_2\Big|_{\mathrm{averaged}}\!\!&=&\!\!
-\ft14[R^{\alpha\beta}(\omega_+)+R^{\alpha\beta}(\omega_-)](T_++T_-)^{\alpha\beta}
+\ft14[R^{\alpha\beta}(\omega_+)-R^{\alpha\beta}(\omega_-)](T_+-T_-)^{\alpha\beta}
\nonumber\\
&&+\ft1{16}[(T_+^\alpha{}_\gamma T_+^\beta{}_\delta+T_-^\alpha{}_\gamma T_-^\beta{}_\delta)
(T_++T_-)^{\alpha\beta}
-(T_+^\alpha{}_\gamma T_+^\beta{}_\delta-T_-^\alpha{}_\gamma T_-^\beta{}_\delta)
(T_+-T_-)^{\alpha\beta}]e^\gamma\wedge e^\delta\nonumber\\
&&+\ft1{16}(T_+^2+T_-^2)(T_++T_-)
\end{eqnarray}
transforms as a doublet under T-duality. Let us first note that all averaged quantities are
derived from  $\left( X_4 (\Omega_+) + X_4(\Omega_-) \right)/2$ and contain only even
powers of the NSNS three-form $H$. Hence $\tilde X_2\big|_{\mathrm{averaged}}$ is
even in the combined power of $h_3$ and $\tilde T$. It naturally splits as 
\beq
\label{eq:ave22}
 {- 8 \pi^2}  \tilde X_2\Big|_{\mathrm{averaged}}  =  {- 8 \pi^2}  \tilde x_2 \Big|_{\mathrm{averaged}}  +\ft1{16}(T_+^2+T_-^2)(T_++T_-) \, ,
\eeq
where ${- 8 \pi^2}  \tilde x_2 \big|_{\mathrm{averaged}}$ are just the first three lines of
(\ref{eq:ave2}).  Both parts have a nice action of T-duality. For the first
\begin{equation}
\mbox{T-duality}: \qquad \fft12[ \tilde x_2 (\omega_+ , T_+)+ \tilde x_2 ( \omega_-, T_-) ]
 \quad\leftrightarrow  \quad - \fft12[\tilde x_2 (\omega_+ , T_+) -  \tilde x_2 ( \omega_-, T_-)] \, ,
\end{equation}
and the averaged and anti-averaged two-forms $\tilde x_2$ form a doublet under T-duality
group. The second part is even more transparent:
\begin{equation}
\mbox{T-duality}:\qquad  (T^2+\tilde T^2) \,T  \quad \leftrightarrow  \quad (T^2+\tilde T^2)\, \tilde T\, .
\end{equation}
Note that the resulting expressions are now odd  in the combined power of $h_3$ and
$\tilde T$.%
\footnote{Such a flip of parity in $H$ in characteristic classes built out of the connection
with torsion plays an important role in D-brane worlvolume couplings to RR fields
\cite{BGR,GGM,GGM2,KM}. These couplings, along with the pull-backs of RR fields wedged with
characteristic polynomials in curvatures, contain partially contracted terms. Terms with an
even/odd number of contractions have even/odd powers of the $B$-field.}

\subsection{The reduction of $X_8$}
\label{sec:x8red}

We now return to ten dimensions and consider the reduction of $X_8(\Omega_+)$.  In
order to do so, we need to consider both the single trace and double
trace pieces.  We define 
\begin{eqnarray}
X_8^S(\Omega_+) &\equiv& \fft{1}{(2 \pi)^4 192} \Tr R(\Omega_+)^4,\nn\\
X_8^D(\Omega_+) &\equiv&  \fft{1}{ 4 (2 \pi)^4 192}  (\Tr R(\Omega_+)^2)^2
=  \fft{1}{192}  X_4 (\Omega_+) \wedge X_4 (\Omega_+),
\end{eqnarray}
and consider the single and double trace pieces separately:
\begin{equation}
\label{eq:X8SD}
X_8(\Omega_\pm)= X^S_8(\Omega_\pm) - X^D_8(\Omega_\pm)\,.
\end{equation}
Before turning to explicit expressions, let us note that, as for $X_4$, we may perform a
horizontal-vertical decomposition 
\begin{equation}
X_8 (\Omega_+)=\tilde X_8+\tilde X_7\wedge e^9,
\end{equation}
where $d \tilde X_8 = \tilde X_7 \wedge T$ and $d \tilde X_7 =0$. As before we define
$\tilde X_6$ by $\tilde X_7=d\tilde X_6$, and take the closed combination
$\tilde X_8-\tilde X_6\wedge T$. Similar decompositions hold for $X^S_8$ and $X^D_8$.

We first consider the single trace part:
\begin{equation}
X^S_8(\Omega_+)=\tilde X^S_8+\tilde X^S_7\wedge e^9,
\end{equation}
where
\begin{eqnarray}
192 (2 \pi)^4  \tilde X^S_8&=&[R^{\alpha\beta}(\omega_+)-\ft14T_+^\alpha T_+^\beta-\ft12T_-^{\alpha\beta}T]
[R^{\beta\gamma}(\omega_+)-\ft14T_+^\beta T_+^\gamma-\ft12T_-^{\beta\gamma}T]\nonumber\\
&&\qquad\times[R^{\gamma\delta}(\omega_+)-\ft14T_+^\gamma T_+^\delta-\ft12T_-^{\gamma\delta}T]
[R^{\delta\alpha}(\omega_+)-\ft14T_+^\delta T_+^\alpha-\ft12T_-^{\delta\alpha}T]\nonumber\\
&&-D_+T_+^\alpha[R^{\alpha\beta}(\omega_+)-\ft14T_+^\alpha T_+^\beta-\ft12T_-^{\alpha\beta}T]
[R^{\beta\gamma}(\omega_+)-\ft14T_+^\beta T_+^\gamma-\ft12T_-^{\beta\gamma}T]
D_+T_+^\gamma\nonumber\\
&&+\ft18D_+T_+^\alpha D_+T_+^\alpha D_+T_+^\beta D_+T_+^\beta.
\label{eq:tX8}
\end{eqnarray}
This expression may be expanded in powers of $T_-$.  By explicit reduction, we find that the
closed combination $\tilde X^S_8-\tilde X^S_6\wedge T$ has no powers of $T_-$.  Instead,
all $T_-$ terms are contained in $\tilde X^S_6$.  What this means is that we may simply set
$T_-=0$ in (\ref{eq:tX8}) to obtain
\begin{eqnarray}
192 (2 \pi)^4  \left(\tilde X^S_8-\tilde X^S_6\wedge T \right) \!\!&=&\!\!
[R^{\alpha\beta}(\omega_+)-\ft14T_+^\alpha T_+^\beta]
[R^{\beta\gamma}(\omega_+)-\ft14T_+^\beta T_+^\gamma]\nn\\
&&\qquad\times[R^{\gamma\delta}(\omega_+)-\ft14T_+^\gamma T_+^\delta]
[R^{\delta\alpha}(\omega_+)-\ft14T_+^\delta T_+^\alpha]\nonumber\\
&&-D_+T_+^\alpha[R^{\alpha\beta}(\omega_+)-\ft14T_+^\alpha T_+^\beta]
[R^{\beta\gamma}(\omega_+)-\ft14T_+^\beta T_+^\gamma]D_+T_+^\gamma\nonumber\\
&&+\ft18D_+T_+^\alpha D_+T_+^\alpha D_+T_+^\beta D_+T_+^\beta.
\label{eq:tX8X6}
\end{eqnarray}
Note that $\tilde X^S_6$ is defined by $\tilde X^S_7=d\tilde X^S_6$.  The explicit form of
$\tilde X^S_6$ is somewhat unwieldy, but may be obtained by subtracting (\ref{eq:tX8X6})
from (\ref{eq:tX8}) and then removing the extra two-form $T$ from the result.  We have
checked that this indeed works. Some useful identities for performing this check are
\begin{equation}
D_+R^{\alpha\beta}(\omega_+)=0,\qquad
D_+D_+T_+^\alpha=R^{\alpha\beta}(\omega)_+T_+^\beta.
\end{equation}

For the double trace (factorized) piece, we first work with
\begin{equation}
X_4(\Omega_+)=\tilde X_4+\tilde X_3\wedge e^9,
\end{equation}
where $X_4(\Omega_+)$ was defined in (\ref{eq:X4Omega+}).
The result was already given in (\ref{eq:X4X2}), but may be rewritten as
\begin{eqnarray}
{- 8 \pi^2}  \tilde X_4&=&
[R^{\alpha\beta}(\omega_+)-\ft14T_+^\alpha T_+^\beta-\ft12T_-^{\alpha\beta}T]
[R^{\beta\alpha}(\omega_+)-\ft14T_+^\beta T_+^\alpha-\ft12T_-^{\beta\gamma}T]
-\ft12D_+T_+^\alpha D_+T_+^\alpha,\nonumber\\
 {- 8 \pi^2} \left(  \tilde X_4-\tilde X_2\wedge T \right) &=&[R^{\alpha\beta}(\omega_+)-\ft14T_+^\alpha T_+^\beta]
[R^{\beta\alpha}(\omega_+)-\ft14T_+^\beta T_+^\alpha]
-\ft12D_+T_+^\alpha D_+T_+^\alpha.
\end{eqnarray}
Finally, with
\begin{equation}
X^D_8(\Omega_+)=\tilde X^D_8+\tilde X^D_7\wedge e^9,
\end{equation}
we have
\begin{eqnarray}
\tilde X^D_8&=&  \fft{1}{192} \,  \tilde X_4\wedge \tilde X_4,\nonumber\\
\tilde X^D_8-\tilde X^D_6\wedge T&=&  \fft{1}{ 192}  (\tilde X_4-\tilde X_2\wedge T)
\wedge(\tilde X_4-\tilde X_2\wedge T).
\end{eqnarray}
Subtracting these expressions gives
\begin{eqnarray}
\tilde X^D_6\wedge T&=&  \fft{1}{ 192} \left( 2\tilde X_4\wedge\tilde X_2\wedge T
-\tilde X_2\wedge\tilde X_2\wedge T\wedge T \right) \nonumber\\
&=&\fft{1}{ 192} \left( 2(\tilde X_4-\tilde X_2\wedge T)\wedge\tilde X_2\wedge T
+\tilde X_2\wedge\tilde X_2\wedge T\wedge T \right) .
\end{eqnarray}
Removing one $T$ gives
\begin{equation}
\tilde X^D_6=\fft{1}{ 192} \left( 2(\tilde X_4-\tilde X_2\wedge T)\wedge\tilde X_2
+\tilde X_2\wedge\tilde X_2\wedge T\right) .
\end{equation}

We can put everything together now using (\ref{eq:X8SD}). Just as for $X_4$, we may
check that the combination $\tilde X_8(\Omega_\pm) -  \tilde X_6(\Omega_\pm) \wedge T $
and, as consequence, the averaged closed eight-form are T-duality invariant. Similarly to
(\ref{eq:ave2}), the six-form $\tilde X_6\big|_{\mathrm{averaged}}
= \left( \tilde X^S_6 - \tilde X^D_6 \right)\big|_{\mathrm{averaged}}$ can again be naturally
split into two parts, both of which transform as part of a doublet under the T-duality
group. In particular, for most of the six-form, T-duality exchanges the averaged and the
(minus) anti-averaged expressions.

\subsection{A doublet of anomalous couplings}
\label{sec:doublet}

We may finally turn to the discussion of the T-duality properties of the coupling
$B_2 \wedge [X_8(\Omega^+) + X_8(\Omega^-)]$. Using the horizontal-vertical
decomposition of the forms discussed above, we may write
\begin{eqnarray}
\label{eq:reduu}
\int_{M_{10}}B \wedge [X_8(\Omega^+) + X_8(\Omega^-)] &=&  \int_{M_9}b_1 \wedge [\tilde X_8(\Omega^+) + \tilde X_8(\Omega^-)] + b_2 \wedge [\tilde X_7(\Omega^+) + \tilde X_7(\Omega^-)] \nonumber \\ 
&=& \int_{M_9}b_1\wedge(\tilde X_8-\tilde X_6\wedge T)\big|_{\mathrm{averaged}}
-h_3\wedge\tilde X_6\big|_{\mathrm{averaged}} \, .
\end{eqnarray}
Since we start from an expression that is invariant under both NSNS gauge transformations
$B \longrightarrow B + d \Lambda_1$ and ten-dimensional diffeomorphisms, the reduction
should yield an expression invariant under NS gauge transformations, nine-dimensional
diffeomorphisms and U(1). The second line in (\ref{eq:reduu}) makes such properties explicit,
notably due to the appearance of the closed eight-form $\tilde X_8-\tilde X_6\wedge T$
in the first term.

The T-duality properties are now equally transparent. Both $b_1$ and
$X_6\big|_{\mathrm{averaged}}$ are doublets under T-duality, while $h_3$ and
$(\tilde X_8-\tilde X_6\wedge T)\big|_{\mathrm{averaged}}$ are invariant. Hence the
entire coupling transforms as a doublet. Namely (\ref{eq:reduu}) is mapped to the IIB version
of the coupling under T-duality. Note that the latter vanishes in the covariant ten-dimensional
IIB action, but due to the contribution of the winding modes on the circle it appears in the
nine-dimensional acton.
Perhaps more covariantly, the coupling can be written on an 11-dimensional manifold
$N_{11}$, where $\partial N_{11} = M_{10}$ (and correspondingly $\partial N_{10} = M_9$)
\begin{equation}
\int_{N_{11}} H_3 \wedge [X_8(\Omega^+) + X_8(\Omega^-)] =  \int_{N_{10}} \tilde T \wedge(\tilde X_8-\tilde X_6\wedge T)\big|_{\mathrm{averaged}}
- d \left( h_3\wedge\tilde X_6\big|_{\mathrm{averaged}} \right) \, .
\end{equation}

So far we have only discussed the CP-odd part of the $\alpha'^3$ correction.
Clearly the CP-even part should behave in a similar way, and we shall discuss these in
Appendix \ref{app:T-CP-even}. Unfortunately these are somewhat heavy (and they also
appear to be less constrained than the couplings discussed in this section), and so we
write down explicitly only the completions of the four-derivative $X_4$ terms and not those
of the eight-derivative $X_8$.  In any case, the complete expression for the eight-derivative
terms is not known to all orders in $H$. The study of the T-duality of the eight-derivative
couplings appears to be the most feasible way of fixing these terms.

\subsection{The correspondence space}
\label{sec:corsp}

It is natural to ask if any of the formal approaches to T-duality are helpful in understanding
the $\alpha'^3$ couplings and fixing their ambiguities.
In this regard, the
correspondence space is a very useful tool for discussions of T-duality. We shall not give
any abstract definitions, but will just look at the present case of interest. If the $H$-flux satisfies
the Bianchi identity $dH=0$ and respects the isometry, i.e.\ has vanishing Lie derivative with
respect to the isometry generator $k$, namely $\mathcal{L}_kH=0$, it produces a closed two
form (with integral periods) given by the contraction $\tilde{T}= \imath_k H$. In fact, if the
space-time is a circle bundle over a base $M_B$, then $\tilde T$ is horizontal, i.e. is a
two-form on $B$, and we can think of the geometrization of $H$ by combining it with the
U(1) bundle over $B$ into a $\mathbb T^2$ bundle. The total space of this principal fibration
will be called the correspondence space, and the exchange symmetry of the two circles will
correspond to T-duality: the curvature two forms $T$ and $\tilde T$ form a doublet under the
T-duality group, and the topological numbers $\left(2 \pi c_1 (M), \int_M H
= \int_{M_B} \tilde T\right)$ for the original background become $\left( \int_{\tilde M} \tilde H
= \int_{M_B} T,  2 \pi c_1(\tilde M)\right)$ \cite{BEM}.  By construction, the quantities computed on this
formal double-fibered space will have the exchange symmetry and hence be T-duality invariant.

Now suppose we do not use the connection with torsion, but instead reduce $X_4$ on
two commuting circles.  The geometry of this reduction should yield expressions with
exchange symmetry between these two circles built in.
Since the reduction is fairly general, we take a metric of the form
\begin{equation}
ds^2=\eta_{\alpha\beta}e^\alpha e^\beta+\sum_i(d\psi_i+\mathcal A^i)^2,
\end{equation}
with an arbitrary number of circles labeled by $i$ and then restrict to $i=2$,  which is the case we want to consider for T-duality. Let
\begin{equation}
e^i=d\psi_i+\mathcal A^i,\qquad de^i=T^i.
\end{equation}
The spin connections are
\begin{equation}
\omega^{\alpha\beta}=\omega^{\alpha\beta}-\ft12T^{i\,\alpha\beta}e^i,\qquad
\omega^{\alpha i}=-\ft12T^{i\,\alpha}{}_\beta e^\beta,\qquad
\omega^{ij}=0.
\end{equation}
This allows us to compute the curvature two-form
\begin{eqnarray}
R^{\alpha\beta}&=&R^{\alpha\beta}-\ft14T^{i\,\alpha}{}_\gamma T^{i\,\beta}{}_\delta
e^\gamma\wedge e^\delta-\ft12T^{i\,\alpha\beta}T^i
-\ft12\nabla_\gamma T^{i\,\alpha\beta}e^\gamma\wedge e^i+\ft14T^{i\,\alpha\gamma}
T^{j\,\gamma\beta}e^i\wedge e^j,\nonumber\\
R^{\alpha j}&=&-\ft12\nabla_\gamma T^{j\,\alpha}{}_\delta e^\gamma\wedge e^\delta
-\ft14T^{j\,\delta}{}_\gamma T^{k\,\alpha\delta}e^\gamma\wedge e^k,\nonumber\\
R^{ij}&=&-\ft14T^{i\,\alpha}{}_\gamma T^j_{\alpha\delta}e^\gamma\wedge e^\delta.
\end{eqnarray}

To keep the formula light, we shall just discuss $X_4$ here; the discussion of the full coupling is very much analogous. We consider $X_4=R^{ab}\wedge R^{ab}$, so that
\begin{equation}
X_4=\tilde X_4+\tilde X_3^ie^i+\ft12\tilde X_2^{ij}e^i\wedge e^j+\ft1{3!}\tilde X_1^{ijk}
e^i\wedge e^j\wedge e^k+\ft1{4!}\tilde X_0^{ijkl}e^i\wedge e^j\wedge e^k\wedge e^l.
\end{equation}
Note that $\tilde X_1^{ijk}$ and $\tilde X_0^{ijkl}$ both vanish in the case we are interested in,
where there are only two U(1)'s.  Closure of $X_4$ then gives
\begin{eqnarray}
&&d\tilde X_4-\tilde X_3^iT^i=0,\qquad d\tilde X_3^i-\tilde X_2^{ij}T^j=0,\qquad
d\tilde X_2^{ij}-\tilde X_1^{ijk}T^k=0,\nonumber\\
&&d\tilde X_1^{ijk}-\tilde X_0^{ijkl}T^l=0,\qquad
d\tilde X_0^{ijkl}=0.
\end{eqnarray}

Explicit reduction gives
\begin{eqnarray}
\label{eq:cor1}
 {- 8 \pi^2} \left(  \tilde X_4-\hat X_2^iT^i \right)&=&R^{\alpha\beta}R^{\alpha\beta}-\ft12R^{\alpha\beta}
T^{i\,\alpha}{}_\gamma T^{i\,\beta}{}_\delta e^\gamma\wedge e^\delta
+\ft12\nabla_\gamma T^{i\,\alpha}{}_\delta\nabla_\epsilon T^{i\,\alpha}{}_\iota
e^\gamma\wedge e^\delta\wedge e^\epsilon\wedge e^\iota,\nonumber\\
\tilde X_3^i&=&d\hat X_2^i+\hat X_1^{ij}T^j,\nonumber\\
\tilde X_2^{ij}&=&d\hat X_1^{ij}+\hat X_0^{ijk}T^k,\nonumber\\
\tilde X_1^{ijk}&=&d\hat X_0^{ijk},\nonumber\\
\tilde X_0^{ijkl}&=&0.
\end{eqnarray}
where
\begin{eqnarray}
\label{eq:cor2}
 {- 8 \pi^2} \, \hat X_2^i&=&[-R^{\alpha\beta}+\ft14T^{j\,\alpha}{}_\gamma T^{j\,\beta}{}_\delta
e^\gamma\wedge e^\delta+\ft14T^{j\,\alpha\beta}T^j]T^{i\,\alpha\beta},\nonumber\\
 {- 8 \pi^2} \, \hat X_1^{ij}&=&-\ft14[T^{i\,\alpha\beta}\nabla_\gamma T^{j\,\alpha\beta}
-T^{j\,\alpha\beta}\nabla_\gamma T^{i\,\alpha\beta}]e^\gamma,\nonumber\\
 {- 8 \pi^2} \, \hat X_0^{ijk}&=&\ft12T^{i\,\alpha\beta}T^{j\,\beta\gamma}T^{k\,\gamma\alpha}.
\end{eqnarray}
Comparison with previous sections (notably Section~\ref{sec:x4red}) shows that the
expressions agree if $h_3 = 0$.
Indeed, let us restrict $i,j =1,2$ and take $(T^1, T^2) = (T, \tilde T)$. Then it is not hard to see
that, provided $h_3=0$, $\tilde X_4-\hat X_2^iT^i$  in (\ref{eq:cor1}) coincides with
$\tilde X_4-\tilde X_2\wedge T\big|_{\mathrm{averaged}}$ in (\ref{eq:ave1}). Moreover, we
can see that $\hat X_2^1$ in (\ref{eq:cor2}) agrees with $\bigl( {- 8 \pi^2}  \tilde x_2
\big|_{\mathrm{averaged}}  +\ft1{16}(T_+^2+T_-^2)(T_++T_-)  \bigr)\big|_{h_3=0}$, and
$\hat X_2^2$ agrees with $\bigl( {- 8 \pi^2}  \tilde x_2 \big|_{\mathrm{anti-averaged}}
+\ft1{16}(T_+^2+T_-^2)(T_+ - T_-)  \bigr)\big|_{h_3=0}$. Here we had to use the vanishing
of $h_3$ which implies the vanishing of either $T$ or $\tilde T$.  (We need to impose
$h_3=db_2-b_1\wedge T = 0=dh_3 = - T \wedge \tilde T$). For $h_3 \neq 0$ one has to worry about the presence
of $\omega_+$ and $\omega_-$ in $\tilde X_4-\hat X_2^iT^i$; for $\tilde X^i_2$ the
mismatch goes beyond the curvature terms.

This is not very surprising. While the correspondence space is a way of geometrizing the
$B$-field, it does not extend to the entire gerbe structure but only the part of it that, due to
the isometry $k$,  reduces to that of an ordinary U(1) bundle.  The symmetry of two
U(1) bundles (and the curvatures $T$ and $\tilde T$) is made explicit, but any twisting
by the horizontal component $h_3$ is at best to be performed by hand. 

We hope that working on the generalized tangent bundle (which does geometrize the entire
$B$-field) will be adequate. Furthermore, one may notice that imposing $h_3=0$ allows us to
construct isomorphisms between the extension of the correspondence space and the
generalized tangent bundle. We hope to return to these questions in the future.

\subsection{Side remark: T-duality invariance of the heterotic Bianchi identity}
\label{sec:hetBI}

In view of how constraining T-duality is, it may seem puzzling that
$\tilde X_4-\tilde X_2\wedge T$ is T-duality invariant even without averaging. We close
this section by discussing the {\sl raison d'\^etre}  of such an ostensibly ``bonus" symmetry.
In fact this invariance (without averaging!) is needed for T-duality invariance of the heterotic
Bianchi identity. Indeed let us note that in $dH=\ft14\alpha'\tr R(\Omega_+)\wedge R(\Omega_+)
-\cdots$, all (even and odd) powers of $\HH$ appear and there is no averaging.

To the best of our knowledge, T-duality invariance of the heterotic Bianchi identity with
the curvature terms included has been discussed only for supersymmetric backgrounds
\cite{EM, Becker:2009df, MMT}. Provided these backgrounds have an isometry, one can show that the
twisted connection is horizontal. Hence $\tr R(\Omega_+)\wedge R(\Omega_+)$ is horizontal
as well and its contraction with the isometry vector vanishes.%
\footnote{The gauge part of the Bianchi identity has received much more attention, at least
when the gauge group is broken to a sum of Abelian factors. We tacitly consider the simplest
case without Wilson lines, and take $\Tr F\wedge F $ to be horizontal as well. }
By decomposing the three form in terms of horizontal and vertical components,
$H = H_3 + H_2 \wedge e$, we get two components of the Bianchi identity:
\bea
d H_3 + H_2 \wedge T &=& \ft14\alpha'\tr R(\Omega_+)\wedge R(\Omega_+)
-\ft14\alpha'\Tr F\wedge F, \nn \\
d H_2 &=& 0.
\eea
We conclude from this that $H_2 = \tilde{T}$ is the curvature of the dual circle bundle,
and arrive at the lower dimensional BI:
\beq
d H_3  = \ft14\alpha'\tr R(\omega_+)\wedge R(\omega_+) -\ft14\alpha'\Tr F\wedge F
 - T \wedge\tilde{T}.
\eeq

Of course, we expect T-duality to hold not only in the supersymmetric backgrounds. Using
again $X_4 (\Omega_+) \equiv  - \fft{1}{8 \pi^2} \tr R(\omega_+)\wedge R(\omega_+)$ and
the notation from Section~\ref{sec:x4red},  we verify that the heterotic Bianchi identity on a
background with an isometry yields:
\bea
d H_3 + H_2 \wedge T &=& - 2 \pi^2 \alpha' \tilde X_4(\Omega_+) -\ft14\alpha'\Tr F\wedge F,\nn\\
d H_2 &=& - 2 \pi^2 \alpha' \tilde X_3^+.
\label{eq:BIred}
\eea
Note that since $H_2$ is not closed, it cannot be exchanged with the curvature of the U(1)
bundle $T$. However, since $\tilde X_3(\Omega_+) = d \, \tilde X_2(\Omega_+)$ is exact, we
can take 
\bea
\label{eq:T-check}
H_2 = \check{T}  - 2 \pi^2 \alpha'  \tilde X_2(\Omega_+),
\eea
and hence define a closed two-form $\check{T}$, which is the curvature of the dual circle
bundle. The first line in (\ref{eq:BIred}) is now
\beq
\label{eq:HetBI}
d H_3 = -2 \pi^2 \alpha'  (\tilde X_4 - T \wedge \tilde X_2)  -\ft14\alpha'\Tr F\wedge F - T \wedge \check{T} \,.
\eeq
In lowest order in $\alpha'$, $ (\tilde X_4 - T \wedge \tilde X_2)$ coincides with the expression
computed in (\ref{eq:X_4-nonave}) which is T-duality invariant.%
\footnote{Note that the closed two-form $\check T$ defined in (\ref{eq:T-check}),  which is
exchanged with the curvature two-form $T$, differs from $\tilde T$ used, e.g., in
Section~\ref{sec:x4red}.}
As a consequence, the reduced heterotic Bianchi identity (\ref{eq:HetBI}) is also T-duality
invariant to order $\alpha'$. Proving the invariance to all orders in $\alpha'$ should require
inclusion of higher order corrections to the Bianchi identity and may also require a higher-order
modification of the T-duality.

\section{Six-dimensional Heterotic/IIA duality}
\label{sec:sixd}

Since the full supersymmetric completion of the eight-derivative terms go up to
$\mathcal O(H^8)$, one would need to compute up to eight-point string amplitudes in order
to completely pin down their form.  Unfortunately, although there does not appear to be
any conceptual issues in working out the eight-point function, technically it appears to be
quite a challenge.  However, by reducing the IIA string to six dimensions on $K3$, we may
perform a test of our conjecture in a simpler yet related setting.

The resulting six-dimensional theory with 16 supercharges first admits four-derivative
corrections of the form $\alpha'R^2$ along with its supersymmetric completion.  Moreover,
the tree-level ten-dimensional heterotic action is known at the four-derivative level
\cite{Metsaev:1987zx,Bergshoeff:1988nn,Bergshoeff:1989de,Chemissany:2007he},
and it may be straightforwardly reduced to six-dimensions on $T^4$.  heterotic/IIA
duality then maps this tree-level heterotic correction to the one-loop IIA correction that
we are interested in.

\subsection{The heterotic theory}

The four-derivative corrections arising from the heterotic string can be written concisely
in terms of a connection with torsion $\Omega_+=\Omega+\fft12H$ and its corresponding
curvature
\begin{equation}
R(\Omega_+)=R(\Omega)+\ft12d\mathcal H+\ft14\mathcal H\wedge\mathcal H,\qquad
\mathcal H^{ab}=H_\mu{}^{ab}dx^\mu.
\end{equation}
The bosonic Lagrangian takes the form
\begin{equation}
e^{-1}\mathcal L=e^{-2\phi}[R+4\partial\phi^2-\ft1{12}H_{\mu\nu\rho}^2
-\ft14\alpha'\Tr F_{\mu\nu}^2+\ft18\alpha'R_{\mu\nu\lambda\sigma}(\Omega_+)
R^{\mu\nu\lambda\sigma}(\Omega_+)],
\label{eq:hetlag}
\end{equation}
where $H$ has a non-trivial Bianchi identity
\begin{equation}
dH=\ft14\alpha'\Tr R(\Omega_+)\wedge R(\Omega_+)-\ft14\alpha'\Tr F\wedge F.
\label{eq:Hbianchi}
\end{equation}
The equations of motion corresponding to (\ref{eq:hetlag}) in linear order in  $\alpha'$ are 
\begin{eqnarray}
R-4\partial\phi^2+4\Box\phi-\ft1{12}H_{\mu\nu\rho}^2-\ft14\alpha'\Tr F_{\mu\nu}^2
+\ft18\alpha'R_{\mu\nu\lambda\sigma}(\Omega_+)
R^{\mu\nu\lambda\sigma}(\Omega_+)&=&0,\nonumber\\
R_{\mu\nu}+2\nabla_\mu\nabla_\nu\phi-\ft14H_{\mu\rho\sigma}H_\nu{}^{\rho\sigma}
-\ft14\alpha'\Tr F_{\mu\rho}F_\nu{}^\rho+\ft14\alpha'R_{\mu\lambda\rho\sigma}(\Omega_+)
R_\nu{}^{\lambda\rho\sigma}(\Omega_+)&=&0,\nonumber\\
d(e^{-2\phi}*H)&=&0,\nonumber\\
e^{2\phi}d(e^{-2\phi}*F)+A\wedge*F-*F\wedge A+F\wedge*H&=&0.
\label{eq:HetEOM}
\end{eqnarray}

Note the appearance of $R(\omp)$ in both the effective action and the Bianchi identity.
Of course, it is not hard to check that the structure of $\alpha'$ terms in the former ensures
that only even powers in $H$ appear. However this is not the case for the latter, and
there is no averaging in $H$. In fact the covariant derivative on the spinors appearing
in the supersymmetry transformations 
\bea
\delta \psi_{\mu} = \left(\partial_{\mu} + \ft14 (\omm)_{\mu} + \cdots\right) \varepsilon
\eea
involves a torsional connection with {\sl opposite} torsion.

The reduction of the heterotic two-derivative action to six dimensions on $\mathbb{T}^4$ is straightforward,
and gives rise to $\mathcal N=(1,1)$ supergravity \cite{Giani:1984dw,Romans:1985tw}
coupled to 20 vector multiplets.  In six dimensions,
there are $4+20$ vectors, with four of them in the gravity multiplet.  The remaining 20 vectors
combine with the 80 scalars living on the coset $SO(4,20)/SO(4)\times SO(20)$ to form
20 vector multiplets.  The bosonic fields in the gravity multiplet are
$(g_{\mu\nu},B_{\mu\nu},(3+1)A_\mu,\phi)$.  However, instead of a complete reduction, we
only consider the six-dimensional metric, antisymmetric tensor and dilaton, as they will
dualize to the NSNS sector of the IIA string.  In this case, we can directly take (\ref{eq:hetlag}) as
a six-dimensional heterotic Lagrangian, while at the same time discarding the heterotic
gauge fields.  To be precise, we write
\begin{equation}
e^{-1}\mathcal L=e^{-2\phi}[R^{\rm het}+4\partial\phi^2-\ft1{12}H_{\mu\nu\rho}^{{\rm het}\,2}
+\ft18\alpha'R^{\rm het}_{\mu\nu\lambda\sigma}(\Omega_+)
R^{{\rm het}\,\mu\nu\lambda\sigma}(\Omega_+)+\cdots],
\label{eq:hetsix}
\end{equation}
where the superscript `het' indicates that these are the heterotic fields.

\subsection{Dualizing to IIA}

At the two-derivative level, $\mathcal N=(1,1)$ supergravity admits two distinct formulations
related by dualization of $H$.  The heterotic version has a non-trivial Bianchi identity
$dH^{\rm het}=-(\alpha'/4)\Tr F\wedge F$, and was denoted $\mathcal N=\tilde 4$ in
\cite{Romans:1985tw}.  In contrast, the IIA version has a trivial Bianchi identity
$dH^{\rm IIA}=0$, and corresponds to $\mathcal N=4$
in \cite{Romans:1985tw}.  Even when including the four-derivative corrections, we see that
the heterotic $H$ equation of motion in (\ref{eq:HetEOM}) remains source-free.  Thus we may
simply extend the heterotic/IIA duality map
\begin{equation}
H^{\rm het}=e^{2\phi}*H^{\rm IIA},\qquad g_{\mu\nu}^{\rm het}=e^{2\phi}g_{\mu\nu}^{\rm IIA},
\qquad\phi=-\varphi^{\rm IIA},
\label{eq:HetIIAmap}
\end{equation}
without additional higher-derivative modifications.  This dualization is designed to exchange
the three-form Bianchi identity with the equation of motion, so that
\begin{equation}
dH^{\rm IIA}=0,\qquad d(e^{-2\varphi}*H^{\rm IIA})=\ft14\alpha'\Tr R^{\rm het}(\Omega_+)
\wedge R^{\rm het}(\Omega_+)-\ft14\alpha'\Tr F\wedge F.
\label{eq:iiaheom}
\end{equation}
(We will discard the $\Tr F\wedge F$ term, but have left it here for completeness.)
Note that, while we could include higher derivative corrections to the map (\ref{eq:HetIIAmap}),
such additions would simply correspond to a field redefinition of the IIA fields.  As we will
see below, the choice (\ref{eq:HetIIAmap}) in fact leads to our desired result for the IIA
theory without further redefinitions.

We now dualize the remaining equations of motion.  At the two derivative level, the
IIA dilaton and Einstein equations are obtained from linear combinations of the
heterotic dilaton and Einstein equations.  We find
\begin{eqnarray}
R^{\rm IIA}-4\partial\varphi^2+4\Box\varphi-\ft1{12}H_{\mu\nu\rho}^{{\rm IIA}\,2}
+\mathcal O(\alpha')&=&0,\nonumber\\
R^{\rm IIA}_{\mu\nu}+2\nabla_\mu\nabla_\nu\varphi-\ft14H^{\rm IIA}_{\mu\rho\sigma}
H_\nu^{{\rm IIA}\,\rho\sigma}+\mathcal O(\alpha')&=&0.
\label{eq:iiaeinseom}
\end{eqnarray}
Note that the covariant derivatives are computed with the IIA metric, and that $\varphi$
is the IIA dilaton.  In order to complete these equations at
the four derivative level, we must work out the $\mathcal O(\alpha')$ terms in both
(\ref{eq:iiaheom}) and (\ref{eq:iiaeinseom}).  Since the higher derivative terms are built
out of $R(\Omega_+)$ we first compute
\begin{eqnarray}
R_{\mu\nu}^{{\rm het}\,ab}
&=&R_{\mu\nu}{}^{ab}-2\delta_{[\mu}^{[a}R_{\nu]}^{b]}+4\delta_{[\mu}^{[a}\partial_{\nu]}\varphi
\partial^{b]}\varphi-\delta_{[\mu}^a\delta_{\nu]}^b\Box\varphi-\ft12H_{[\mu}{}^{ac}H_{\nu]}{}^{cb}
\nonumber\\
&&+\nabla_{[\mu}*H_{\nu]}{}^{ab}+2*H_{\mu\nu}{}^{[a}\partial^{b]}\varphi
+2\delta_{[\mu}^{[a}*H_{\nu]}{}^{b]\alpha}\partial_\alpha\varphi.
\label{eq:Rhet}
\end{eqnarray}
Note that we have used the lowest order equations of motion to simplify this result.
In addition, all terms on the right hand side are IIA fields, although from now on we
drop the `IIA' notation to avoid cluttering the equations.
The first and second lines of this expression have opposite parity.

The heterotic dilaton equation of motion in (\ref{eq:HetEOM}) involves the square of the
heterotic Riemann tensor.  Using (\ref{eq:Rhet}), and the lowest-order equations of
motion, we find (after much tedious manipulations)
\begin{equation}
e^{4\phi}R_{\mu\nu\rho\sigma}^{\rm het}(\Omega_+)^2=\ft12R_{\mu\nu\rho\sigma}(\Omega_+)^2
+\ft12E_4-\Box R(\Omega_+)
+\ft13\partial_\mu H^2\partial^\mu\varphi-\ft13H^2\Box\varphi
+2H^2_{\mu\nu}\nabla^\mu\nabla^\nu\varphi,
\label{eq:iiarsq}
\end{equation}
where
\begin{eqnarray}
E_4=-\ft18\epsilon_6\epsilon_6 R(\Omega_+)^2
&\equiv&-\ft18\epsilon_{\alpha\beta\mu_1\cdots\mu_4}\epsilon^{\alpha\beta\nu_1\cdots\nu_4}
R^{\mu_1\mu_2}{}_{\nu_1\nu_2}(\Omega_+)R^{\mu_3\mu_4}{}_{\nu_3\nu_4}(\Omega_-)
\nonumber\\
&=&R_{\mu\nu\rho\sigma}(\Omega_+)R^{\rho\sigma\mu\nu}(\Omega_+)
-4R_{\mu\nu}(\Omega_+)R^{\nu\mu}(\Omega_+)+R(\Omega_+)^2.
\label{eq:E4def}
\end{eqnarray}
Note that the dilaton factor is consistent with this being a one-loop term from the IIA point of view.
The right hand side of (\ref{eq:iiarsq}) was worked out using
\begin{eqnarray}
R_{\mu\nu}{}^{\lambda\sigma}(\Omega_+)&=&
R_{\mu\nu}{}^{\lambda\sigma}+\ft12(\nabla_\mu H_\nu{}^{\lambda\sigma}
-\nabla_\nu H_\mu{}^{\lambda\sigma})+\ft14(H_\mu{}^{\lambda\alpha}H_{\nu\alpha}{}^\sigma
-H_\nu{}^{\lambda\alpha}H_{\mu\alpha}{}^\sigma),\nonumber\\
R_{\mu\nu}(\Omega_+)&=&R_{\mu\nu}-\ft12\nabla^\alpha H_{\alpha\mu\nu}-\ft14H^2_{\mu\nu},
\nonumber\\
R(\Omega_+)&=&R-\ft14H^2.
\label{eq:Rplusexpl}
\end{eqnarray}
Furthermore, the Ricci tensor and Ricci scalar with torsion can be rewritten using the lowest-order
equations of motion
\begin{eqnarray}
R_{\mu\nu}(\Omega_+)&=&-2\nabla_\mu\nabla_\nu\varphi
-H_{\mu\nu}{}^\alpha\partial_\alpha\varphi,\nonumber\\
R(\Omega_+)&=&-2\Box\varphi.
\label{eq:RcRseom}
\end{eqnarray}

For the $H^{\rm IIA}$ equation of motion, we work out $\Tr R^{\rm het}\wedge R^{\rm het}$ where
$R^{\rm het}$ is given in (\ref{eq:Rhet}).  This will consist of two pieces, a CP-odd piece from
the first line of (\ref{eq:Rhet}) wedged into itself plus the second line wedged into itself, and a
CP-even piece from the cross-term.  Again using the lowest order equations of motion, we find
the CP-odd piece to be
\begin{equation}
R^{{\rm het}\,ab}_{[\mu\nu}R_{\lambda\sigma]}^{{\rm het}\,ab}\big|_{\rm CP\mhyphen odd}=
(R_{[\mu\nu}{}^{ab}+\ft12H_\mu{}^{ac}H_\nu{}^{cb})(R_{\lambda\sigma]}{}^{ab}
+\ft12H_\lambda{}^{ad}H_{\sigma]}{}^{db})+(\nabla_{[\mu}H_\nu{}^{ab})
(\nabla_\lambda H_{\sigma]}{}^{ab}).
\label{eq:RwRodd}
\end{equation}
However, this can be written more instructively as the `averaged' expression for $\Tr R\wedge R$
\begin{equation}
R^{{\rm het}\,ab}_{[\mu\nu}R_{\lambda\sigma]}^{{\rm het}\,ab}\big|_{\rm CP\mhyphen odd}=\ft12
[R_{[\mu\nu}{}^{ab}(\Omega_+)R_{\lambda\sigma]}{}^{ab}(\Omega_+)
+R_{[\mu\nu}{}^{ab}(\Omega_-)R_{\lambda\sigma]}{}^{ab}(\Omega_-)].
\end{equation}
The remaining CP-even cross-term is somewhat unusual.  It can be written as a divergence
\begin{equation}
\ft1{4!}\epsilon_{\rho\delta}{}^{\mu\nu\lambda\sigma}
R^{{\rm het}\,ab}_{[\mu\nu}R_{\lambda\sigma]}^{{\rm het}\,ab}\big|_{\rm CP\mhyphen even}
=\nabla^\lambda T_{\rho\delta\lambda},
\end{equation}
where $T_{\rho\delta\lambda}$ is completely antisymmetric
\begin{eqnarray}
T_{\rho\delta\lambda}&=&
\ft12R_{\mu\nu[\rho\delta}H_{\lambda]}{}^{\mu\nu}
-R_{\mu[\rho}H_{\delta\lambda]}{}^\mu+\partial_{[\rho}\varphi\nabla^\mu H_{\delta\lambda]\mu}
-2\nabla^\mu(H_{\mu[\rho\delta}\partial_{\lambda]}\varphi)
\nonumber\\
&&\qquad+\ft1{24}(2H_\rho{}^{ab}H_\delta{}^{bc}H_\lambda{}^{ca}+\ft13H^2H_{\rho\delta\lambda})
+\ft13\partial\varphi^2H_{\rho\delta\lambda}.
\label{eq:ttens}
\end{eqnarray}

Collecting the above results, we may write down the four-derivative equations of motion in the
IIA duality frame.  For $H$, we have the Bianchi identity and equation of motion
\begin{eqnarray}
\label{eq:iiaHeom}
dH&=&0,\nonumber\\
\nabla^\lambda[e^{-2\varphi}H_{\rho\delta\lambda}+\ft32\alpha'T_{\rho\delta\lambda}]
&=&-\ft32\alpha'\ft1{4!}\epsilon_{\rho\delta}{}^{\mu\nu\lambda\sigma}
\ft12[R_{[\mu\nu}{}^{ab}(\Omega_+)R_{\lambda\sigma]}{}^{ab}(\Omega_+)
+R_{[\mu\nu}{}^{ab}(\Omega_-)R_{\lambda\sigma]}{}^{ab}(\Omega_-)].\nonumber\\
\end{eqnarray}
For the dilaton, we have
\begin{equation}
e^{-2\varphi}(R+4\square\varphi-4\partial\varphi^2-\ft1{12}H^2)=0.
\label{eq:iiadileom}
\end{equation}
Note in particular that this dilaton equation is unaffected by the one-loop $\alpha'$ correction.
Finally, the trace of the Einstein equation is
\begin{equation}
e^{-2\varphi}(R+2\square\varphi-\ft14H^2)=\ft18\alpha'[
\ft12R_{\mu\nu\rho\sigma}(\Omega_+)^2+\ft12E_4-\square R(\Omega_+)
+\ft13\partial_\mu H^2\partial^\mu\varphi-\ft13H^2\square\varphi+2H^2_{\mu\nu}
\nabla^\mu\nabla^\nu\varphi].
\label{eq:triiaeins}
\end{equation}
For completeness, we would also want the uncontracted Einstein equation, which involves
working out the uncontracted $R^{\rm het}_{\mu\lambda\rho\sigma}(\Omega_+)
R^{{\rm het}\,\lambda\rho\sigma}_\nu(\Omega_+)$.  Although we have only worked this out
implicitly using Maple and the dualization map (\ref{eq:HetIIAmap}), this was sufficient
for us to uniquely determine the IIA Lagrangian.

\subsection{Finding a IIA Lagrangian}

Given the equations of motion (\ref{eq:iiaHeom}), (\ref{eq:iiadileom}) and (\ref{eq:triiaeins}),
as well as the uncontracted Einstein equation, we now wish to obtain an effective IIA action.
Since the CP-even and CP-odd terms are distinct, it is possible to work out the effective
action in both sectors independently.

\subsubsection{The CP-even sector}

We start with the CP-even sector and work
systematically by writing out the most general four-derivative Lagrangian for the metric,
antisymmetric tensor and dilaton.  Up to integration by parts, we write
\begin{eqnarray}
e^{-1}\delta\mathcal L_{\rm CP\mhyphen even}
&=&\alpha_1R_{\mu\nu\lambda\sigma}^2+\alpha_2R_{\mu\nu}^2+\alpha_3R^3
+\beta_1R_{\mu\nu\lambda\sigma}H^{\mu\nu\alpha}H^{\lambda\sigma}{}_\alpha
+\beta_2R_{\mu\nu}H^{\mu\alpha\beta}H^\nu{}_{\alpha\beta}+\beta_3RH^2\nonumber\\
&&+\beta_4\nabla_\mu H_{\nu\alpha\beta}\nabla^\nu H^{\mu\alpha\beta}
+\gamma_1H_{\mu\nu\lambda}H^{\mu\alpha}{}_\beta H^{\nu\beta}{}_\gamma
H^{\lambda\gamma}{}_\alpha
+\gamma_2(H^2_{\mu\nu})^2+\gamma_3(H^2)^2\nonumber\\
&&+\delta_1R_{\mu\nu}\partial^\mu\varphi\partial^\nu\varphi+\delta_2R\partial\varphi^2
+\delta_3(\Box\varphi)^2+\delta_4(\partial\varphi^2)^2-\epsilon_1\Box\phi\partial\varphi^2
-\epsilon_2\Box\varphi\nonumber\\
&&+\kappa_1H^2\partial\varphi^2+\kappa_2H^2_{\mu\nu}\partial^\mu\varphi\partial^\nu\varphi
-\kappa_3H^2\Box\varphi-\kappa_4H^2_{\mu\nu}\nabla^\mu\nabla^\nu\varphi.
\label{eq:Lparam}
\end{eqnarray}
Note that we have not yet used any lowest order equations of motion to simplify the above.

We can match the equations of motion derived from this Lagrangian with those obtained
from dualization.  In practice, we work with the six-dimensional dyonic string solution
\cite{Duff:1995yh}, and match coefficients symbolically using Maple with grtensor.  The
result is reasonably straightforward, and we find (up to an overall normalization)
\begin{eqnarray}
&&\alpha_1=1,\kern3.8em\alpha_2=-2+\kappa_2+\kappa_4,\kern3.3em
\alpha_3=\ft12-\ft16\kappa_1-\ft16\kappa_2+\kappa_3+\ft1{36}\delta_4,\nonumber\\
&&\beta_1=-\ft14\kappa_2,\qquad\beta_2=-\ft12-\ft14\kappa_2-\ft12\kappa_4,\qquad
\beta_3=\ft1{12}+\ft14\kappa_1+\ft1{12}\kappa_2-\ft12\kappa_3,\qquad
\beta_4=\ft12+\ft14\kappa_2,\nonumber\\
&&\gamma_1=\ft1{24},\kern3.5em\gamma_2=\ft1{16}\kappa_2+\ft1{16}\kappa_4,\kern3.9em
\gamma_3=\ft1{288}-\ft5{96}\kappa_1-\ft1{96}\kappa_2+\ft1{16}\kappa_3,\nonumber\\
&&\delta_1=-4\kappa_2-4\kappa_4,\qquad\delta_2=-4\kappa_1-\ft13\delta_4,\kern1.9em
\delta_3=\ft{10}3\kappa_1+\ft{10}3\kappa_2+4\kappa_3+4\kappa_4+\ft{25}{36}\delta_4,\nonumber\\
&&\epsilon_1=8\kappa_1+\ft53\delta_4,\kern2.9em
\epsilon_2=-\ft43\kappa_1-\ft43\kappa_2-4\kappa_3-2\kappa_4-\ft5{18}\delta_4.
\label{eq:Lcoefs}
\end{eqnarray}
There are five undetermined parameters, which we have taken to be $\{\kappa_1,\kappa_2,
\kappa_3,\kappa_4,\delta_4\}$.  However these parameters are unphysical, and may be
removed by appropriate shifts.

In particular, this Lagrangian is only determined up to expressions involving squares of the
equations of motion.  To be specific, we define the dilaton, antisymmetric tensor and Einstein
equations
\begin{eqnarray}
E^\varphi&=&R-4\partial\varphi^2+4\Box\varphi-\ft1{12}{H^2},\nonumber\\
E^B_{\mu\nu}&=&\nabla^\lambda H_{\mu\nu\lambda}-2H_{\mu\nu\lambda}\partial^\lambda\varphi,
\nonumber\\
E_{\mu\nu}&=&R_{\mu\nu}+2\nabla_\mu\nabla_\nu\varphi-\ft14H^2_{\mu\nu}.
\end{eqnarray}
Since there are five quadratic combinations
\begin{equation}
(E^\varphi)^2,\quad E^\varphi E^\mu_\mu,\quad (E^\mu_\mu)^2,\quad
(E^B_{\mu\nu})^2,\quad (E_{\mu\nu})^2.
\end{equation}
there are precisely five free parameters in the recreated Lagrangian.  Note that this
freedom is not associated with field redefinitions.  The field redefinition freedom is the ability
to shift, say the dilaton by an appropriate function of the fields
\begin{equation}
\varphi\to\varphi+\delta\varphi(g_{\mu\nu},B_{\mu\nu},\varphi).
\end{equation}
This changes the Lagrangian according to
\begin{equation}
\mathcal L\to\mathcal L+\delta\mathcal L=\mathcal L+\fft{\delta\mathcal L}{\delta\varphi}\delta\varphi
=\mathcal L+E^\varphi\delta\varphi,
\end{equation}
where $E^\varphi$ is just the lowest order dilaton equation of motion.  (We have assumed
integration by parts when necessary.)  Of course, this works for any of the fields.  So what
we see is that field redefinitions allow us to shift the Lagrangian by terms linear in the
lowest order equations of motion multiplied by an arbitrary combination of the fields and
their derivatives.

Examination of (\ref{eq:Lcoefs}) indicates that there is a `preferred' combination of terms in
the four-derivative Lagrangian that is independent of the dilaton.  The dilaton independent
Lagrangian is simply obtained by setting the five undetermined parameters to zero,
{\it i.e.}~by setting $\kappa_i=0$ and $\delta_4=0$.  This gives the CP-even four-derivative
Lagrangian
\begin{eqnarray}
e^{-1}\delta\mathcal L_{\rm CP\mhyphen even}
&=&R_{\mu\nu\lambda\sigma}^2-2R_{\mu\nu}^2+\ft12R^2-\ft12R_{\mu\nu}
H^{2\,\mu\nu}+\ft1{12}RH^2+\ft12\nabla_\mu H_{\nu\alpha\beta}\nabla^\nu H^{\mu\alpha\beta}
\nonumber\\
&&+\ft1{24}H_{\mu\nu\lambda}H^{\mu\alpha}{}_\beta H^{\nu\beta}{}_\gamma
H^{\lambda\gamma}{}_\alpha+\ft1{288}(H^2)^2.
\label{eq:iia4dlag}
\end{eqnarray}

Since we claim the natural curvature quantities are computed out of the connection with torsion,
$\Omega_+=\Omega+\fft12H$, we wish to rewrite the Lagrangian in terms of
$R_{\mu\nu\rho\sigma}(\Omega_+)$ and its contractions.  In particular, motivated by the string
amplitude results, we anticipate an expression of the form $t_4t_4R^2+E_4$, where the
first term arises from the even-even spin structure sector, and where $E_4$ defined in
(\ref{eq:E4def}) arises from the odd-odd spin structure sector.  For $t_4t_4R^2$, we have
\begin{equation}
t_4t_4R(\Omega_+)^2\equiv R_{\mu\nu\rho\sigma}(\Omega_+)^2
=R_{\mu\nu\rho\sigma}^2+\nabla_\mu H_\nu{}^{ab}
\nabla_\nu H_\mu{}^{ab}-R_{\mu\nu\lambda\sigma}H^{\mu\lambda a}H^{\nu\sigma a}
+\ft18(H^2_{\mu\nu})^2-\ft18H^4,
\label{eq:t4t4Rplus2}
\end{equation}
where
\begin{equation}
H^4\equiv H_{\mu\nu\rho}H^{\mu ab}H^{\nu bc}H^{\rho ca},
\end{equation}
while for $E_4$, we find (allowing for integration by parts)
\begin{equation}
E_4=R_{\mu\nu\rho\sigma}^2-4R_{\mu\nu}^2+(R-\ft14H^2)^2
-3R_{\mu\nu\rho\sigma}H^{\mu\rho a}H^{\nu\sigma a}+3R_{\mu\nu}H^{2\,\mu\nu}
-\ft18(H^2_{\mu\nu})^2-\ft18H^4.
\label{eq:e4e4Rplus2}
\end{equation}
Inserting (\ref{eq:t4t4Rplus2}) and (\ref{eq:e4e4Rplus2}) into (\ref{eq:iia4dlag}), and choosing
to use curvatures built out of $\Omega_+$ then gives
\begin{eqnarray}
2e^{-1}\delta\mathcal L_{\rm CP\mhyphen even}&=&R_{\mu\nu\rho\sigma}(\Omega_+)^2+E_4
+4R_{\mu\nu\rho\sigma}(\Omega_+)H^{\mu\rho a}H^{\nu\sigma a}
-4R_{\mu\nu}(\Omega_+)H^{2\,\mu\nu}+\ft23R(\Omega_+)H^2\nonumber\\
&&+\ft19(H^2)^2-\ft23H^4.
\label{eq:iiaapm}
\end{eqnarray}

While we would like to attribute the first two terms in (\ref{eq:iiaapm}) to the even-even
and odd-odd spin structure sectors, respectively, the remaining terms may appear to be
an unexpected addition to the four-derivative Lagrangian.  Note, however, that the
three-point amplitude for the IIA string compactified on $K3$ was computed in 
\cite{Gregori:1997hi}, where it was found that the even-even and odd-odd sectors
gave identical contributions at the level of the three-point function
\begin{equation}
\left.\mathcal A\right|_{\mathrm{3-pt}}=R_{\mu\nu\rho\sigma}^2+\ft12\nabla_\mu H_\nu{}^{ab}
\nabla_\nu H_\mu{}^{ab}
=R_{\mu\nu\rho\sigma}^2+R_{\mu\nu\rho\sigma}H^{\mu\rho a}H^{\nu\sigma a}+\cdots.
\end{equation}
Here we have made use of the on-shell relation
\begin{equation}
\nabla_\mu H_\nu{}^{ab}\nabla_\nu H_\mu{}^{ab}
=2R_{\mu\nu\lambda\sigma}H^{\mu\lambda a}H^{\nu\sigma a}+\cdots,
\end{equation}
which holds at the level of the three-point function.  Comparing this with (\ref{eq:t4t4Rplus2})
indicates that $t_4t_4R(\Omega_+)^2$ is sufficient for matching the even-even sector at
this level.  However, $E_4$
given in (\ref{eq:e4e4Rplus2}) by itself is incomplete for the odd-odd sector.  Instead, the
odd-odd contribution must have the form
\begin{equation}
E_4+4R_{\mu\nu\rho\sigma}(\Omega_+)H^{\mu\rho a}H^{\nu\sigma a}+\cdots
=R_{\mu\nu\rho\sigma}^2+R_{\mu\nu\rho\sigma}H^{\mu\rho a}H^{\nu\sigma a}+\cdots,
\end{equation}
to agree with the three-point function.  This suggests that the second and third terms in
(\ref{eq:iiaapm}) both arise in the odd-odd sector.  The remaining terms do not contribute
to three-point functions, and hence cannot be fixed at this level.

We note, however, that the explicit $R(\Omega_+)H^2$ terms in (\ref{eq:iiaapm}) can
in fact be written more compactly using the identities
\begin{eqnarray}
\epsilon_6\epsilon_6H^2R(\Omega_+)&\equiv&
\epsilon_{\alpha\mu_0\cdots\mu_4}\epsilon^{\alpha\nu_0\cdots\nu_4}
H^{\mu_1\mu_2}{}_{\nu_0}H_{\nu_1\nu_2}{}^{\mu_0}R^{\mu_3\mu_4}{}_{\nu_3\nu_4}(\Omega_+)
\nonumber\\
&=&-24R_{\mu\nu\rho\sigma}(\Omega_+)H^{\mu\rho a}H^{\nu\sigma a}
+24R_{\mu\nu}(\Omega_+)H^{2\,\mu\nu}-4R(\Omega_+)H^2-2(H^2_{\mu\nu})^2+4H^4,
\nonumber\\
\label{eq:e5e5H2Rplus}
\end{eqnarray}
and
\begin{equation}
\epsilon_6\epsilon_6H^4\equiv
\epsilon_{\alpha\beta\mu_1\cdots\mu_4}\epsilon^{\alpha\beta\nu_1\cdots\nu_4}
H^{\mu_1\mu_2a}H_{\nu_1\nu_2a}H^{\mu_3\mu_4b}H_{\nu_3\nu_4b}
=-8(H^2)^2+24(H^2_{\mu\nu})^2.
\label{eq:e4e4H4}
\end{equation}
The final result for the CP-even sector is then
\begin{equation}
2e^{-1}\mathcal L_{\rm CP\mhyphen even}=t_4t_4R(\Omega_+)^2
-\ft18\epsilon_6\epsilon_6R(\Omega_+)^2-\ft16\epsilon_6\epsilon_6H^2R(\Omega_+)
-\ft1{72}\epsilon_6\epsilon_6H^4.
\label{eq:CPelag}
\end{equation}
This is now written in a form where we expect the first term to arise from the even-even sector
and the remaining three to arise from the odd-odd sector.  We will demonstrate that this is in
fact the case below when we investigate the closed string four-point function.

\subsubsection{The CP-odd sector}

Turning now to the CP-odd sector, we take a slightly different approach to constructing the
effective Lagrangian.  Instead of writing out all possible four-derivative terms, we conjecture
that the purely gravitational correction $B\wedge\Tr R^2$ has a natural extension to the
`averaged' form
\begin{eqnarray}
e^{-1}\mathcal L_{\rm CP\mhyphen odd}
&=&\ft18\epsilon^{\alpha\beta\mu\nu\rho\sigma}B_{\alpha\beta}\ft12
[R_{\mu\nu}{}^{ab}(\Omega_+)R_{\rho\sigma}{}^{ab}(\Omega_+)+
R_{\mu\nu}{}^{ab}(\Omega_-)R_{\rho\sigma}{}^{ab}(\Omega_-)]\nonumber\\
&=&e^{-1}\ft12[\mathcal L^{(+)}+\mathcal L^{(-)}].
\label{eq:CPoLag}
\end{eqnarray}
This preserves gauge invariance under transformations of $B$ since $\Tr R^2\to
\Tr R(\Omega_+)^2$ shifts by an exact form.
We consider this Lagrangian to be a function of both $B_{\mu\nu}$ and
$\Omega_{+\,\mu}{}^{ab}$, so that the variation of $\mathcal L^{(+)}$ is
\begin{equation}
e^{-1}\delta\mathcal L^{(+)}=\ft18\epsilon^{\alpha\beta\mu\nu\rho\sigma}
R_{\mu\nu}{}^{ab}(\Omega_+)R_{\rho\sigma}{}^{ab}(\Omega_+)\delta B_{\alpha\beta}
+\ft14\epsilon^{\alpha\beta\mu\nu\rho\sigma}B_{\alpha\beta}R_{\mu\nu}{}^{ab}(\Omega_+)
\delta R_{\rho\sigma}{}^{ab}(\Omega_+).
\end{equation}
When averaged, the first term gives us the expected contribution to the right hand side of
the $H$ equation of motion (\ref{eq:iiaHeom}).  However the second term is extra, and we
would like to demonstrate that it vanishes when averaged.  To do so, we first compute the
variation of Riemann
\begin{equation}
\delta R_{\mu\nu}{}^{ab}(\Omega_+)=2(\nabla_{[\mu}\delta\Omega_{+\,\nu]}{}^{ab}
+\ft12H_{[\mu}{}^{ac}\delta\Omega_{+\,\nu]}{}^{cb}-\ft12H_{[\mu}{}^{bc}\delta\Omega_{+\,\nu]}{}^{ca}).
\label{eq:deltaRp}
\end{equation}
and integrate by parts to handle the first term in (\ref{eq:deltaRp}).  The result is
\begin{equation}
e^{-1}\delta\mathcal L^{(+)}_{\rm extra}
=-\ft12\epsilon^{\alpha\beta\mu\nu\rho\sigma}[\ft13H_{\rho\alpha\beta}
R_{\mu\nu}{}^{ab}(\Omega_+)+B_{\alpha\beta}\nabla_\rho R_{\mu\nu}{}^{ab}(\Omega_+)
+B_{\alpha\beta}R_{\mu\nu}{}^{cb}(\Omega_+)H_\rho{}^{ac}]\delta\Omega_{+\,\sigma}{}^{ab}.
\end{equation}

We now use the identity
\begin{equation}
\nabla_{[\rho}R_{\mu\nu]}{}^{ab}(\Omega_+)=-R_{[\rho\mu}{}^{c[b}H_{\nu]}{}^{a]c}
-H_{[\mu}{}^{c[a}\nabla_\nu H_{\rho]}{}^{b]c},
\end{equation}
to obtain
\begin{equation}
e^{-1}\delta\mathcal L^{(+)}_{\rm extra}
=-\ft12\epsilon^{\alpha\beta\mu\nu\rho\sigma}[\ft13H_{\rho\alpha\beta}
R_{\mu\nu}{}^{ab}(\Omega_+)+\ft12B_{\alpha\beta}H_\mu{}^{ac}H_\nu{}^{cd}H_\rho{}^{db}]
\delta\Omega_{+\,\sigma}{}^{ab}.
\end{equation}
Note that $H_{[\mu}{}^{ac}H_\nu{}^{cd}H_{\rho]}{}^{db}$ is symmetric under $a\leftrightarrow b$
interchange, so the second term vanishes.  As a result, we are left without any bare
$B_{\alpha\beta}$ terms, and we have simply
\begin{equation}
e^{-1}\delta\mathcal L^{(+)}_{\rm extra}
=-\ft16\epsilon^{\alpha\beta\mu\nu\rho\sigma}H_{\rho\alpha\beta}
R_{\mu\nu}{}^{ab}(\Omega_+)\delta\Omega_{+\,\sigma}{}^{ab}.
\end{equation}
Of course, we still would like to show that this vanishes when averaged.

For simplicity, we focus on the $H$ equation of motion.  In this case, we take
\begin{equation}
\delta\Omega_{+\,\mu}{}^{ab}\to\ft12\delta H_\mu{}^{ab}
=\ft12e^{\nu a}e^{\rho b}\delta H_{\mu\nu\rho}.
\label{eq:dwtodH}
\end{equation}
Furthermore, averaging picks out terms odd in $H$.  After substituting in (\ref{eq:Rplusexpl}) for
$R_{\mu\nu}{}^{ab}(\Omega_+)$, we end up with
\begin{equation}
e^{-1}\delta\mathcal L_{\rm extra}
=-\ft1{12}\epsilon^{\alpha\beta\mu\nu\rho\sigma}H_{\rho\alpha\beta}\nabla_\mu
H_\nu{}^{ab}\delta H_\sigma{}^{ab}.
\end{equation}
We conjecture that this term vanishes using the lowest order equations of motion.
In particular
\begin{equation}
\epsilon^{\alpha\beta\mu\nu\rho[\sigma}H_{\rho\alpha\beta}\nabla_\mu H_\nu{}^{ab]}=0,
\end{equation}
for the dyonic string solution.  Assuming this holds, we have thus demonstrated that the
CP-odd Lagrangian (\ref{eq:CPoLag}) indeed reproduces the right hand side of the
$H$ equation of motion (\ref{eq:iiaHeom}).

Combining the CP-even and CP-odd sectors, we have obtained the one-loop four-derivative
contribution to the NSNS sector of the six-dimensional IIA Lagrangian using heterotic/IIA duality.
The result is
\begin{eqnarray}
e^{-1}\mathcal L_{\rm IIA}&=&e^{-2\varphi}[R+4\partial\varphi^2-\ft1{12}H_{\mu\nu\rho}^2]
\nonumber\\
&&+\ft1{16}\alpha'[t_4t_4R(\Omega_+)^2
-\ft18\epsilon_6\epsilon_6R(\Omega_+)^2-\ft16\epsilon_6\epsilon_6H^2R(\Omega_+)
-\ft1{72}\epsilon_6\epsilon_6H^4]\nonumber\\
&&+\ft1{32}\alpha'\epsilon^{\alpha\beta\mu\nu\rho\sigma}B_{\alpha\beta}\ft12
[R_{\mu\nu}{}^{ab}(\Omega_+)R_{\rho\sigma}{}^{ab}(\Omega_+)+
R_{\mu\nu}{}^{ab}(\Omega_-)R_{\rho\sigma}{}^{ab}(\Omega_-)],
\label{eq:iiafin}
\end{eqnarray}
where $\varphi$ is the IIA dilaton, and where the CP-even expressions are defined in
(\ref{eq:t4t4Rplus2}), (\ref{eq:E4def}), (\ref{eq:e4e4Rplus2}), (\ref{eq:e5e5H2Rplus})
and (\ref{eq:e4e4H4}).  As a point of comparison, the supersymmetric completion of
$R^2$ in the context of the $(1,0)$ theory was investigated in
\cite{Bergshoeff:1986vy,Bergshoeff:1986wc,Bergshoeff:1987rb,Bergshoeff:2012ax}.
A complete result
was obtained for the supersymmetrization of $t_4t_4R^2$, while only a partial result
was given for the supersymmetrization of the Gauss-Bonnet combination (corresponding
to $\epsilon_6\epsilon_6R^2$).  Considering only bosonic fields, the supersymmetric
combinations took the form
\begin{eqnarray}
t_4t_4R^2\quad&\to&\quad
t_4t_4R(\Omega_-)^2+B\wedge R(\Omega_-)\wedge R(\Omega_-),\nn\\
\epsilon_6\epsilon_6R^2\quad&\to&\quad
\epsilon_6\epsilon_6R(\Omega_+)^2+B\wedge R(\Omega_+)\wedge R(\Omega_+)+\cdots,
\label{eq:11Rsq}
\end{eqnarray}
where the ellipsis denotes additional terms involving $H$ that were not determined.
Although the Lagrangian (\ref{eq:iiafin}) pertains to the $(1,1)$ theory (despite being
incomplete in the RR sector), it is reassuring to see that it agrees with (\ref{eq:11Rsq}).
In fact, we suggest that the additional $\epsilon\epsilon H^2R$ and $\epsilon\epsilon H^4$
terms in (\ref{eq:iiafin}) complete the missing terms in (\ref{eq:11Rsq}).

\subsection{The one-loop four-point function in six dimension}

Although we have taken an indirect route to the IIA Lagrangian, (\ref{eq:iiafin}), it ought to
be possible to obtain it directly from a string worldsheet computation.  This was in fact
done at the level of the three-point amplitude in \cite{Gregori:1997hi}.  It is important to
note, however, that the three-point computation cannot determine any of the $H^4$
terms nor any terms involving the Ricci tensor or curvature scalar.  For those terms, one
would have to go to the level of the four-point function.

We focus on the kinematical structure of the six-dimensional one-loop four-point function in
the odd-odd spin structure sector.  It turns out that this is sufficient for the identification of the
$\epsilon\epsilon$ terms in (\ref{eq:iiafin}), thus allowing us to avoid a rather tedious
computation of the full one-loop integrals.  As in (\ref{eq:5pointodd}), we take the first
vertex operator in the $(-1,-1)$ picture, and the remaining three in the $(0,0)$ picture.
The amplitude is then of the form
\begin{eqnarray}
\mathcal A_{\rm o\mhyphen o}&\sim&\theta^{(1)}_{\mu_1\nu_1}\cdots\theta^{(4)}_{\mu_4\nu_4}
\Biggl\langle\psi\cdot\partial X(0)\psi^{\mu_1}
\prod_{i=2}^4(i\partial X^{\mu_i}+\ft12\alpha'k_i\cdot\psi\psi^{\mu_i})\nn\\
&&\kern6.5em\tilde\psi\cdot\bar\partial X(0)\tilde\psi^{\nu_1}
\prod_{i=2}^4(-i\bar\partial X^{\nu_i}+\ft12\alpha'k_i\cdot\tilde\psi\tilde\psi^{\nu_i})
\prod_{i=1}^4e^{ik_i\cdot X}\Biggr\rangle_{\!\!\rm odd}\!\!,
\label{eq:A4oo}
\end{eqnarray}
where the polarizations may be decomposed according to (\ref{eq:thetapol}).

This amplitude can be expanded in the number of worldsheet fermions.  Since we consider
the odd spin structure in six dimensions, we must soak up six fermion zero modes.  This
gives four distinct contributions, arising from taking either six or eight fermions on the left
in combination with six or eight on the right.  The six fermion expressions are of the form
\begin{equation}
\mathcal A_6\sim\langle\partial X^\alpha(0)\partial X^\mu\rangle
\langle\psi_\alpha\psi^\mu(k\cdot\psi\psi^\mu)(k\cdot\psi\psi^\mu)\rangle,
\end{equation}
while the eight fermion expressions are of the form
\begin{equation}
\mathcal A_8\sim
\langle\partial X^\alpha(0)\rangle
\langle\psi_\alpha\psi^\mu(k\cdot\psi\psi^\mu)(k\cdot\psi\psi^\mu)(k\cdot\psi\psi^\mu)\rangle.
\end{equation}

We begin with the contribution from six fermions on the left and six on the right, keeping in
mind that we focus only on the kinematics.  Taking, {\it e.g.}, both $i\partial X^{\mu_2}$ and
$-i\bar\partial X^{\nu_2}$  from the second vertex operator, we have a contribution of the
form
\begin{equation}
\mathcal A_{6\bar6}\sim
\epsilon_{\alpha\mu_1k_3\mu_3k_4\mu_4}\epsilon_{\beta\nu_1k_3\nu_3k_4\nu_4}
\left\langle\partial X^\alpha(0)\partial X^{\mu_2}(z_2)\bar\partial X^\beta(0)
\bar\partial X^{\nu_2}(\bar z_2)\prod e^{ik_i\cdot X(z_i,\bar z_i)}\right\rangle,
\label{eq:A66def}
\end{equation}
where we have omitted the polarization tensors for brevity.  Note that we have introduced
a shorthand notation for momenta $k_i$ contracted against the $\epsilon$-tensor.  The
bosonic amplitude is complicated by possible zero mode contractions,
$\langle\partial X^\mu\bar\partial X^\nu\rangle\sim\eta^{\mu\nu}$,
as well as contractions against the exponentials, $\langle\partial X^\mu(z_i)
e^{ik_j\cdot X(z_j)}\rangle\sim k^\mu_j\partial \Delta(z_{ij})\langle e^{ik_j\cdot X(z_j)}
\rangle$.  This amplitude starts at $\mathcal O(k^4)$ with a pair of zero mode contractions
and continues up to $\mathcal O(k^8)$ when all the $\partial X^\mu$ and $\bar\partial X^\mu$
terms contract against the exponentials.  The general expression is of the form
\begin{eqnarray}
\mathcal A_{6\bar6}&\sim&
\epsilon_{\alpha\mu_1k_3\mu_3k_4\mu_4}\epsilon_{\beta\nu_1k_3\nu_3k_4\nu_4}
\Bigl[\eta^{\alpha\beta}\eta^{\mu_2\nu_2}+\eta^{\alpha\nu_2}\eta^{\beta\mu_2}
+\sum_{i,j}\Bigl(
\eta^{\alpha\beta}k_i^{\mu_2}k_j^{\nu_2}\partial\Delta(z_{2i})\bar\partial\Delta(\bar z_{2j})\nn\\
&&+\eta^{\mu_2\nu_2}k_i^\alpha k_j^\beta\partial\Delta(z_i)\bar\partial\Delta(\bar z_j)
+\eta^{\alpha\nu_2}k_i^{\mu_2}k_j^\beta\partial\Delta(z_{2i})\bar\partial\Delta(\bar z_j)
+\eta^{\beta\mu_2}k_i^\alpha k_j^{\nu_2}\partial\Delta(z_i)\bar\partial\Delta(\bar z_{2j})\Bigr)\nn\\
&&+\Bigl(\eta^{\alpha\mu_2}\partial\partial\Delta(z_2)+\sum_{i,j}k_i^\alpha k_j^{\mu_2}
\partial\Delta(z_i)\partial\Delta(z_{2j})\Bigr)\Bigl(\eta^{\beta\nu_2}\bar\partial\bar\partial
\Delta(\bar z_2)+\sum_{i,j}k_i^\beta k_j^{\nu_2}\bar\partial\Delta(\bar z_i)\bar\partial
\Delta(\bar z_{2j})\Bigr)\Bigl]\nn\\
&&\times\prod_{i<j}\chi(z_{ij},\bar z_{ij})^{s_{ij}},
\label{eq:A66}
\end{eqnarray}
where $s_{ij}=k_i\cdot k_j$.  The first two terms in the square brackets originate from
a pair of zero mode contractions, the sum from a single zero mode contraction, and
the final product in the square brackets from the factorized left and right contractions.

Since we are only interested in the four-derivative contribution to the effective action,
and since there are already four momentum factors contracted against the $\epsilon$-tensors,
we may guess that we could simply ignore all the $k$-dependent terms inside the
square brackets and also take $s_{ij}\to0$ in the final line of (\ref{eq:A66}).  This would be
the case, except for the fact that there are potential short distance singularities as $z_{ij}\to0$
on the string worldsheet.  The additional non-vanishing contributions arise from worldsheet
integrals of the form
\begin{equation}
\int\fft{d^2z}{2\tau_2}\partial\Delta(z_{ij})\bar\partial\Delta(\bar z_{ij})\chi(z_{ij},\bar z_{ij})^{s_{ij}}
\sim\fft1{s_{ij}}.
\label{eq:momden}
\end{equation}
As a result, we end up with an expression of the form
\begin{equation}
\mathcal A_{6\bar6}\Big|_{k^4}\sim
\epsilon_{\alpha\mu_1k_3\mu_3k_4\mu_4}\epsilon_{\beta\nu_1k_3\nu_3k_4\nu_4}
\left[\eta^{\alpha\beta}\eta^{\mu_2\nu_2}+\eta^{\alpha\nu_2}\eta^{\beta\mu_2}
+\sum_{ij}\eta^{\alpha\beta}\fft{k_i^{\mu_2}k_j^{\nu_2}}{s_{ij}}
-\fft12\eta^{\alpha\mu_2}\eta^{\beta\nu_2}\right].
\label{eq:A66fin}
\end{equation}
Although the relative factors between the terms may be fixed by performing the worldsheet
integration, in practice we determined them empirically by matching with the effective action.
Finally, recall that in deriving this expression, we singled out the second vertex operator
in (\ref{eq:A66def}).  The full amplitude is obtained by summing over permutations of vertex
operators $2$, $3$ and $4$ independently on both the left and the right.

The next contribution we consider is for eight fermions on the left and six on the right.  Taking
eight fermions on the left is unique, while on the right we once again single out
$-i\bar\partial X^{\nu_2}$ from the second vertex operator.  This gives
\begin{equation}
\mathcal A_{8\bar6}\sim
\epsilon_{\beta\nu_1k_3\nu_3k_4\nu_4}
\left\langle\partial X^\alpha(0)\bar\partial X^\beta(0)
\bar\partial X^{\nu_2}(\bar z_2)\prod e^{ik_i\cdot X(z_i,\bar z_i)}\right\rangle
\left\langle\psi^\alpha(0)\psi^{\mu_1}(z_1)\prod_{i=2}^4k_i\cdot\psi(z_i)
\psi^{\mu_i}(z_i)\right\rangle,
\end{equation}
where we have performed the fermion zero mode contraction in the right sector.  The
boson contractions give
\begin{eqnarray}
\mathcal A_{8\bar6}^B&\sim&
\sum_i\Bigl[\Bigl(\eta^{\alpha\beta}k_i^{\nu_2}\bar\partial\Delta(\bar z_{2i})
+\eta^{\alpha\nu_2}k_i^\beta\bar\partial\Delta(\bar z_i)\Bigr)\nn\\
&&\qquad+k_i^\alpha\partial\Delta(z_i)\Bigl(\eta^{\beta\nu_2}\bar\partial\bar\partial\Delta(\bar z_2)
+\sum_{j,k}k_j^\beta k_k^{\nu_2}\bar\partial\Delta(\bar z_j)\bar\partial\Delta(\bar z_{2k})\Bigr)
\Bigr]\prod_{i<j}\chi(z_{ij},\bar z_{ij})^{s_{ij}},\qquad
\label{eq:A86B}
\end{eqnarray}
while the fermion contractions give
\begin{eqnarray}
\mathcal A_{8\bar6}^F&\sim&
\eta^{\alpha\mu_1}\epsilon_{k_2\mu_2k_3\mu_3k_4\mu_4}\partial\Delta(z_1)
+\sum_{i=2}^4\bigl(\delta^{\alpha\mu_i}\epsilon_{\mu_1k_ik_j\mu_jk_k\mu_k}
-k_i^\alpha\epsilon_{\mu_1\mu_ik_j\mu_jk_k\mu_k}\bigr)\partial\Delta(z_i)\nn\\
&&+\sum_{i=2}^4\bigl(k_i^{\mu_1}\epsilon_{\alpha\mu_ik_j\mu_jk_k\mu_k}-\eta^{\mu_1\mu_i}
\epsilon_{\alpha k_ik_j\mu_jk_k\mu_k}\bigr)\partial\Delta(z_{1i})
+\sum_{2\le i<j\le4}\bigl(-k_i\cdot k_j\epsilon_{\alpha\mu_1\mu_i\mu_jk_k\mu_k}\nn\\
&&+k_i^{\mu_j}\epsilon_{\alpha\mu_1\mu_ik_jk_k\mu_k}
+k_j^{\mu_i}\epsilon_{\alpha\mu_1k_i\mu_jk_k\mu_k}
-\eta^{\mu_i\mu_j}\epsilon_{\alpha\mu_1k_ik_jk_k\mu_k}\bigr)\partial\Delta(z_{ij}).
\label{eq:A86F}
\end{eqnarray}
In the above, the indices $i$, $j$ and $k$ take on the distinct values $2$, $3$ and $4$.

Combining (\ref{eq:A86B}) and (\ref{eq:A86F}) gives an expression that starts at $\mathcal O(k^6)$.
Hence any contribution at the four-derivative level must originate from momentum denominators
of the form (\ref{eq:momden}).  We find
\begin{eqnarray}
\mathcal A_{8\bar6}\Big|_{k^4}\!\!&\sim&\!\!
\epsilon_{\alpha\nu_1k_3\nu_3k_4\nu_4}\Bigl[
-\fft{k_1^{\nu_2}}{s_{12}}\bigl(k_2^{\mu_1}\epsilon_{\alpha\mu_2k_3\mu_3k_4\mu_4}
-\eta^{\mu_1\mu_2}\epsilon_{\alpha k_2k_3\mu_3k_4\mu_4}\bigr)\nn\\
&&+\fft{k_3^{\nu_2}}{s_{23}}\bigl(-k_2\cdot k_3\epsilon_{\alpha\mu_1\mu_2\mu_3k_4\mu_4}
+k_2^{\mu_3}\epsilon_{\alpha\mu_1\mu_2k_3k_4\mu_4}
+k_3^{\mu_2}\epsilon_{\alpha\mu_1k_2\mu_3k_4\mu_4}
-\eta^{\mu_2\mu_3}\epsilon_{\alpha\mu_1k_2k_3k_4\mu_4}\bigr)\nn\\
&&+\fft{k_4^{\nu_2}}{s_{24}}\bigl(-k_2\cdot k_4\epsilon_{\alpha\mu_1\mu_2\mu_4k_3\mu_3}
+k_2^{\mu_4}\epsilon_{\alpha\mu_1\mu_2k_4k_3\mu_3}
+k_4^{\mu_2}\epsilon_{\alpha\mu_1k_2\mu_4k_3\mu_3}
-\eta^{\mu_2\mu_4}\epsilon_{\alpha\mu_1k_2k_4k_3\mu_3}\bigr)\Bigr].\nn\\
\label{eq:A86}
\end{eqnarray}
Note that the full contribution will include permutations of $2$, $3$ and $4$ on the right.
In addition, the amplitude $\mathcal A_{6\bar 8}$ with six fermions on the left and eight
on the right is simply related by interchange of $\{\mu_i\}\leftrightarrow\{\nu_i\}$ in the
above expression.

Finally, there is the contribution with eight fermions on both the left and the right
\begin{eqnarray}
\mathcal A_{8\bar8}&\sim&\left\langle\partial X^\alpha(0)\bar\partial X^\beta(0)
\prod e^{ik_i\cdot X(z_i,\bar z_i)}\right\rangle
\left\langle\psi^\alpha(0)\psi^{\mu_1}(z_1)\prod_{i=2}^4k_i\cdot\psi(z_i)
\psi^{\mu_i}(z_i)\right\rangle\nn\\
&&\kern5em\times\left\langle\tilde\psi^\beta(0)\tilde\psi^{\mu_1}(\bar z_1)
\prod_{i=2}^4k_i\cdot\tilde\psi(\bar z_i)\tilde\psi^{\mu_i}(\bar z_i)\right\rangle.
\end{eqnarray}
At the lowest order, only the bosonic zero mode contraction is important.  The amplitude
is then given by a product of (\ref{eq:A86F}) with its conjugate.  Again, we identify only the
pole terms of the form (\ref{eq:momden}).  The result is
\begin{eqnarray}
\mathcal A_{8\bar8}\Big|_{k^4}\!\!&\sim&\!\!
\sum_{i=2}^4\fft1{s_{1i}}\bigl(k_i^{\mu_1}\epsilon_{\alpha\mu_ik_j\mu_jk_k\mu_k}
-\eta^{\mu_1\mu_i}\epsilon_{\alpha k_ik_j\mu_jk_k\mu_k}\bigr)
\bigl(k_i^{\nu_1}\epsilon_{\alpha\nu_ik_j\nu_jk_k\nu_k}
-\eta^{\nu_1\nu_i}\epsilon_{\alpha k_ik_j\nu_jk_k\nu_k}\bigr)\nn\\
&&\kern-2em+\!\!\sum_{2\le i<j\le4}\fft1{s_{ij}}
\bigl(-k_i\cdot k_j\epsilon_{\alpha\mu_1\mu_i\mu_jk_k\mu_k}
+k_i^{\mu_j}\epsilon_{\alpha\mu_1\mu_ik_jk_k\mu_k}
+k_j^{\mu_i}\epsilon_{\alpha\mu_1k_i\mu_jk_k\mu_k}
-\eta^{\mu_i\mu_j}\epsilon_{\alpha\mu_1k_ik_jk_k\mu_k}\bigr)\nn\\
&&\times\bigl(-k_i\cdot k_j\epsilon_{\alpha\nu_1\nu_i\nu_jk_k\nu_k}
+k_i^{\nu_j}\epsilon_{\alpha\nu_1\nu_ik_jk_k\nu_k}
+k_j^{\nu_i}\epsilon_{\alpha\nu_1k_i\nu_jk_k\nu_k}
-\eta^{\nu_i\nu_j}\epsilon_{\alpha\nu_1k_ik_jk_k\nu_k}\bigr).
\label{eq:A88}
\end{eqnarray}
This amplitude is complete, and in particular no sum over permutations is needed, as
the eight fermion terms in the expansion of (\ref{eq:A4oo}) are unique.

\subsubsection{Recreating the effective Lagrangian}

At the four-derivative level, the odd-odd four-point amplitude is given by the sum of
$\mathcal A_{6\bar6}$, $\mathcal A_{8\bar6}$, $\mathcal A_{6\bar 8}$ and
$\mathcal A_{8\bar8}$, given by (\ref{eq:A66fin}), (\ref{eq:A86}) and its conjugate and
(\ref{eq:A88}).  This amplitude captures the scattering of gravitons, antisymmetric
tensors and dilatons, depending on how the polarizations $\theta_{\mu\nu}^{(i)}$
are chosen.  We are primarily interested in the scattering of four gravitons,
of two gravitons and two antisymmetric tensors, and of four antisymmetric tensors:
\begin{equation}
h\mathrm-h\mathrm-h\mathrm-h,\quad h\mathrm-h\mathrm-b\mathrm-b,\quad
b\mathrm-b\mathrm-b\mathrm-b.
\end{equation}
(Amplitudes with an odd number of antisymmetric tensors automatically vanish.)

We now take an effective Lagrangian of the form
\begin{equation}
\mathcal L=\mathcal L_{\rm tree}+\mathcal L_{\rm loop}^{\rm o\mhyphen o},
\label{eq:Ltreeloop}
\end{equation}
where $\mathcal L_{\rm tree}$ is given by (\ref{eq:Ltree0}) and
$\mathcal L_{\rm loop}^{\rm o\mhyphen o}$
is given by the parameterization (\ref{eq:Lparam}).  For simplicity, we remove the dilaton
terms in $\mathcal L_{\rm loop}^{\rm o\mhyphen o}$ by setting $\delta_i=\epsilon_i=\kappa_i=0$.  
This assumes, of course, that the dilaton couplings are absent in the string frame, and is motivated
by the result of heterotic dualization.  We will see that this assumption is justified in the end by our
ability to self-consistently obtain an effective Lagrangian reproducing the string amplitude.

At the four-point four-derivative level, the scattering amplitude receives contributions from
$s$, $t$ and $u$
channel exchange diagrams as well as from a four-point contact interaction.  The exchange
diagrams involve a three-point interaction from $\mathcal L_{\rm tree}$ connected to a
three-point interaction from $\mathcal L_{\rm loop}^{\rm o\mhyphen o}$  by either a graviton,
antisymmetric tensor or dilaton propagator.  The four-point contact term comes only from
$\mathcal L_{\rm loop}^{\rm o\mhyphen o}$.
Recreating the effective Lagrangian by matching the amplitude computed from
(\ref{eq:Ltreeloop}) with the string amplitude is a rather tedious exercise.  So instead we have
taken a shortcut by matching amplitudes involving a specific basis of momenta and polarizations
with the assistance of Mathematica.  As a technical note, while (\ref{eq:Ltreeloop}) is given
in the string frame, the actual scattering amplitude is computed in the Einstein frame, as that
corresponds directly with the choice of polarization tensors (\ref{eq:thetapol}) used in the
string vertex operators.

Up to an overall normalization, we find that the effective Lagrangian that generates the
odd-odd string amplitude is
\begin{eqnarray}
e^{-1}\delta\mathcal L_{\rm loop}^{\rm o\mhyphen o}
&=&R_{\mu\nu\lambda\sigma}^2-4R_{\mu\nu}^2+R^2
+\ft12R_{\mu\nu\lambda\sigma}H^{\mu\nu a}H^{\lambda\sigma a}
-R_{\mu\nu}H^{2\,\mu\nu}+\ft16RH^2\nonumber\\
&&+\ft1{144}(H^2)^2-\ft18(H^2_{\mu\nu})^2+\ft5{24}H^4.
\end{eqnarray}
We may now compare this with the CP-even four-derivative Lagrangian obtained by
dualizing the heterotic string.  The most straightforward comparison is with (\ref{eq:iia4dlag})
and (\ref{eq:t4t4Rplus2}), and the result is
\begin{equation}
e^{-1}\mathcal L_{\rm loop}^{\rm o\mhyphen o}=2e^{-1}\mathcal L_{\rm CP\mhyphen even}
-R_{\mu\nu\lambda\sigma}(\Omega_+)^2,
\end{equation}
or equivalently
\begin{equation}
2e^{-1}\mathcal L_{\rm CP\mhyphen even}=t_4t_4R(\Omega_+)^2
+e^{-1}\mathcal L_{\rm loop}^{\rm o\mhyphen o}.
\end{equation}
Since $\mathcal L_{\rm loop}^{\rm o\mhyphen o}$ here corresponds to the odd-odd sector,
we conclude that
$t_4t_4R(\Omega_+)^2$ arises from the even-even sector.  Furthermore, this justifies the
identification of the remaining terms in (\ref{eq:CPelag}) as arising from the odd-odd sector.

\section{$H^2 R^3$  and $H^4 R^2$ terms in ten dimensions}
\label{sec:morecor}

At this point, it is worth recalling what we have learned.  Based on a combination of four and five
point function results, we have conjectured in Section~\ref{sec:stam} that the one-loop
eight-derivative effective Lagrangian for
the type II string in ten dimensions takes the form given in (\ref{eq:olLaghat}) in the NSNS sector.
In particular, for all except the odd-odd spin structure sector, the Lagrangian is obtained from
the purely gravitational correction by the replacement $R\to R(\Omega_+)$, where $\Omega_+$
is the connection with torsion defined in (\ref{eq:Omegatorsion}).  We have given evidence for
this conjecture by considering T-duality in Section~\ref{sec:tdual} and six-dimensional
heterotic/IIA duality in Section~\ref{sec:sixd}.

In fact, the six-dimensional four-derivative Lagrangian, (\ref{eq:iiafin}), provides additional
information about the five and six-point function results in ten dimensions, albeit up to some
ambiguities.  To see this, we recall that (\ref{eq:iiafin}) is obtained by $K3$ compactification
of the IIA string.  This allows us to lift the six-dimensional four-point function to a ten-dimensional
six-point function with two legs on $K3$.

Before considering the lift, it is more straightforward to start with the reduction of
(\ref{eq:olLaghat}) on a background taken to be the direct product of $K3$ with a
six-dimensional spacetime.  In this case, the ten-dimensional Riemann tensor with
torsion, $R(\Omega_+)$, splits into separate internal and external parts.  Furthermore,
since there is no internal $H$-flux, the internal Riemann is torsion-free.  In this case,
only the factorized parts of (\ref{eq:olLaghat}) contribute to four-derivative couplings in
six dimensions.  The reduction follows by making use of the topological data of $K3$
\begin{eqnarray}
-\fft1{24\pi^2}\int_{K3} \tr R\wedge R&=&\sigma(K3)=-16,\nn\\
\fft1{32\pi^2}\int_{K3}d^4x\sqrt{h}(R_{\mu\nu\rho\sigma}^2-4R_{\mu\nu}^2+R^2)&=&\chi(K3)=24.
\end{eqnarray}
We find
\begin{eqnarray}
\mathcal L_{\rm loop}^{\alpha'^3}\Big|_{K3}
&=&(2\pi\sqrt{\alpha'})^4\Bigl[\sqrt{-g}\fft{\alpha'}{16}\left(t_4t_4R(\Omega_+)^2
-\ft18\epsilon_6\epsilon_6R(\Omega_+)^2-\ft16\epsilon_6\epsilon_6H^2R(\Omega_+)^2
+\cdots\right)\nn\\
&&\kern6em+\fft{\alpha'}4B_2\wedge\fft12\left(\tr R(\Omega_+)^2+\tr R(\Omega_-)^2\right)\Bigr].
\end{eqnarray}
Note that the $H^2(\nabla H)^2R(\Omega_+)$ term in ten dimensions does not survive the
$K3$ reduction.  The dimensionful factor out front may be absorbed in the six-dimensional
gravitational constant along with the definition of the six-dimensional dilaton.  The terms in
the square brackets then match their six-dimensional counterparts given in (\ref{eq:iiafin}),
except for the $\epsilon_6\epsilon_6H^4$ term, which was unspecified in ten dimensions.

The reason the $\epsilon_6\epsilon_6H^4$ term was not seen is that it uplifts to an
expression of the form $\epsilon_{10}\epsilon_{10}H^4R^2$, which can only be probed
by six or higher point functions.  Of course, we can directly lift (\ref{eq:iiafin}) on $K3$,
and thereby make a prediction for the six-point function.  However, there is a slight
complication involved, and that is that there are several possible ways to write out the
six-dimensional $H^4$ term with two $\epsilon$ tensors.  In particular, there are six
natural combinations
\begin{eqnarray}
\mathcal A_1&=&\epsilon_{\alpha\beta\mu_1\cdots\mu_4}\epsilon^{\alpha\beta\nu_1\cdots\nu_4}
H^{\mu_1}{}_{\nu_1a}H^{\mu_2}{}_{\nu_2}{}^a H^{\mu_3}{}_{\nu_3b}
H^{\mu_4}{}_{\nu_4}{}^b
=-2[(H^2)^2-2(H^2_{\mu\nu})^2-2H^4],\nn\\
\mathcal A_2&=&\epsilon_{\alpha\beta\mu_1\cdots\mu_4}\epsilon^{\alpha\beta\nu_1\cdots\nu_4}
H^{\mu_1\mu_2a}H_{\nu_1\nu_2a}H^{\mu_3\mu_4b}H_{\nu_3\nu_4b}
=-8[(H^2)^2-3(H^2_{\mu\nu})^2],\nn\\
\mathcal A_3&=&\epsilon_{\alpha\mu_0\mu_1\cdots\mu_4}\epsilon^{\alpha\nu_0\nu_1\cdots\nu_4}
H^{\mu\nu}{}_{\nu_0}H_{\nu_1\nu_2}{}^{\mu_0}H^{\mu_3}{}_{\nu_3a}H^{\mu_4}{}_{\nu_4}{}^b
=-2[(H^2)^2-2(H^2_{\mu\nu})^2-2H^4],\nn\\
\mathcal A_4&=&\epsilon_{\alpha\mu_0\mu_1\cdots\mu_4}\epsilon^{\alpha\nu_0\nu_1\cdots\nu_4}
H^{\mu\nu}{}_{\nu_0}H_{\nu_1\nu_2}{}^{\mu_0}H^{\mu_3\mu_4a}H_{\nu_3\nu_4a}
=-4[(H^2)^2-5(H^2_{\mu\nu})^2+4H^4],\nn\\
\mathcal A_5&=&\epsilon_{\alpha\mu_0\mu_1\cdots\mu_4}\epsilon^{\alpha\nu_0\nu_1\cdots\nu_4}
H^{\mu_0\mu_1\mu_2}H_{\nu_0\nu_1\nu_2}H^{\mu_3}{}_{\nu_3a}H^{\mu_4}{}_{\nu_4}{}^a
=-6[(H^2)^2-6(H^2_{\mu\nu})^2+6H^4],\nn\\
\mathcal A_6&=&\epsilon_{\alpha\mu_0\mu_1\cdots\mu_4}\epsilon^{\alpha\nu_0\nu_1\cdots\nu_4}
H^{\mu_0\mu_1\mu_2}H_{\nu_0\nu_1\nu_2}H^{\mu_3\mu_4a}H_{\nu_3\nu_4a}
=-12[(H^2)^2-3(H^2_{\mu\nu})^2].
\end{eqnarray}
We have defined the $\epsilon_6\epsilon_6H^4$ term in (\ref{eq:e4e4H4}) to match
$\mathcal A_2$.  However, it could equally well have come from other contractions, such as
\begin{equation}
\ft23\mathcal A_6,\qquad\ft23(4\mathcal A_1+\mathcal A_4),\qquad
\ft13(9\mathcal A_1+\mathcal A_5),\qquad\ft23(4\mathcal A_3+\mathcal A_4),\qquad
\ft13(9\mathcal A_3+\mathcal A_5),
\end{equation}
or any linear combinations thereof.  All six possible terms, $\mathcal A_1,\ldots,\mathcal A_6$,
have a natural lift to ten dimensions by taking $\epsilon_6\to\epsilon_{10}$ and multiplying
by $R^{\mu_5\mu_6}{}_{\nu_5\nu_6}(\Omega_+)R^{\mu_7\mu_8}{}_{\nu_7\nu_8}(\Omega_+)$.
Because there are multiple possible contractions between $H$ and Riemann in ten dimensions,
we expect these lifts to be distinct.  Hence there is an ambiguity at the level of $H^4R^2$ that
would be resolved by explicit computation of the six point function.

Schematically, we expect the lift of (\ref{eq:iiafin}) to take the form
\begin{equation}
2e^{-1}\delta\mathcal L^{\mbox{lift}} = t_8 t_8 R(\omp)^4 +\epsilon\epsilon R(\omp)^4
 + \alpha_n H^{2n} R^{4-n} (\omp) + \cdots,
\label{eq:iiaapm-lift}
\end{equation}
where the kinematic structure of the last term is left unspecifie and the ellipsis indicates
possible gradient and Ricci terms.  The $H^2R^3$ term is fully specified by the five-point
function, and is given in (\ref{eq:olLaghat}), based on the result of \cite{Peeters:2001ub}.
As we have just seen, the $H^4R^2$ term is partially determined by the uplift of the
six-dimensional Lagrangian.  However, an ambiguity remains as the lift is not unique.
Finally, since there is no internal $H$-flux, the six-dimensional result cannot make
predictions about the structure of $H^6 R$ and $H^8$ terms.
Finally, since this lift of the six-dimensional Lagrangian confirms the presence of these
additional $H^{2n}R^{4-n}$ terms in the odd-odd spin structure sector, is suggests that
there is either some subtlety in the computation of the five-point function in the
Green-Schwarz formalism or in properly identifying the effective Lagrangian in
\cite{Richards:2008sa}, which did not find these additional terms.

\section{To eleven dimensions and \ldots back}
\label{sec:morecor11}

The one-loop $R^4$ couplings are intimately related to M5 anomalies, and their lift to
eleven-dimensions has a number of important applications.%
\footnote{While the supersymmetric completion of the eleven-dimensional $R^4$ term
may be studied directly
\cite{Howe:2003cy,Cederwall:2004cg,Peeters:2005tb,Rajaraman:2005ag,Hyakutake:2006aq,Hyakutake:2007sm},
the lift from ten dimensions provides a complementary approach as well as potential
new insight into the web of dualities.}
Due to the fact that the
generalized curvature terms are even in $H$, the modified couplings can lifted to
eleven dimensions. Roughly, since the $H$'s always come in pairs, the new
eleven-dimensional indices coming from $H_{\mu \nu \rho} \rightarrow
\hat{G}_{\mu \nu \rho \lambda}$%
\footnote{We shall hat all the eleven-dimensional quantities.}
also come in pairs and can be contracted with each other.  We shall do this lift by considering
the simplest situation when the eleven-dimensional spacetime is a direct product
$M_{11} = M_{10} \times S^1$, and $\hat{G}_{\mu \nu \rho 11}$ is the only non-vanishing
component of the four-form $\hat G$. However, having obtained the full eleven-dimensional
coupling this way, we may subsequently perform the reduction on a nontrivially fibered
circle (non-trivial graviphoton) and with generic $G$, yielding both the NSNS three-form
$H$ as well as the RR four-form $F_4$. This way we would, in principle, obtain the full
eight-derivative one-loop coupling in IIA involving all RR fields.

Let us briefly recall the story without $B_2$ (or $\hat{C}_3$). Taking the eleven-dimensional
metric to be
\beq
\label{g11d}
d \hat{s}^2_{11} = e^{-\phi/3} \left( d s^2_{10} + e^{\phi} (d \psi + A)^2 \right)
\eeq
yields the IIA action in the string frame. Note that the tree-level $R^4$ terms do not survive
the eleven-dimensional decompactification limit, while the one-loop terms lift.
When $M_{11} = M_{10} \times S^1$ the reduction is well understood, and the ten- and
eleven-dimensional  terms are related as
\beq
\fft{\alpha'^3}{2 \kappa^2_{10}} \int_{M_{10}}  B_2 \wedge X_8 (R) \quad \leftrightarrow
\quad \fft{(4\pi \kappa^2_{11})^{2/3}}{\kappa^2_{11}} \int_{M_{11}} \hat{C}_3 \wedge X_8 (\hat{R}) \, .
\eeq
Note that the radius of the circle does not need to be taken constant, since due to the
conformal invariance of $X_8$ the dilation drops out of the ten-dimensional coupling.
The reduction of the eleven dimensional completion of
$(t_8t_8-\fft18 \epsilon_{10}\epsilon_{10})R^4 $ is more subtle, as these terms are not conformally
invariant. However, one can show that the expression with complicated dilation dependance
obtained via the direct reduction is on-shell equivalent to the standard one-loop term without dilation, as discussed in \cite{FT}. We shall return to the dilation shortly. 
The reduction with nontrivial graviphoton has been discussed in \cite{LM}. 

Before considering the lift of the ten-dimensional term to eleven dimensions, let us look at the
four-form $\hat G$. Define the O(1,10)-valued one-form
\beq
\hat{\GG}_1^{abc} = 4 \hat{G}_{\mu \nn \rho \lambda} d \hat{x}^{\mu} \hat{e}^{a \nu}  \hat{e}^{b \rho}  \hat{e}^{c \lambda} \, .
\eeq
With the standard reduction $\hat{G}_4 = F_4 + H_3 \wedge e$ (with $e= d \psi + A$, and
the induced ten-dimensional Bianchi identities $d H_3 = 0$ and $d F_4 - H_3 \wedge F_2 = 0$)
and using the metric (\ref{g11d}),  we can see 
\beq
\hat{\GG}_1^{abc}  \quad \longrightarrow  \quad (e^{\phi/2} \FF^{abc}; \HH^{ab}) \, ,
\eeq
where $\FF$ and $\HH$ are O(1,9)-valued one-forms. Of course the latter is familiar and
enters the connection with torsion $\Omega_\pm =\Omega \pm \ft12\HH$. It is important
that the $\HH$ that arises from the reduction of $\hat{\GG}$ comes without powers of
the dilation, and hence {\sl all}  NSNS sector terms produced from reduction of the modified
eleven-dimensional $R^4$ terms are at one loop.

It is useful to write down the explicit expressions for the characteristic class  computed from the
connection with torsion $\Omega_{\pm}=\Omega \pm \fft12 \HH$, where
$\mathcal H^{ab}=H_\mu{}^{ab}dx^\mu$. Of course the shift in the connection results in a
shift of the characteristic class by an exact term.  In particular
\begin{equation}
p_1(\Omega_{\pm}) = - \frac{1}{8 \pi^2} \left( \tr R^2 + d \left( \pm \tr \HH R
+ \frac{1}{4} \tr \HH \nabla \HH \pm \frac{1}{12} \tr \HH^3 \right)\right) \, ,
\end{equation}
and similarly for $p_2$ (although for $p_2$ the shift term is very long and not very enlightening). 
We can write down the relevant averaged quantity
\beq
\overline{X}_4 =  \frac{1}{2} \left( p_1(\omp) + p_1(\omm) \right) = - \frac{1}{32 \pi^2} \left( \tr R^2 +   \frac{1}{4} d \,  \tr \HH \nabla \HH \right) \, .
\label{eq:x4shift}
\eeq
This four-form appears in the six-dimensional couplings and is to completed in the
CP-even sector by (\ref{eq:CPelag}) or equivalently (\ref{eq:iia4dlag}).
The ten dimensional couplings involve the following eight-form:
\begin{eqnarray}
\label{eq:x8shift}
\overline{X}_{8} &=& \fft1{ 192 (2\pi)^4}\bigg[  \left( \tr R^4 -\fft1{4}(\tr R^2)^2 \right) + \nn \\
&&\quad d \,\, \bigg( \fft1{2} \tr \left( \HH \nabla \HH R^2 + \HH R \nabla \HH R
+ \HH R^2 \nabla \HH \right) -\fft1{8} \left( \tr R^2 \, \tr \HH \nabla \HH
+ 2 \tr \HH R \, \tr R \nabla \HH \right)    \nn \\
&&\qquad \fft1{16 }  \tr  \left( 2 \HH^3 (\nabla \HH R + R \nabla \HH) + \HH R \HH^2 \nabla \HH
+ \HH \nabla \HH \HH^2 R \right)\nn\\
&& \qquad - \fft1{2}    \left( \tr \HH \nabla \HH \,  \tr R \HH^2 +  \tr R \nabla \HH \,  \tr \HH^3
+ \tr \nabla \HH \HH^2 \, \tr \HH R \right)    + \nn \\
&&\qquad \fft1{32 } \tr \nabla \HH \HH^5   -\fft1{192}  \tr \nabla \HH \HH^2 \, \tr \HH^3  + 
 \fft1{16 }  \tr \HH (\nabla \HH)^3  -\fft1{64} \tr \HH \nabla \HH \, \tr (\nabla \HH)^2  \bigg) \bigg].
\end{eqnarray}

In order to discuss the lift to eleven dimensions (or as a toy example from six to seven
dimensions) we think of the simplest situation when $M_{D+1} = M_D \times S^1$, i.e.\ the
higher-dimensional  spacetime is a direct product  of the lower-dimensional one with a
circle at fixed radius, and $H_{\mu \nu \rho} = \hG_{\mu \nu \rho 11}$ and
$\hG_{\mu \nu \rho \lambda} =0$.  One can simply add an extra summation index on $\HH$
and then check that (\ref{eq:x4shift}) and (\ref{eq:x8shift}) are correctly reproduced from
higher-dimensional couplings.%
\footnote{Crucially $\hGG^{ab 11} \rightarrow \HH^{ab}$ has the right scaling to keep the
shifted terms at one-loop.}
Note that taking $M_{11}$ to be a generic circle bundle and taking generic $G_4$ we
recover the full IIA set of our eight-derivative couplings at one loop, with RR fields (the
three-form and the ten-dimensional graviphoton) included.

We start with the lift of the six-dimensional CP-odd term.
The seven-dimensional lift of the coupling involving the four-form  (\ref{eq:x4shift}) is rather straightforward.%
\footnote{There is a slight catch --- differently from the eleven-dimensional lift,
there should also be four-derivative couplings involving the vectors. (Recall that the
$O(4, 20)/O(4) \times O(20)$ moduli space in six dimensions lifts to seven-dimensional
$O(3,19)/O(3) \times O(19)$.). We will be ignoring these here.} 
\beq
\fft 14 B_2 \wedge \overline{X}_4 \quad \longrightarrow \quad - \frac{1}{32 \pi^2} \quad C_3 \wedge
\left( \tr R^2 -    \frac{1}{12} d \left( \GG^{abc} \wedge (\nabla \GG)^{abc} \right) \right).
\label{eq:iia4dlagxx}
\eeq
Even the lifting of its CP-even completion is not too difficult. The action (\ref{eq:iia4dlag}) lifts
to
\begin{eqnarray}
e^{-1}\delta\mathcal L^{\rm{lift}}&=&R_{\mu\nu\lambda\sigma}^2-2R_{\mu\nu}^2+\ft12R^2
-\ft1{6} R_{\mu\nu}G^{2\,\mu\nu}+\ft1{48}RG^2
+\ft1{6} \nabla_\mu G_{\nu\alpha\beta \gamma }\nabla^\nu G^{\mu\alpha\beta \gamma }\nn\\
&&+\ft1{48}G_{\mu\nu\lambda\rho}G^{\mu \rho \alpha}{}_\beta
G^{\nu \sigma \beta}{}_\gamma G^{\lambda\gamma}{}_{\alpha \sigma}+\ft1{288\cdot12}(G^2)^2
- \ft1{216}(G_{\mu \nu})^2+ (\mbox{eom})^2.
\label{eq:iia4dlagx}
\end{eqnarray}

The eleven-dimensional lift of (\ref{eq:x8shift}) is going to be a little more troublesome, and we
do not fully know its CP-even completion so will not attempted the lift. We note that there
are again ambiguities associated with this lifting as the number of possible terms in ten
dimensions is larger that their ten-dimensional descendants.  Consider, e.g.
\begin{eqnarray}
&&  \tr \HH (\nabla \HH)^3 - \fft1{4} \tr \HH \nabla \HH \, \tr (\nabla \HH)^2   \quad \longrightarrow 
\nonumber\\
&&\qquad A \hGG^{abc} \hnabla \hGG^{bde}  \hnabla \hGG^{cdf} \hnabla \hGG^{efa} 
+ B \hGG^{abc} \hnabla \hGG^{bcd}  \hnabla \hGG^{def} \hnabla \hGG^{efa}
+  C \hGG^{abe} \hnabla \hGG^{abe}  \hnabla \hGG^{cdf} \hnabla \hGG^{cdf} .\qquad
\label{eq:h4lift}
\end{eqnarray}
The first term reduces only to a single-trace contribution in ten dimensions (if one keeps
track of only terms made of $\HH$), and the last term reduces to a double-trace, while the
term in the middle can give rise to both types of ten-dimensional terms. Similarly,
\begin{eqnarray}
 && \tr \nabla \HH \HH^5   -\fft1{6}  \tr \nabla \HH \HH^2 \, \tr \HH^3   \quad \longrightarrow 
\nonumber\\
&&\qquad  \hnabla \hGG^{abg} \hGG^{ghc} \hGG^{cdh}  \bigg[ A'  \hGG^{hde}  \hGG^{efi} \hGG^{ifa}
+ B'  \hGG^{dei}  \hGG^{efh} \hGG^{ifa} \bigg] +
 \nonumber\\
&&\qquad C'  \hnabla \hGG^{abg} \hGG^{bch}  \hGG^{cdg} \hGG^{dei}   \hGG^{efh} \hGG^{ifa}
+ D'  \hnabla \hGG^{abi} \hGG^{bci}  \hGG^{cag} \hGG^{deh}   \hGG^{efh} \hGG^{fdg} ,
\label{eq:h6lift}
\end{eqnarray}
with the first line reducing to single-trace piece in ten-dimensions, the last term to a double trace
and the first term on the second line to a mixed piece.
Once more, getting the full IIA coupling is now just a mater of reduction. Note that by
construction all couplings will have even powers of RR fields, as they should.

Although the full eleven-dimensional lift is rather tedious, we may highlight the terms that
will arise up to cubic order in the three-form $\hat C$.  Starting in the CP-odd sector, we lift
$B_2\wedge\overline X_8$, were we only explicitly retain the first two lines of (\ref{eq:x8shift}).
The result is
\begin{eqnarray}
B_2\wedge\overline X_8&\longrightarrow&\fft1{192(2\pi)^4}\Bigl[
\hat C\wedge\left(\tr\hat R^4-\ft14(\tr\hat R^2)^2\right)\nn\\
&&\kern4em+\hat G\wedge\Bigl(\ft14\left(
\hat R^{ab}\hat R^{bc}\hat\mathcal G^{cde}\hat\nabla\hat\mathcal G^{dae}
+2\hat R^{ab}\hat\mathcal G^{bce}\hat R^{cd}\hat\nabla\hat\mathcal G^{dae}
+\hat R^{ab}\hat R^{bc}\hat\nabla\hat\mathcal G^{cde}\hat\mathcal G^{dae}\right)\nn\\
&&\kern7em-\ft1{24}\left(\tr\hat R^2\wedge\hat\mathcal G^{abe}\hat\nabla\hat\mathcal G^{bae}
+6\hat R^{ab}\hat\mathcal G^{bae}\hat R^{cd}\hat\mathcal G^{dce}\right)+\cdots\Bigr)\Bigr].
\label{eq:CPolift}
\end{eqnarray}

In the CP-even sector, we need only work to quadratic order in $\hat C$.  For the even-even
sector, we have, schematically
\begin{equation}
t_8t_8R(\Omega_+)^4=t_8t_8(R^4+6(\nabla H)^2R^2+2H^2R^3+\cdots).
\end{equation}
The lift of $t_8t_8(\nabla H)^2R^2$ is, unfortunately, not as straightforward as it may appear.
A simple attempt at the lift would be to take
\begin{eqnarray}
&&t_8^{\mu_1\cdots\mu_8}t^8_{\nu_1\cdots\nu_8}
\nabla_{\mu_1}H_{\mu_2}{}^{\nu_1\nu_2}\nabla_{\mu_3}H_{\mu_4}{}^{\nu_3\nu_4}
R_{\mu_5\mu_6}{}^{\nu_5\nu_6}R_{\mu_7\mu_8}{}^{\nu_7\nu_8}\nn\\
&&\qquad\longrightarrow\quad
t_8^{\mu_1\cdots\mu_8}t^8_{\nu_1\cdots\nu_8}
\nabla_{\mu_1}\hat G_{\mu_2}{}^{\nu_1\nu_2a}\nabla_{\mu_3}\hat G_{\mu_4}{}^{\nu_3\nu_4a}
\hat R_{\mu_5\mu_6}{}^{\nu_5\nu_6}\hat R_{\mu_7\mu_8}{}^{\nu_7\nu_8}.
\end{eqnarray}
However, the reduction would generate terms with not just $a=11$ (as intended),
but also $\mu_2=\mu_4=11$ and $\nu_{1\mathrm{\;or\;}2}=\nu_{3\mathrm{\;or\;}4}=11$.
These undesirable additions can be avoided by expanding out the $t_8t_8$ contractions
in full, and then lifting each individual term in the expansion.  This, however, leads to a
proliferation of terms.  We will not attempt to write out this lift, but we instead note that a
direct eleven-dimensional construction of the $(\hat\nabla\hat G)^2\hat R^2$ terms,
which we denote by $\Delta\mathcal L\left((\hat\nabla\hat G)^2\hat R^2\right)$, is
given in \cite{Deser:1998jz,Deser:2000xz,Deser:2005kb,Peeters:2005tb}.  As
discussed in \cite{Peeters:2005tb}, the circle reduction of
$\Delta\mathcal L\left((\hat\nabla\hat G)^2\hat R^2\right)$ matches $t_8t_8(\nabla H)^2R^2$.
(Moreover, when compactified on $K3$,  it agrees with the $(\nabla G)^2$ term
in the seven dimensional action (\ref{eq:iia4dlagx}).)
Hence this expression is indeed the desired lift (up to a few ambiguities that are fixed by the
five-point function).  The lift of $t_8t_8 H^2 R^3$ is more direct, and we find
\begin{equation}
t_8t_8R(\Omega_+)^4\longrightarrow
t_8t_8\hat R^4+\Delta\mathcal L\left((\hat\nabla\hat G)^2\hat R^2\right)
-t_8^{\mu_1\cdots\mu_8}t^8_{\nu_1\cdots\nu_8}\hat G_{\mu_1}{}^{\nu_1ab}
\hat G_{\mu_2}{}^{\nu_2ab}\hat R_{\mu_3\mu_4}{}^{\nu_3\nu_4}
\hat R_{\mu_5\mu_6}{}^{\nu_5\nu_6}\hat R_{\mu_7\mu_8}{}^{\nu_7\nu_8}+\cdots.
\label{eq:eelift}
\end{equation}

For the lift of the odd-odd part of the CP-even sector, it is best to start from (\ref{eq:ooLag0}),
which keeps all the $H$ terms  explicit.  (Moreover this expression demonstrates that
there are no $\nabla H$ terms contributing at quadratic order.)  The lift then takes the form
\begin{eqnarray}
-\ft18\epsilon_{10}\epsilon_{10}(R^4+\ft23H^2R^3+\cdots)&\longrightarrow&
-\ft1{24}\epsilon_{\alpha\beta\gamma\mu_1\cdots\mu_8}
\epsilon^{\alpha\beta\gamma\nu_1\cdots\nu_8}
\hat R^{\mu_1\mu_2}{}_{\nu_1\nu_2}\hat R^{\mu_3\mu_4}{}_{\nu_3\nu_4}
\hat R^{\mu_5\mu_6}{}_{\nu_5\nu_6}\hat R^{\mu_7\mu_8}{}_{\nu_7\nu_8}\nn\\
&&-\ft1{96}\epsilon_{\alpha\mu_1\cdots\mu_{10}}
\epsilon^{\alpha\nu_1\cdots\nu_{10}}
\hat G^{\mu_1\mu_2}{}_{\nu_1\nu_2}\hat G^{\mu_3\mu_4}{}_{\nu_3\nu_4}
\hat R^{\mu_5\mu_6}{}_{\nu_5\nu_6}\hat R^{\mu_7\mu_8}{}_{\nu_7\nu_8}
\hat R^{\mu_9\mu_{10}}{}_{\nu_9\nu_{10}}\nn\\
&&+\cdots.
\label{eq:oolift}
\end{eqnarray}

We now collect (\ref{eq:CPolift}), (\ref{eq:eelift}) and (\ref{eq:oolift}) to give the eleven-dimensional
Lagrangian
\bea
{2\kappa_{11}^2}\ \mathcal L_{\rm M} &=&\hat R*1-\ft12 \hG \wedge *\hG
-\ft16 \hat C \wedge \hG \wedge \hG
+(4\pi\kappa_{11}^2)^{2/3}\left( \sqrt{-g}(t_8t_8 - \ft1{24} \epsilon_{11}\epsilon_{11})
\hat R^4+ \hat C\wedge X_{8}(\hat \Omega)  \right)   \nn \\
&&+\Delta \mathcal L\left( (\hat \nabla \hG)^2 \hat R^2 \right)
+  (4\pi\kappa_{11}^2)^{2/3} \sqrt{-g}\biggl[  -t_8t_8\hat G^2\hat R^3 - 
\ft1{96} \epsilon_{11}\epsilon_{11} \hG^2 \hat R^3 \biggr]\nn\\
&& + (4\pi\kappa_{11}^2)^{2/3} \fft{\hG}{192(2\pi)^4} \wedge \biggl[\ft14\left(
\hat R^{ab}\hat R^{bc}\hat\mathcal G^{cde}\hat\nabla\hat\mathcal G^{dae}
+2\hat R^{ab}\hat\mathcal G^{bce}\hat R^{cd}\hat\nabla\hat\mathcal G^{dae}
+\hat R^{ab}\hat R^{bc}\hat\nabla\hat\mathcal G^{cde}\hat\mathcal G^{dae}\right)\nn\\
&&\kern11em-\ft1{24}\left(\tr\hat R^2\wedge\hat\mathcal G^{abe}\hat\nabla\hat\mathcal G^{bae}
+6\hat R^{ab}\hat\mathcal G^{bae}\hat R^{cd}\hat\mathcal G^{dce}\right) \biggr]  \nn \\ 
&& +  \mathcal O(\hG^4) \, ,
\label{11dG^3}
\eea
where $t_8t_8\hat G^2\hat R^3$ and $\epsilon_{11}\epsilon_{11} \hG^2 \hat R^3$ are implicitly
defined in (\ref{eq:eelift}) and (\ref{eq:oolift}), respectively.  The relevance of this Lagrangian
will be discussed shortly. The eight-derivative terms in the first line are standard
\cite{DLM, Green:1997di}. The difficulties in the direct eleven-dimensional  construction of
the $\hG^2 \hat R^3$ terms are discussed in \cite{Hyakutake:2007sm}.

\section{Discussion}

We shall conclude by providing a partial list of questions that was not addressed in this paper.

\vskip 0.3cm
\noindent {\bf Dilaton:}
We have mostly ignored the dilaton in our discussion, and believe that our expressions are
correct. This point of view is vindicated in part by the discussion of Section~\ref{sec:sixd},
where the absence of the dilaton in the complete six-dimensional action (\ref{eq:iiafin}) is a
result of the dualisation (and not simple a ``result" of ignoring the dilaton). It would have been
nice to be able prove this in full generality. One could have imagined the correct dilation
dependance to come from a reduction from M-theory to type IIA. Let us recall what is known
about this without modifying the $\alpha'^3$ terms by $H$. It is not hard to show that, as far as
the one-loop topological part goes, the reduction  $\hat C_3 \wedge X_8(\hat\Omega)
\longrightarrow
B_2 \wedge X_8 (\Omega) $ produces no dilaton-dependent terms. Such terms naively arise in
reducing the eleven-dimensional $(t_8t_8 -  \fft 1{24} \epsilon_{11} \epsilon_{11})\hat R^4$ terms
from eleven to ten. However they can be shown to be on-shell equivalent to the standard
$(t_8t_8 -  \fft 1 8 \epsilon_{10} \epsilon_{10})R^4$ one-loop terms without any dilation
dependance. The mixed terms involving eleven-dimensional curvature and four-form flux
$G$ discussed in Section~\ref{sec:morecor11} do yield dilaton-dependent terms upon
reduction. We have not verified that, as expected, this dependance can be eliminated
on-shell so that the action (\ref{eq:olLaghat}) receives no contributions from the dilation.

\vskip 0.3cm
\noindent {\bf Higher orders in $\alpha'$:}
We have worked here at lowest relevant order in $\alpha'$.  One of the results of the
six-dimensional calculation of Section~\ref{sec:sixd} is that the classical duality map
between heterotic and type IIA theories (\ref{eq:HetIIAmap}) receives no modifications
at order $\alpha'$. On the contrary, it seems such modifications will be needed for the
T-duality transformation in heterotic strings in order to be able to verify the correct
T-duality transformation of the heterotic Bianchi identity to all orders in $\alpha'$. In
many ways it seems like T-duality might be the most tangible way of computing
couplings at higher orders in $\alpha'$.

Higher orders in $\alpha'$ also arise in the low energy expansion of the string scattering
amplitudes.  While we have focused on the $\alpha'^3$ coupling, which requires an
eight-point function to pin down the $H^8$ term, there have been complementary studies
that stick with the four-point function, but work to all orders in $\alpha'$
\cite{Kehagias:1997cq,Kehagias:1997jg,Green:1999pv,Policastro:2006vt,Green:2008uj,Policastro:2008hg}.

\vskip 0.3cm
\noindent {\bf Fixing the ambiguities and generalized geometry:}
In two instances, namely in Section~\ref{sec:morecor} and Section~\ref{sec:morecor11},
we have obtained some information about the higher-dimensional couplings from
lower-dimensional ones. As we discussed, these lifts are ambiguous. In particular, up to
this ambiguity, we have discussed $H^2R^3$ and $H^4 R^2$ terms here and have left
$H^6 R$ and $H^8$ terms completely unfixed. Most likely T-duality provides a more
realistic calculation path towards fixing these terms than direct string seven- and
eight-point calculations. For the
eleven-dimensional couplings of Section~\ref{sec:morecor11}, supersymmetry is {\it a priori}
the only available tool
\cite{Howe:2003cy,Cederwall:2004cg,Peeters:2005tb,Rajaraman:2005ag,Hyakutake:2006aq,Hyakutake:2007sm}.
In this context, not to mention the obvious self-interest,  a formal
understanding of the couplings in terms of generalised geometry might be very interesting.
Indeed, on the generalised tangent bundle $E$
\begin{equation}
0 \, {\longrightarrow} \, T^*M  \, {\longrightarrow} \, E \,  {\longrightarrow} \, TM \, {\longrightarrow} 
 \, 0,
\end{equation}
the torsionful connections $\ompm$ appear
naturally, and one is led to asking how much generalised geometry captures the systematics
of the string perturbation theory, and not just classical supersymmetry. Hopefully we have
provided here enough information about the higher-derivative couplings to use as a
meaningful guideline in search of  ``characteristic" higher-curvature objects in generalised
geometry.%
\footnote{An obvious, though likely naive, obstacle is that on $E$ differently from Riemannian geometry the no-torsion condition does not lead to an unambiguous generalized Riemann tensor \cite{CCW}. This causes no problems as far as the two-derivative supergravity is concerned.

Note also that $E$ captures the information about the metric and the $B$-field, and in order to cover the full NSNS sector one has to work on $E \otimes L$ where $L$ is a trivial line bundle with sections $\mbox{exp}(-\phi) \in L$. Hence thinking about the relation of the $\alpha'^3$ coupling in generalized geometry will most likely also provide a natural way of incorporating the dilation in the couplings.}
Finding these is probably the most practical way of completing (\ref{eq:olLaghat}).

\vskip 0.3cm
\noindent {\bf Theories with eight supercharges:}
Finally, another venue of research is to study Calabi-Yau reductions of the modified
couplings and their effect in $\mathcal{N} = 2$ four-dimensional theories (and related six-dimensional $(1,0)$ theories%
\footnote{For the $(1,0)$ theory, only the special case where the gravitational terms do
not enter the Bianchi identity was considered so far  \cite{Bergshoeff:1986vy,Bergshoeff:1986wc,Bergshoeff:1987rb,Bergshoeff:2012ax}.}).
As already remarked, 
not all the ambiguities in $\alpha_n H^{2n} R^{4-n} (\omp) $ have been fixed for $n \geq 2$.
However, most of these terms are not relevant for Calabi-Yau compactifications. In fact,  for
$\mathcal N = 2$ actions arising from reductions without internal fluxes,  for the couplings
involving up to four derivatives, only terms at most cubic in $H$ are relevant.  In other words,
the action in (\ref{eq:olLaghat}) should suffice. It is known that
$R^2$ terms coupled to scalars in vector multiplets descend from ten-dimensional $R^4$
terms. Our ten-dimensional couplings seem to predict four-derivative couplings quadratic in $H$.
However the $B$-field in Calabi-Yau reductions of the IIA theory is a part of a hypermultiplet,
and hence the study of the modified couplings in the Calabi-Yau reductions can provide
interesting knowledge about the interactions between the hyper and vector multiplets
beyond the well-known two-derivative factorization of the respective moduli spaces.  In this
context, the fact that the eleven-dimensional action up to cubic order in three-form $\hat C$,
(\ref{11dG^3}), is known without ambiguities becomes very important.


\subsubsection*{Acknowledgments}

Useful discussions with  M. Duff, H. Elvang, M. Gualtieri, N. Hitchin, A. Kahle, I. Melnikov,
D. Robbins, P. Vanhove and D. Waldram are gratefully acknowledged. 
RM would like to thank the Simons Center for Geometry and Physics, the Isaac Newton Institute
and the Yukawa Institute for Theoretical Physics for hospitality during the course of this work.
This work was supported in part by the US Department of Energy under grant
DE-SC0007859 (JTL) and by Agence Nationale de la Recherche under grant 12-BS05-003-01 (RM).


\appendix


\section{Conventions, definitions and some useful formulae}
\label{app:conv}

We collect here some of our conventions and useful formulae.

\vskip 0.3cm
\noindent
$\rhd$ {\bf Connections and curvatures:} The  connection with torsion is defined as $\Omega_{\pm }=\Omega \pm \ft12\mathcal H$, which reads in components
\begin{equation}
\Omega_{\pm \,\mu}{}^{\alpha\beta}=\Omega_\mu{}^{\alpha\beta} \pm \ft12H_\mu{}^{\alpha\beta}.
\end{equation}
The curvature computed out of $\Omega_{\pm}$ is then
\begin{equation}
R(\Omega_{\pm})=R \pm \ft12d\mathcal H+\ft14\mathcal H\wedge\mathcal H,\qquad
\mathcal H^{\alpha\beta}=H_\mu{}^{\alpha\beta}dx^\mu,
\end{equation}
or in components
\begin{equation}
R(\Omega_{\pm})_{\mu\nu}{}^{\alpha\beta}=R_{\mu\nu}{}^{\alpha\beta}
 \pm \nabla_{[\mu}H_{\nu]}{}^{\alpha\beta}+\ft12H_{[\mu}{}^{\alpha\gamma}H_{\nu]\gamma}{}^\beta.
\end{equation}
where $R_{\mu\nu}{}^{\alpha\beta}$ denotes the Riemann tensor. In all quantities computed from connection with torsion we write the argument  $\ompm$. The quantities without an argument by default use the Levi-Civita connection.
Note that $R(\Omega_{\pm})_{\mu\nu}{}^{\alpha\beta}$ satisfies a useful Bianchi identity
(due to closure of $H$):
\beq
R(\Omega_+)_{\mu\nu \alpha\beta} = R(\Omega_-)_{\alpha\beta\mu \nu} \,.
\eeq
Finally, we use $\bar R (\ompm)$ for  the linearised curvature.

\vskip 0.3cm
\noindent
$\rhd$ {\bf Backgrounds with isometry:} For testing the T-duality properties, we shall consider 
U(1) fibered backgrounds: $U(1) \hookrightarrow M \stackrel{\pi}{\longrightarrow}M_B$. The curvature of  the globally defined smooth one-form $e$  is denoted by $T$:  $d e = \pi^* T$ and $T \in H^2(M_B, \mathbb{Z})$. For $H$ respecting the isometry,  we have
\begin{equation}
H_3=h_3+\tilde T\wedge e,
\end{equation}
where $h_3=db_2-b_1\wedge T$, so that $dh_3=-\tilde T\wedge T$.

The connection  $\Omega_{\pm}$ can be written in terms of horizontal quantities:
\begin{eqnarray}
\Omega_{\pm}^{\alpha\beta}&=&\omega_{\pm}^{\alpha\beta}-\ft12 (T \mp \tilde T)^{\alpha\beta}e,\nonumber\\
\Omega_{\pm}^{\alpha9}&=&-\ft12 (T \pm \tilde T)^\alpha{}_\gamma e^\gamma,
\end{eqnarray}
 where $\omega_+=\omega+\fft12 h$. 
\begin{equation}
R^{\alpha\beta}(\omega_{\pm})=R^{\alpha\beta} \pm \ft12\nabla_\gamma h_\delta{}^{\alpha\beta}
e^\gamma\wedge e^\delta+\ft14h_\gamma{}^{\alpha\epsilon}h_{\delta\epsilon}{}^\beta
e^\gamma\wedge e^\delta.
\end{equation}
And the Bianchi identity for $R^{\alpha\beta}(\omega_+)$:
\beq
R(\omega_+)_{\mu\nu \alpha\beta} = R(\omega_-)_{\alpha\beta\mu \nu}  + \nabla_{[\alpha} h_{\beta \mu \nu]} =   R(\omega_-)_{\alpha\beta\mu \nu}  -  T_{[\alpha \beta} \tilde T_{ \mu \nu]} \,.
\eeq

\vskip 0.3cm
\noindent
$\rhd$ {\bf Odd-odd sector:}  Throughout the text we use a shorthand notation for quantities
such as $\epsilon_{10}\epsilon_{10}R^4 $. These expressions use the covariant epsilon tensor
so that e.g.
\beq
\epsilon_{10} \epsilon_{10}R^4= \epsilon_{\alpha\beta\mu_1\cdots\mu_8}\epsilon^{\alpha\beta\nu_1\cdots\nu_8}
R^{\mu_1\mu_2}{}_{\nu_1\nu_2} R^{\mu_3\mu_4}{}_{\nu_3\nu_4}
R^{\mu_5\mu_6}{}_{\nu_5\nu_6}R^{\mu_7\mu_8}{}_{\nu_7\nu_8} \,.
\eeq
Note that a pair of indices is contracted between two epsilons. It is not hard to check that
\beq
\epsilon_{10} \epsilon_{10}R^4(\omp)  = \epsilon_{10} \epsilon_{10}R^4 (\omm)
\eeq

In the odd-odd sector one also encounters other structures, such as 
\beq
\epsilon_{10} \epsilon_{10} H^2 R^3 = \epsilon_{\alpha\mu_0\mu_1\cdots\mu_8}\epsilon^{\alpha\nu_0\nu_1\cdots\nu_8}
H^{\mu_1\mu_2}{}_{\nu_0}H_{\nu_1\nu_2}{}^{\mu_0}
R^{\mu_3\mu_4}{}_{\nu_3\nu_4}
R^{\mu_5\mu_6}{}_{\nu_5\nu_6}R^{\mu_7\mu_8}{}_{\nu_7\nu_8},
\eeq
with only one index contracted between two epsilons. Even if we use a similar shorthand for
the terms with higher powers of $H$, their structure is less algorithmic, and specific contractions
are spelled out in the body of the paper where needed.

\vskip 0.3cm
\noindent
$\rhd$ {\bf Even-even sector ($t_8$ tensor):} A similar shorthand appears in the even-even
sector expressions:
\beq
t_8 t_8 R^4= t_{\mu_1\cdots\mu_8}t_{\nu_1\cdots\nu_8}
R^{\mu_1\mu_2}{}_{\nu_1\nu_2} R^{\mu_3\mu_4}{}_{\nu_3\nu_4}
R^{\mu_5\mu_6}{}_{\nu_5\nu_6}R^{\mu_7\mu_8}{}_{\nu_7\nu_8} \,.
\eeq
The $t_8$ tensor is antisymmetric within a pair of indices and is symmetric under exchange of
a pair of indices:
\beq
t_{\mu_1\mu_2 \cdots\mu_8} =  - t_{\mu_2 \mu_1 \mu_3\cdots\mu_8}  =  t_{\mu_3 \mu_4 \mu_1\mu_2 \mu_5  \cdots\mu_8}\, .
\eeq
This leads to the identity
\beq
t_8 t_8 R^4 (\omp) = t_8 t_8 R^4(\omm).
\eeq
The explicit expression for $t_8$ quartic in the metric is not very useful, but it can be defined via:
\beq
 t_8 M^4 =  24 \left(\tr M^4 - \ft14 (\tr M^2)^2\right),
\eeq
for any antisymmetric matrix $M$. It is also related to $\epsilon_8$ via some
$\Gamma$-matrix algebra:
\bea
\fft{1}{2^4}  \epsilon_{i_1 \cdots i_8}  R_{\mu_1\mu_2} \left(\Gamma^{\mu_1\mu_2}\right)^{i_1i_2}
\cdots R_{\mu_7\mu_8} \left(\Gamma^{\mu_7\mu_8}\right)^{i_7i_8}
&=& \fft 12  \epsilon^{\mu_1\cdots\mu_8} R_{\mu_1\mu_2} R_{\mu_3\mu_4} R_{\mu_5\mu_6}
R_{\mu_7\mu_8} \nonumber \\
&& +  \, t_8^{\mu_1\cdots\mu_8} R_{\mu_1\mu_2} R_{\mu_3\mu_4} R_{\mu_5\mu_6}
R_{\mu_7\mu_8}
\eea

Finally, there is a relation between $\epsilon_{10}\epsilon_{10}$ and $t_8 t_8$ structures:
\bea
\fft 18 \epsilon_{10} \epsilon_{10}R^4   &= & t_8 t_8 R^4  + 192 \,  \tr R_{\mu \nu} R_{\rho \lambda}  \tr R_{\mu \rho} R_{\nu \lambda}  \nonumber\\
 && - 768 \, \tr R^{\mu \nu \rho \lambda} R_{\mu}\,^{\sigma} \,_{\rho} \, ^{\delta} R_{\sigma} \,^{\iota} \, _{\nu} \,^{\kappa} R_{\delta \iota \lambda \kappa} + \mbox{Ricci terms} \,.
\eea
Note that the former does not contain any Ricci-like term. The internal contractions on the
four-index curvature tensor appear only inside $\epsilon_{10} \epsilon_{10}R^4$.

\section{T-duality (revisited) and CP-even terms}
\label{app:T-CP-even}

For the most part, and certainly ``conceptually",  we have been treating all $\mathcal O (\alpha'^3)$
corrections as a single object. However, the discussion of T-duality has been a notable exception;
the discussion of Section~\ref{sec:tdual} covers only the CP-odd part, i.e. the
even-odd + odd-even spin structure sectors. The reason the CP-even part has been relegated
to the poor relative status to be placed in an appendix is simple. The expressions are big,
cumbersome and fairly unenlightening. Nevertheless, for sake of completeness, if for nothing
else, here we do discuss of the CP-even terms, which consist of the even-even and odd-odd
sectors. To keep the expressions manageable, we shall just present in detail only the
six-dimensional four-derivative couplings here.  Similar ideas for using T-duality in the CP-even
sector to constrain the $B$-field and dilaton contributions to the ten-dimensional eight-derivative
Lagrangian were considered in \cite{Garousi:2012yr,Garousi:2012jp}.\footnote{In \cite{McOrist:2012yc} a combination of (different and multiple) dualities is argued to lead also to a dual for  $C_3 \wedge X_8$. Such couplings are beyond the scope of our present consideration. }

\subsection{Circle reduction of CP-even terms}

As shown in Section~\ref{sec:doublet}, the CP-odd anomalous couplings in a background with
an isometry transforms as a part of a doublet under T-duality. The six dimensional version of
the coupling (\ref{eq:reduu}) reads as:
\beq
\label{eq:reduu6}
\int_{M_{6}} B \wedge [X_4(\Omega^+) + X_4(\Omega^-)] =  \int_{M_5 }b_1\wedge(\tilde X_4-\tilde X_2\wedge T)\big|_{\mathrm{averaged}}
-h_3\wedge\tilde X_2\big|_{\mathrm{averaged}} \, .
\eeq
Here we shall focus on its completion:
\beq
2e^{-1}\delta\mathcal L_{\rm CP\mhyphen even}=e^{-1}\mathcal L_{\rm loop}^{\rm e\mhyphen e}
+e^{-1}\mathcal L_{\rm loop}^{\rm o\mhyphen o},
\label{eq:Lloopeeoo}
\eeq
where
\begin{eqnarray}
e^{-1}\mathcal L_{\rm loop}^{\rm e\mhyphen e}&=&t_4t_4R(\Omega_+)^2
 =  R_{\mu\nu\rho\sigma}(\Omega_+)^2,\nn\\
e^{-1}\mathcal L_{\rm loop}^{\rm o\mhyphen o}&=&-\ft18\epsilon_6\epsilon_6
\left(R(\Omega_+)^2+\ft43H^2R(\Omega_+)+\ft19H^4\right)\nn\\
&=& E_4+\ft23R(\Omega_+)H^2
+4R_{\mu\nu\rho\sigma}(\Omega_+)H^{\mu\rho a}H^{\nu\sigma a}
-4R_{\mu\nu}(\Omega_+)H^{2\,\mu\nu}+\ft19(H^2)^2-\ft23H^4 \,.
\end{eqnarray}
This is the form of the six-dimensional  four-derivative Lagrangian given in (\ref{eq:iiaapm}).

We shall now compactify on a circle and examine the properties of (\ref{eq:Lloopeeoo})
under T-duality.  We use the conventions of Section~\ref{sec:tdual} and begin by making
the observation that (\ref{eq:riemanns1}) may be written as
\begin{eqnarray}
R_{\alpha\beta}{}^{\gamma\delta}(\Omega_+)&=&R_{\alpha\beta}{}^{\gamma\delta}(\omega_+)
-\ft14(T_{+\alpha}{}^\gamma T_{+\beta}{}^\delta-T_{+\alpha}{}^\delta T_{+\beta}{}^{\gamma}
+T_{-\alpha\beta}T_-^{\gamma\delta})
-\ft14T_{+\alpha\beta}T_-^{\gamma\delta},\nonumber\\
R_{\alpha9}{}^{\gamma\delta}(\Omega_+)&=&-\ft12D_\alpha(\omega_+)
T_-^{\gamma\delta},\nonumber\\
R_{\alpha\beta}{}^{\gamma9}(\Omega_+)&=&-\ft12D^\gamma(\omega_-)T_{+\alpha\beta},
\nonumber\\
R_{\alpha9}{}^{\gamma9}(\Omega_+)&=&\ft14T_{+\alpha\delta}T_-^{\gamma\delta},
\label{eq:Rmred}
\end{eqnarray}
while (\ref{eq:riccis1}) may be written as
\begin{eqnarray}
R_{\alpha\beta}(\Omega_+)&=&R_{\alpha\beta}(\omega_+)-\ft14(T^2_{+\alpha\beta}
+T^2_{-\alpha\beta}),\nonumber\\
R_{\alpha9}(\Omega_+)&=&-\ft12D^\gamma(\omega_-)T_{+\gamma\alpha},\nonumber\\
R_{9\alpha}(\Omega_+)&=&-\ft12D^\gamma(\omega_+)T_{-\gamma\alpha},\nonumber\\
R_{99}(\Omega_+)&=&\ft14(T_+T_-).
\end{eqnarray}

We first examine $\mathcal L_{\rm loop}^{\rm e\mhyphen e}$.  The circle reduction proceeds
by taking
\begin{equation}
R_{ABCD}(\Omega_+)^2=R_{\alpha\beta\gamma\delta}(\Omega_+)^2
+2R_{\alpha9\gamma\delta}(\Omega_+)^2+2R_{\alpha\beta\gamma9}(\Omega_+)^2
+4R_{\alpha9\gamma9}(\Omega_+)^2.
\end{equation}
Using the Riemann reduction (\ref{eq:Rmred}), integrating by parts and simplifying, we find
\begin{eqnarray}
\label{eq:evevT}
e^{-1}\mathcal L_{\rm loop}^{\rm e\mhyphen e}&=&\ft12[R_{\alpha\beta\gamma\delta}(\omega_+)^2
+R_{\alpha\beta\gamma\delta}(\omega_-)^2]+[(D^\alpha(\omega_+)T_{-\alpha\lambda})^2
+(D^\alpha(\omega_-)T_{+\alpha\lambda})^2]\nonumber\\
&&-\ft14[R_{\alpha\beta\gamma\delta}(\omega_+)
T_+^{\alpha\beta}T_-^{\gamma\delta}+R_{\alpha\beta\gamma\delta}(\omega_-)
T_-^{\alpha\beta}T_+^{\gamma\delta}]\nonumber\\
&&-[R_{\alpha\beta}(\omega_+)T_-^{2\,\alpha\beta}
+R_{\alpha\beta}(\omega_-)T_+^{2\,\alpha\beta}]\nonumber\\
&&+\ft12h_{\alpha\beta\lambda}
[T_+^{\gamma\lambda}D_\gamma(\omega_-)T_+^{\alpha\beta}
-T_-^{\gamma\lambda}D_\gamma(\omega_+)T_-^{\alpha\beta}]
-\ft18[(h_{\alpha\beta\lambda}T_+^{\alpha\beta})^2+(h_{\alpha\beta\lambda}T_-^{\alpha\beta})^2]
\nonumber\\
&&+\ft14h_{\alpha\beta\lambda}h_{\gamma\delta}{}^\lambda[T_+^{\alpha\gamma}T_+^{\beta\delta}
+T_-^{\alpha\gamma}T_-^{\beta\delta}]-\ft18[(T_+^2)^2+(T_-^2)^2]+\ft38[(T^2_{+\mu\nu})^2
+(T^2_{-\mu\nu})^2]\nonumber\\
&&-\ft58T_{-\alpha\beta}T_+^{\beta\gamma}T_{-\gamma\delta}T_+^{\delta\alpha}
+\ft1{16}T_+^2T_-^2+\ft14T^2_{+\mu\nu}T_-^{2\,\mu\nu}+\ft7{16}(T_+T_-)^2\nonumber\\
&&+\ft1{16}(T_+T_-)[T_+^2+T_-^2]+\ft18[T^2_{+\mu\nu}+T^2_{-\mu\nu}]
T_+^{\mu\lambda}T_-^\nu{}_\lambda.
\end{eqnarray}
Note that we are ignoring all scalars in this circle reduction.

We may pause and draw some conclusions already. Since (\ref{eq:evevT}) contains the entire
even-even sector, it should by itself transform properly under T-duality.  Indeed, it is conveniently
written in terms of T-duality invariants $R (\omega_{\pm})$ and $T_+$ and an anti-invariant $T_-$.
Note that  (\ref{eq:evevT}) has only even combined powers of $h_3$ and $\tilde T$.  It  is naturally
decomposed as
\begin{equation}
\mathcal L_{\rm loop}^{\rm e\mhyphen e}=
\mathcal L_{\rm loop}^{\rm e\mhyphen e}\big|_{\rm inv}
+\mathcal L_{\rm loop}^{\rm e\mhyphen e}\big|_{\rm anti\mhyphen inv},
\end{equation}
where the first term contains even powers of $T_-$ and second contains odd powers.
As in (\ref{eq:reduu6}), it is a part of a T-duality doublet (the other part being
$\mathcal L_{\rm loop}^{\rm e\mhyphen e}\big|_{\rm inv}
-\mathcal L_{\rm loop}^{\rm e\mhyphen e}\big|_{\rm anti\mhyphen inv}$).
At the level of the three-point function (which we discuss below), we have
\begin{eqnarray}
e^{-1}\mathcal L_{\rm loop}^{\rm e\mhyphen e}\big|_{\rm inv}&=&
\ft12[R_{\alpha\beta\gamma\delta}(\omega_+)^2
+R_{\alpha\beta\gamma\delta}(\omega_-)^2]+\cdots,\nn\\
e^{-1}\mathcal L_{\rm loop}^{\rm e\mhyphen e}\big|_{\rm anti\mhyphen inv}&=&
-\ft14[R_{\alpha\beta\gamma\delta}(\omega_+)
T_+^{\alpha\beta}T_-^{\gamma\delta}+R_{\alpha\beta\gamma\delta}(\omega_-)
T_-^{\alpha\beta}T_+^{\gamma\delta}]+\cdots.
\label{eq:eeinvanti}
\end{eqnarray}

We now turn to the odd-odd spin structure sector.  From (\ref{eq:Lloopeeoo}), we see that
there are six terms to reduce.  In order to reduce $E_4$, we first obtain
\begin{eqnarray}
R_{ABCD}(\Omega_+)R^{CDAB}(\Omega_+)&=&R_{\alpha\beta\gamma\delta}(\omega_+)
R^{\alpha\beta\gamma\delta}(\omega_-)
-[R_{\alpha\beta\gamma\delta}(\omega_+)T_+^{\alpha\gamma}T_+^{\beta\delta}+
R_{\alpha\beta\gamma\delta}(\omega_-)T_-^{\alpha\gamma}T_-^{\beta\delta}]\nonumber\\
&&-\ft14[R_{\alpha\beta\gamma\delta}(\omega_+)
T_-^{\alpha\beta}T_+^{\gamma\delta}+R_{\alpha\beta\gamma\delta}(\omega_-)
T_+^{\alpha\beta}T_-^{\gamma\delta}]\nonumber\\
&&+D_\alpha(\omega_+)T_{-\beta\gamma}D^\alpha(\omega_-)T_+^{\beta\gamma}
-\ft18[(h_{\alpha\beta\lambda}T_+^{\alpha\beta})^2+(h_{\alpha\beta\lambda}T_-^{\alpha\beta})^2]
\nonumber\\
&&+\ft14h_{\alpha\beta\lambda}h_{\gamma\delta}{}^\lambda[T_+^{\alpha\gamma}T_+^{\beta\delta}
+T_-^{\alpha\gamma}T_-^{\beta\delta}]+\ft7{32}[(T_+^2)^2+(T_-^2)^2]-\ft3{16}(T_+T_-)^2
\nonumber\\
&&-\ft5{16}[(T^2_{+\mu\nu})^2+(T^2_{-\mu\nu})^2]+T_+T_-T_+T_-
+\ft18[T^2_{+\mu\nu}+T^2_{-\mu\nu}]T_+^{\mu\lambda}T_-^\nu{}_\lambda\nonumber\\
&&+\ft1{16}[T_+^2+T_-^2](T_+T_-),\nonumber\\
R_{AB}(\Omega_+)R^{BA}(\Omega_+)&=&R_{\alpha\beta}(\omega_+)R^{\beta\alpha}(\omega_+)
-\ft12R_{\alpha\beta}(\omega_+)[T_+^{2\,\alpha\beta}+T_-^{2\,\alpha\beta}]\nonumber\\
&&+\ft12D^\alpha(\omega_-)T_{+\alpha\lambda}D^\beta(\omega_+)T_{-\beta}{}^\lambda
+\ft1{16}[(T^2_{+\mu\nu})^2+(T^2_{-\mu\nu})^2]\nonumber\\
&&+\ft18T^2_{+\mu\nu}T_-^{2\,\mu\nu}+\ft1{16}(T_+T_-)^2,\nonumber\\
R(\Omega_+)R(\Omega_+)&=&R(\omega_+)^2-\ft12R(\omega_+)[T_+^2+T_-^2]
+\ft12R(\omega_+)(T_+T_-)\nonumber\\
&&+\ft1{16}[(T_+^2)^2+(T_-^2)^2]+\ft18T_+^2T_-^2+\ft1{16}(T_+T_-)^2
-\ft18[T_+^2+T_-^2](T_+T_-).\nonumber\\
\end{eqnarray}
This can be combined to give
\begin{eqnarray}
E_4&=&R_{\alpha\beta\gamma\delta}(\omega_+)R^{\alpha\beta\gamma\delta}(\omega_-)
-4R_{\alpha\beta}(\omega_+)R^{\beta\alpha}(\omega_+)+R(\omega_+)^2\nonumber\\
&&-[R_{\alpha\beta\gamma\delta}(\omega_+)T_+^{\alpha\gamma}T_+^{\beta\delta}+
R_{\alpha\beta\gamma\delta}(\omega_-)T_-^{\alpha\gamma}T_-^{\beta\delta}]
-\ft14[R_{\alpha\beta\gamma\delta}(\omega_+)
T_-^{\alpha\beta}T_+^{\gamma\delta}+R_{\alpha\beta\gamma\delta}(\omega_-)
T_+^{\alpha\beta}T_-^{\gamma\delta}]\nonumber\\
&&+2R_{\alpha\beta}(\omega_+)[T_+^{2\,\alpha\beta}+T_-^{2\,\alpha\beta}]
-\ft12R(\omega_+)[T_+^2+T_-^2]+\ft12R(\omega_+)(T_+T_-)\nonumber\\
&&+D_\alpha(\omega_+)T_{-\beta\gamma}D^\alpha(\omega_-)T_+^{\beta\gamma}
-2D^\alpha(\omega_-)T_{+\alpha\lambda}D^\beta(\omega_+)T_{-\beta}{}^\lambda
-\ft18[(h_{\alpha\beta\lambda}T_+^{\alpha\beta})^2+(h_{\alpha\beta\lambda}T_-^{\alpha\beta})^2]
\nonumber\\
&&+\ft14h_{\alpha\beta\lambda}h_{\gamma\delta}{}^\lambda[T_+^{\alpha\gamma}T_+^{\beta\delta}
+T_-^{\alpha\gamma}T_-^{\beta\delta}]+\ft9{32}[(T_+^2)^2+(T_-^2)^2]-\ft38(T_+T_-)^2
\nonumber\\
&&-\ft9{16}[(T^2_{+\mu\nu})^2+(T^2_{-\mu\nu})^2]+T_+T_-T_+T_-+\ft18T_+^2T_-^2
-\ft12T^2_{+\mu\nu}T_-^{2\,\mu\nu}+\ft18[T^2_{+\mu\nu}+T^2_{-\mu\nu}]
T_+^{\mu\lambda}T_-^\nu{}_\lambda\nonumber\\
&&-\ft1{16}[T_+^2+T_-^2](T_+T_-)
\end{eqnarray}

The additional terms in (\ref{eq:Lloopeeoo}) may be reduced as well.  We find
\begin{eqnarray}
R_{ABCD}(\Omega_+)H^{ACE}H^{BDE}&=&\ft12[R_{\alpha\beta\gamma\delta}(\omega_+)+
R_{\alpha\beta\gamma\delta}(\omega_-)]h^{\alpha\gamma\lambda}h^{\beta\delta}{}_\lambda
\nonumber\\
&&+\ft14[R_{\alpha\beta\gamma\delta}(\omega_+)T_+^{\alpha\gamma}T_+^{\beta\delta}+
R_{\alpha\beta\gamma\delta}(\omega_-)T_-^{\alpha\gamma}T_-^{\beta\delta}]\nonumber\\
&&-\ft14[R_{\alpha\beta\gamma\delta}(\omega_+)T_+^{\alpha\gamma}T_-^{\beta\delta}+
R_{\alpha\beta\gamma\delta}(\omega_-)T_-^{\alpha\gamma}T_+^{\beta\delta}]\nonumber\\
&&-\ft18[(h_{\alpha\beta\lambda}T_+^{\alpha\beta})^2+(h_{\alpha\beta\lambda}T_-^{\alpha\beta})^2]
-\ft14h_{\alpha\beta\lambda}h_{\gamma\delta}{}^\lambda T_+^{\alpha\gamma}T_-^{\beta\delta}
\nonumber\\
&&+\ft12h_{\alpha\beta\gamma}(T_+-T_-)^{\gamma\lambda}[D_\alpha(\omega_+)T_{-\beta\lambda}
+D_\alpha(\omega_-)T_{+\beta\lambda}]\nonumber\\
&&-\ft1{16}[(T_+^2)^2+(T_-^2)^2]+\ft1{16}[T_+^2+T_-^2](T_+T_-)-\ft14T_+T_-T_+T_-\nonumber\\
&&+\ft1{16}[(T^2_{+\mu\nu})^2+(T^2_{-\mu\nu})^2]
+\ft1{16}[T^2_{+\mu\nu}+T^2_{-\mu\nu}]T_+^{\mu\lambda}T_-^\nu{}_\lambda,\nonumber\\
R_{AB}(\Omega_+)H^{2\,AB}&=&R_{\alpha\beta}(\omega_+)h^{2\,\alpha\beta}
+\ft14(2R_{\alpha\beta}(\omega_+)-h^2_{\alpha\beta})[T_+^{2\,\alpha\beta}+T_-^{2\,\alpha\beta}]
\nonumber\\
&&-\ft12R_{\alpha\beta}(\omega_+)[T_+^{\alpha\lambda}T_-^\beta{}_\lambda
+T_-^{\alpha\lambda}T_+^\beta{}_\lambda]\nonumber\\
&&-\ft14h_{\alpha\beta\gamma}(T_+-T_-)^{\beta\gamma}
[D_\lambda(\omega_-)T_+^{\lambda\alpha}+D_\lambda(\omega+)T_-^{\lambda\alpha}]
\nonumber\\
&&-\ft18[(T^2_{+\mu\nu})^2+(T^2_{-\mu\nu})^2]-\ft14T^2_{+\mu\nu}T_-^{2\,\mu\nu}
+\ft14[T^2_{+\mu\nu}+T^2_{-\mu\nu}]T_+^{\mu\lambda}T_-^\nu{}_\lambda\nonumber\\
&&+\ft1{16}[T_+^2+T_-^2](T_+T_-)-\ft18(T_+T_-)^2,\nonumber\\
R(\Omega_+)H^2&=&R(\omega_+)h^2+\ft14(3R(\omega_+)-h^2)[T_+^2+T_-^2]
-\ft14(6R(\omega_+)-h^2)(T_+T_-)\nonumber\\
&&-\ft3{16}[(T_+^2)^2+(T_-^2)^2]-\ft38T_+^2T_-^2-\ft38(T_+T_-)^2
+\ft9{16}[T_+^2+T_-^2](T_+T_-),\nonumber\\
(H^2)^2&=&(h^2)^2+\ft32h^2[T_+^2+T_-^2]-3h^2(T_+T_-)+\ft9{16}[(T_+^2)^2+(T_-^2)^2]
\nonumber\\
&&+\ft98T_+^2T_-^2+\ft94(T_+T_-)^2-\ft94[T_+^2+T_-^2](T_+T_-),\nonumber\\
H^4&=&h^4+\ft32h_{\alpha\beta\lambda}h_{\gamma\delta}{}^\lambda
[T_+^{\alpha\gamma}T_+^{\beta\delta}+T_-^{\alpha\gamma}T_-^{\beta\delta}]
-3h_{\alpha\beta\lambda}h_{\gamma\delta}{}^\lambda T_+^{\alpha\gamma}T_-^{\beta\delta}
\nonumber\\
&&+\ft3{16}[(T^2_{+\mu\nu})^2+(T^2_{-\mu\nu})^2]+\ft34T^2_{+\mu\nu}T_-^{2\,\mu\nu}
-\ft34[T^2_{+\mu\nu}+T^2_{-\mu\nu}]T_+^{\mu\lambda}T_-^\nu{}_\lambda\nonumber\\
&&+\ft38T_+T_-T_+T_-.
\end{eqnarray}

Adding the above together, we find
\begin{eqnarray}
e^{-1}\mathcal L_{\rm loop}^{\rm o\mhyphen o}
&=&R_{\alpha\beta\gamma\delta}(\omega_+)R^{\alpha\beta\gamma\delta}(\omega_-)
-4R_{\alpha\beta}(\omega_+)R^{\beta\alpha}(\omega_+)+R(\omega_+)^2
+\ft23R(\omega_+)h^2\nonumber\\
&&+2[R_{\alpha\beta\gamma\delta}(\omega_+)+
R_{\alpha\beta\gamma\delta}(\omega_-)]h^{\alpha\gamma\lambda}h^{\beta\delta}{}_\lambda
-4R_{\alpha\beta}(\omega_+)h^{2\,\alpha\beta}+\ft14(h^2)^2-\ft23h^4\nonumber\\
&&-\ft14[R_{\alpha\beta\gamma\delta}(\omega_+)
T_-^{\alpha\beta}T_+^{\gamma\delta}+R_{\alpha\beta\gamma\delta}(\omega_-)
T_+^{\alpha\beta}T_-^{\gamma\delta}]\nonumber\\
&&-[R_{\alpha\beta\gamma\delta}(\omega_+)T_+^{\alpha\gamma}T_-^{\beta\delta}+
R_{\alpha\beta\gamma\delta}(\omega_-)T_-^{\alpha\gamma}T_+^{\beta\delta}]
-\ft12R(\omega_+)(T_+T_-)\nonumber\\
&&+2R_{\alpha\beta}(\omega_+)[T_+^{\alpha\lambda}T_-^\beta{}_\lambda
+T_-^{\alpha\lambda}T_+^\beta{}_\lambda]-\ft16h^2(T_+T_-)
+D_\alpha(\omega_+)T_{-\beta\gamma}D^\alpha(\omega_-)T_+^{\beta\gamma}\nonumber\\
&&-2D^\alpha(\omega_-)T_{+\alpha\lambda}D^\beta(\omega_+)T_{-\beta}{}^\lambda
+2h_{\alpha\beta\gamma}[T_+^{\gamma\lambda}D_\alpha(\omega_+)T_{-\beta\lambda}
-T_-^{\gamma\lambda}D_\alpha(\omega_-)T_{+\beta\lambda}]\nonumber\\
&&+h_{\alpha\beta\gamma}[T_+^{\beta\gamma}
D_\lambda(\omega+)T_-^{\lambda\alpha}-
T_-^{\beta\gamma}D_\lambda(\omega_-)T_+^{\lambda\alpha}]
-\ft18[(h_{\alpha\beta\lambda}T_+^{\alpha\beta})^2+(h_{\alpha\beta\lambda}T_-^{\alpha\beta})^2]
\nonumber\\
&&+\ft14h_{\alpha\beta\lambda}h_{\gamma\delta}{}^\lambda[T_+^{\alpha\gamma}T_+^{\beta\delta}
+T_-^{\alpha\gamma}T_-^{\beta\delta}]
+h_{\alpha\beta\lambda}h_{\gamma\delta}{}^\lambda T_+^{\alpha\gamma}T_-^{\beta\delta}
+\ft3{32}[(T_+^2)^2+(T_-^2)^2]\nonumber\\
&&-\ft18(T_+T_-)^2-\ft3{16}[(T^2_{+\mu\nu})^2+(T^2_{-\mu\nu})^2]
+\ft14T_+T_-T_+T_-\nonumber\\
&&-\ft18[T^2_{+\mu\nu}+T^2_{-\mu\nu}]T_+^{\mu\lambda}T_-^\nu{}_\lambda
+\ft1{16}[T_+^2+T_-^2](T_+T_-).
\end{eqnarray}
This complicated expression is again a sum of T-duality invariants and anti-invariants, and
hence the coupling is a doublet in the same sense as in the other sectors.  For later reference,
we note that
\begin{eqnarray}
e^{-1}\mathcal L_{\rm loop}^{\rm o\mhyphen o}\big|_{\rm inv}&=&
-\ft14[R_{\alpha\beta\gamma\delta}(\omega_+)
T_-^{\alpha\beta}T_+^{\gamma\delta}+R_{\alpha\beta\gamma\delta}(\omega_-)
T_+^{\alpha\beta}T_-^{\gamma\delta}]\nonumber\\
&&-[R_{\alpha\beta\gamma\delta}(\omega_+)T_+^{\alpha\gamma}T_-^{\beta\delta}+
R_{\alpha\beta\gamma\delta}(\omega_-)T_-^{\alpha\gamma}T_+^{\beta\delta}]+\cdots\nn\\
e^{-1}\mathcal L_{\rm loop}^{\rm o\mhyphen o}\big|_{\rm anti\mhyphen inv}&=&
R_{\alpha\beta\gamma\delta}(\omega_+)R^{\alpha\beta\gamma\delta}(\omega_-)+\cdots.
\label{eq:ooinvanti}
\end{eqnarray}
The reason the invariant and anti-invariant expressions are flipped is due to the GSO projection,
with IIA having $\mathcal L_{\rm loop}^{\rm o\mhyphen o}\sim -\fft18\epsilon\epsilon R^2$ and
IIB having $\mathcal L_{\rm loop}^{\rm o\mhyphen o}\sim +\fft18\epsilon\epsilon R^2$.

We now comment on the overall T-duality structure of the CP-even terms.  Since the even-even
and odd-odd contributions transform independently under T-duality, the expression
$(t_8t_8R^4\mp\ft18\epsilon_{10}\epsilon_{10}R^4)$ transforms properly under T-duality
for either relative sign.  We have been discussing only one-loop type IIA terms here, and
hence we have a relative minus between two terms.  Recall that eight-derivative couplings at
$\mathcal O(\alpha'^3)$ with a  relative $+$ (and an absence of the CP-odd term) appear in
the tree-level type IIA action, and in type IIB (both at tree-level and one-loop). The map
\begin{equation}
\delta\mathcal L_{\rm CP\mhyphen even}\big|_{\rm inv}
+\delta\mathcal L_{\rm CP\mhyphen even}\big|_{\rm anti\mhyphen inv}
\quad\leftrightarrow\quad
\delta\mathcal L_{\rm CP\mhyphen even}\big|_{\rm inv}
-\delta\mathcal L_{\rm CP\mhyphen even}\big|_{\rm anti\mhyphen inv}
\end{equation}
is the map between IIA and IIB variables. Let us also recall that the tree-level terms are weighted
by a T-duality invariant $\sqrt{-g} \exp(-2 \phi)$. For the one-loop terms we should weigh the parts
of the doublet by $R_{10}$ and $1/R_{10}$.  We will take a closer look at this below.

Finally lets us recall that the nine-dimensional theory with maximal supersymmetry has an
$SL(2, \mathbb{Z})$ R-symmetry, and its three one-forms form a singlet and a doublet
under that group. In (\ref{eq:reduu}), $b_1$ (which is identified locally with $B_{\mu 9} d x^{\mu}$
for IIA reduction and with $g_{\mu 9} d x^{\mu}$ for IIB) is the $SL(2, \mathbb{Z})$ singlet. Hence
the whole set of $\mathcal O(\alpha'^3)$ couplings discussed in this paper is also an R-symmetry
singlet. The $SL(2, \mathbb{Z})$ doublet part will have no CP-odd couplings and a relative $+$
between even-even and odd-odd contributions. Similarly, in eight dimensions the two different
factors in the U-duality group have different $\mathcal O(\alpha'^3)$ coupling associated with
them. The $SL(2, \mathbb{Z})$ part will have a doublet of anomalous CP-odd couplings and a
relative minus between even-even and odd-odd contributions, while the $SL(3, \mathbb{Z})$
part has no CP-odd coupling and a relative plus.%
\footnote{This distinction is there already for the perturbative $SL(2, \mathbb{Z}) \times
SL(2, \mathbb{Z})$ part of the U-duality group.}
For dimension seven and lower we lose the distinction between these two structures, essentially
between $J_0(\omp)$ and $J_1(\omp)$ defined in (\ref{superinv}), and the fact that the U-duality
group becomes semi-simple is consistent with the structure of the quantum corrections.

\subsection{A look at the string three-point function}

As we have seen above, the CP-even expressions split into T-duality invariants and anti-invariants.
We now examine how this is realized on the string world sheet.  For simplicity, we examine the
one-loop three-point function for Type II strings compactified on $K3\times S^1$.
We start with the even-even spin structure amplitude
\begin{equation}
\mathcal A_{\rm e\mhyphen e}\sim\theta_{\mu_1\nu_1}\theta_{\mu_2\nu_2}\theta_{\mu_3\nu_3}
\times
{[i\partial X^{\mu_1}+\ft{\alpha'}2k_1\cdot\psi\psi^{\mu_1}]\atop
[i\bar\partial X^{\nu_1}+\ft{\alpha'}2k_1\cdot\tilde\psi\tilde\psi^{\nu_1}]}
{[i\partial X^{\mu_2}+\ft{\alpha'}2k_2\cdot\psi\psi^{\mu_2}]\atop
[i\bar\partial X^{\nu_2}+\ft{\alpha'}2k_2\cdot\tilde\psi\tilde\psi^{\nu_2}]}
{[i\partial X^{\mu_3}+\ft{\alpha'}2k_3\cdot\psi\psi^{\mu_3}]\atop
[i\bar\partial X^{\nu_3}+\ft{\alpha'}2k_3\cdot\tilde\psi\tilde\psi^{\nu_3}]}.
\end{equation}
In order to get a non-vanishing contribution at the four-derivative level, we need to take two fermion
pairs from the left-movers and two fermion pairs from the right-moves.  Since this already brings
in four momenta factors, the remaining factor must be a bosonic zero mode contraction.  In
other words, we have schematically
\begin{equation}
\mathcal A\sim\langle\partial X\bar\partial X\rangle\langle k\cdot\psi\psi k\cdot\psi\psi\rangle
\langle k\cdot\tilde\psi\tilde\psi k\cdot\tilde\psi\tilde\psi\rangle.
\end{equation}

For three gravitons, we end up with the usual expression (this time in five dimensions)
\begin{equation}
\mathcal A_{h\mhyphen h\mhyphen h}\sim t_4t_4R^2.
\end{equation}
This expression has the same form at both small and large radius.  It is easy to see that
the amplitude for an odd number of vectors vanish.  Hence the only remaining amplitude
to consider is that of one graviton and two vectors.  If we let
\begin{equation}
g_{\mu9}=g_{9\mu}=A_\mu,\qquad B_{\mu9}=-B_{9\mu}=\tilde A_\mu,\qquad
A_\mu^\pm=A_\mu\pm\tilde A_\mu,
\end{equation}
the the polarization tensors for the vectors have the form
\begin{equation}
\theta_{\mu9}=A^+_\mu,\qquad\theta_{9\mu}=A^-_\mu.
\end{equation}
We now consider
\begin{eqnarray}
\mathcal A_{h\mhyphen A^+\mhyphen A^+}&\sim&\langle\partial X\bar\partial X\rangle
\langle k\cdot\psi\psi k\cdot\psi\psi\rangle
\langle k_2\cdot\tilde\psi\tilde\psi^9 k_3\cdot\tilde\psi\tilde\psi^9\rangle\nonumber\\
&\sim&-k_2\cdot k_3\langle\partial X\bar\partial X\rangle
\langle k\cdot\psi\psi k\cdot\psi\psi\rangle
\langle \tilde\psi\tilde\psi\rangle\langle\tilde\psi^9\tilde\psi^9\rangle,
\end{eqnarray}
where we have labeled the circle direction explicitly.  Since on-shell three-particle kinematics
gives $k_i\cdot k_j=0$, we see that this amplitude vanishes.  There is, however, a
non-vanishing amplitude for
\begin{equation}
\mathcal A_{h\mhyphen A^+\mhyphen A^-}\sim\langle\partial X^9\bar\partial X^9\rangle
\langle k\cdot\psi\psi k\cdot\psi\psi\rangle
\langle k\cdot\tilde\psi\tilde\psi k\cdot\tilde\psi\tilde\psi\rangle.
\end{equation}
This involves a bosonic zero mode contraction on the circle
\begin{equation}
\langle\partial X^9\bar\partial X^9\rangle=-\sigma\fft{\alpha'}{8\pi\tau_2},
\qquad\mbox{where}\qquad\sigma=\cases{
+1&$R\to\infty$,\cr
-1&$R\to0$.}
\end{equation}

Putting the $h\mhyphen h\mhyphen h$ and $h\mhyphen A^+\mhyphen A^-$ amplitudes together,
we find
\begin{equation}
\mathcal A_{\rm e\mhyphen e}\sim t_4t_4R^2+\sigma R_{\mu\nu\rho\sigma}
T_+^{\mu\nu}T_-^{\rho\sigma},
\end{equation}
where $T^\pm=dA^\pm$.  (We are ignoring precise numerical factors here.)  To be explicit
(and in a schematic notation), we have, for the even-even spin structure
\begin{center}
\begin{tabular}{ll}
IIA&IIB\\
\hline
$R^A\to\infty:\quad t_4t_4R^2+RT^A_+T^A_-$&
$R^B\to\infty:\quad t_4t_4R^2+RT^B_+T^B_-$\\
$R^A\to0:~\,\quad t_4t_4R^2-RT^A_+T^A_-$&
$R^B\to0:~\,\quad t_4t_4R^2-RT^B_+T^B_-$\\
\end{tabular}
\end{center}
We see that this is T-duality covariant under the relation
\begin{equation}
T_+^A=T_+^B,\qquad T_-^A=-T_-^b,\qquad R^A=\alpha'/R^B.
\end{equation}
In particular, $t_4t_4R^2$ is invariant, and $RT_+T_-$ is anti-invariant, in agreement with
(\ref{eq:eeinvanti}).

Turning now to the odd-odd spin structure, we put the first vertex operator in the $(-1,-1)$
picture and write
\begin{equation}
\mathcal A_{\rm o\mhyphen o}\sim\theta_{\mu_1\nu_1}\theta_{\mu_2\nu_2}\theta_{\mu_3\nu_3}
\times
{[\psi\cdot\partial X]\psi^{\mu_1}\atop
[\tilde\psi\cdot\bar\partial X]\tilde\psi^{\nu_1}}
{[i\partial X^{\mu_2}+\ft{\alpha'}2k_2\cdot\psi\psi^{\mu_2}]\atop
[i\bar\partial X^{\nu_2}+\ft{\alpha'}2k_2\cdot\tilde\psi\tilde\psi^{\nu_2}]}
{[i\partial X^{\mu_3}+\ft{\alpha'}2k_3\cdot\psi\psi^{\mu_3}]\atop
[i\bar\partial X^{\nu_3}+\ft{\alpha'}2k_3\cdot\tilde\psi\tilde\psi^{\nu_3}]}.
\end{equation}
(The $\psi\cdot\partial X$ is a picture changing operator insertion.)  Here we have to soak up
six fermion zero modes on each side of the string, and the first non-vanishing contribution
will have four momentum factors and will contain $\epsilon\epsilon$.  For three graviton
scattering in five dimensions, the $\psi^9$ zero mode can only come from the PCO.  As a
result, we find schematically
\begin{equation}
\mathcal A_{h\mhyphen h\mhyphen h}\sim\sigma\epsilon\epsilon R^2,
\end{equation}
where the sign factor $\sigma=\pm1$ comes from the $\langle\partial X^9\bar\partial X^9\rangle$
zero mode contraction between the two PCO's.

For one graviton and two vectors, the circle directions must be on opposite sides of the string
(otherwise we would have two 9's on a single $\epsilon$).  This furthermore ensures that the
PCO's will not involve any $\partial X^9$ factors.  Hence we find
\begin{equation}
\mathcal A_{h\mhyphen A^+\mhyphen A^-}\sim h_{\mu_1\nu_1}A_{\mu_2}^+A_{\nu_3}^-
\epsilon^{\mu_1k_2\mu_2k_3\alpha9}\epsilon^{\nu_1k_2k_3\nu_3}{}_{\alpha9}
\sim h_{\mu_1\nu_1}A_{\mu_2}^+A_{\nu_3}^-
k_2^{\mu_1}k_3^{\nu_1}k_3^{\mu_2}k_2^{\nu_3}.
\end{equation}
This has the same structure as the even-even contribution, except that it does not have the
$\sigma$ sign factor.

We now see that the odd-odd spin structure contributes
\begin{equation}
\mathcal A_{o-o}\sim\sigma\epsilon\epsilon R^2
+R_{\mu\nu\rho\sigma}T_+^{\mu\nu}T_-^{\rho\sigma}.
\end{equation}
In order to see the implications for the IIA and IIB strings, we must take the different GSO
projections into account.  In particular, this gives an extra sign for the IIB string.  As a result,
we have
\begin{center}
\begin{tabular}{ll}
IIA&IIB\\
\hline
$R^A\to\infty:\quad~~~\>\epsilon\epsilon R^2+RT^A_+T^A_-$&
$R^B\to\infty:\quad-\epsilon\epsilon R^2-RT^B_+T^B_-$\\
$R^A\to0:~\,\quad-\epsilon\epsilon R^2+RT^A_+T^A_-$&
$R^B\to0:~\,\quad~~\,\epsilon\epsilon R^2-RT^B_+T^B_-$\\
\end{tabular}
\end{center}
which we can verify is properly T-duality covariant.  (Recall, also, that both $t_4t_4R^2$ and
$\epsilon\epsilon R^2$ are identical at the Riemann-squared level.  We are also ignoring
the five-dimensional antisymmetric tensor field.)  This agrees with the odd-odd expresion
(\ref{eq:ooinvanti}).

Finally, in addition to providing a consistency check of T-duality, this calculation demonstrates that,
at the three-point function level, the vector fields must enter in the $T_+T_-$ combination,
and not as $T_+T_++T_-T_-$.


\end{document}